\documentclass[fleqn,usenatbib]{mnras}

\usepackage[T1]{fontenc}
\usepackage[spanish, english]{babel,varioref}
\usepackage{graphicx}	
\usepackage{amsmath}	
\usepackage{amssymb}	
\usepackage{multicol}        
\usepackage{bm}		
\usepackage{pdflscape}	
\usepackage{ulem} 
\usepackage{tabularx}
\usepackage{longtable}
\usepackage{float}
\usepackage[flushleft]{threeparttable}
\usepackage{adjustbox}
\usepackage{subfig}
\newcommand{\dens}{$n_{\rm e}$}
\newcommand{\temp}{$T_{\rm e}$}
\newcommand{\tempnii}{$T_{\rm e}$([\ion{N}{ii}])}
\newcommand{\tempoiii}{$T_{\rm e}$([\ion{O}{iii}])}

\newcommand{\hii}{\ion{H}{ii} }

\newcommand{\Lam}{$\lambda$}

\usepackage[T1]{fontenc}
\usepackage{ae,aecompl}

\usepackage{newtxtext,newtxmath}
\usepackage{xspace}

\title[ICFs for giant HII regions]{Ionization correction factors and dust depletion patterns in giant \hii regions
}

\author[Amayo et al.]{A. Amayo,$^{1}$\thanks{Contact e-mail: \href{mailto:amedina, gdelgado@astro.unam.mx}{amedina, gdelgado@astro.unam.mx}}
G. Delgado-Inglada,$^{1}$
G. Stasi\'nska$^{2}$
\\
$^{1}$Instituto de Astronom\'ia, Universidad Nacional Aut\'onoma de M\'exico, Ap. 70-264, 04510, CDMX, Mexico\\
$^{2}$LUTH, Observatoire de Paris, CNRS, Universit\'e Paris Diderot; Place Jules Janssen, F-92190 Meudon, France\\
}

\date{Last updated ...; in original form ...}
\pubyear{2020}

\begin{document}
\label{firstpage}
\pagerange{\pageref{firstpage}--\pageref{lastpage}}
\maketitle

\begin{abstract}
We provide new ionization correction factors (ICFs) for carbon, nitrogen, neon, sulfur, chlorine, and argon in giant \hii regions. The ICFs were computed using the most representative photoionization models from a large initial grid. The models were selected using an observational sample of 985 giant \hii regions (GHR)
in spiral galaxies and blue compact galaxies (BCG). The observational sample was also used to assign a weight to each model describing how well it agrees with observations in the [\ion{O}{iii}]/H$\beta$ versus [\ion{N}{ii}]/H$\alpha$ diagram. In addition to the ICFs we provide, for the first time,
analytical expressions for their formal uncertainties. We use our ICFs to compute the abundances of nitrogen, neon, sulfur, and argon in our samples. Our abundances are robust within the adopted framework, but may require revision in the case of important changes in atomic data or in the spectral energy distribution of the ionizing radiation in \hii regions. 
Considering the abundance patterns we obtained for the BCG sample (abundances for the GHR sample are less reliable) we find that oxygen is depleted  into dust grains at a rate increasing with metallicity and reaching 0.12 dex at solar abundances. The discussion of possible depletion of sulfur and argon requires considering recent Type Ia Supernova yields, which are still uncertain.
\end{abstract}




\section{Introduction}

\hii regions are clouds of gas ionized by the ultraviolet radiation of massive O-B-type stars. From the emission line spectrum of \hii regions one can determine their chemical composition which traces the present-day abundances in the gas-phase of the interstellar medium (ISM) in galaxies. After correction for depletion of some elements into dust grains, the study of chemical abundances in \hii regions allows one to determine radial variation of abundance ratios across galaxies, constrain galactic chemical evolution models and test galactic formation scenarios \citep{Pagel2009, Esteban2015, Carigi2019}. 

The total abundance of a particular element in the gaseous phase is computed by adding up the ionic abundances of all the ions. When some of the ions are not observed, either because they have very weak emission lines or because their lines are emitted in an unobserved part of the spectrum, ionization correction factors (ICFs) are needed. The first ICFs for \hii regions were proposed based on similarities between the ionization potentials of ions \citep[e.g.,][]{Peimbert&Costero1969,PeimbertTorres1977}. Such ICFs may not be adequate in some cases since the ionization structure does not depend only on ionization potentials but also on the physical processes inside the nebula and on the SED of the ionizing radiation field. The ICFs derived from photoionization models are, in principle, more suitable since they include all the physics involved in the ionized nebulae \citep[e.g.,][]{Stasinska1978, Mathis1985, Kingsburgh1994, Izotov2006, PerezMontero2007}. However, they depend on the atomic physics and the input assumptions (geometry, ionizing source, etc.).

The paper is organized as follows. In Sect. \ref{sec:obs} we present the observational data that we have chosen a) to help defining the photoionization model grid, b) to test our ICFs, and c) to infer general depletion patterns occurring in \hii regions. In Sect.\ref{sec:models} we present the grid of photoionization models used to compute our ICFs. In Sect. \ref{sec:ICFs} we give polynomial expressions for the ICFs proposed for carbon, nitrogen, neon, sulfur, chlorine, and argon and their uncertainties. In Sect. \ref{sec:abunds} we compute the abundances for the objects in our observational samples as well as the associated uncertainties, including uncertainties estimated for the ICFs. 
In Sect. \ref{sec:discussion} we discuss the resulting gas-phase abundance patterns and evaluate the validity of our ICFs, when possible. In Sect. \ref{sec:depletion} we present the depletion patterns deduced from our study. A summary is given in Sect. \ref{sec:conclusions}.

\section{Observational sample}
\label{sec:obs}

Our observational sample consists of 985 giant \hii regions comprising giant \hii regions in spiral galaxies (the GHR sample) and giant \hii regions in blue compact galaxies (the BCG sample).
The sample is first used to select a grid of photoionization models that best represent the observations (Section~\ref{sec:models}) and to assign a weight to the models (Section~\ref{sec:weights}) when using them to compute the ICFs.

The BCG sample consists of 140 low metallicity blue compact galaxies from which 108 were taken from \citet{Izotov2007} and 32 were added by him throughout the years. The sample from \citet{Izotov2007} contains 93 spectra observed by \citet{Izotov2004} with the Kitt Peak 4m-telescope (with an spectral resolution of $\sim$7 {\AA}) and 15 objects from the DR5 of the Sloan Digital Sky Server (SDSS, with an spectral resolution between $\sim$2 and $\sim$5 {\AA}). The objects from the SDSS were selected by \citet{Izotov2007} using criteria that take into account the signal-to-noise ratio, the H$\beta$ equivalent width, and the H$\beta$ flux. The spectral range covered by the observations is $\sim$3500--9200 {\AA} ensuring that the emission lines needed to compute the physical conditions and ionic abundances are observed in most of the cases. Although \citet{Izotov2007} reported the intensities of [\ion{O}{iii}] $\lambda$5007, we did not use these lines  because they likely have saturation problems. We used the  [\ion{O}{iii}] $\lambda$4959 lines instead as they give the same information without being affected by saturation.

The GHR sample contains 845 giant \hii regions from the compilation by \citet{Vale2016}, hereafter VA16,
to which we added the observations by \citet{Croxall2015, Croxall2016}. 
Most of the observations gathered by VA16 were made with the 8m-Very Large Telescope (VLT) and the 10m-Keck Telescope, while the rest were made with medium to big telescopes, ranging from the 4m-Kitt Peak telescope up to the 8.2m-Subaru array telescopes.
In cases where the intensity of H$\alpha$ was not available, VA16 obtained it using the theoretical ratio H$\alpha$/H$\beta$ $=$ 2.86. From the sample gathered by VA16 we excluded seven objects whose positions in the [\ion{O}{iii}]/H$\beta$ versus [\ion{N}{ii}]/H$\alpha$ BPT diagram \citep{Baldwin} indicate that  they are likely not ionized by recently born stars. These objects are NGC\,300\,7, NGC\,0598\,B90, NGC\,4395-003-003, NGC\,5236\,p5, NGC\,5236\,p36, M31\,H7 and NGC\,1512\,15 \citep[observed by][respectively]{BresolinGieren2009, Bresolin2010, vanZee1998, BresolinRyan2009, Zurita2012, Bresolin2012}.

The observations from Croxall et al. were all made with the Multi Object Double Spectrographs on the 8.4m-Large Binocular Telescope Observatory. Following the suggestions of the authors, we removed objects that show signatures of extra ionizing sources in their spectra, such as shock ionization and/or supernovae remnants features.
These objects are: NGC\,5194+2.5+9.5, NGC\,5194-6.9+20.8, NGC\,5194+30.2+2.2, NGC\,5194+13.3-141.3, NGC\,5457-250.8-52.0, NGC\,5457+650.1+270.7, NGC\,5457+299.1+464.0, NGC\,5457-345.5+273.8. Although \citet{Croxall2016} also suggest to remove NGC\,5457+44.7+153.7 due to its location on the [\ion{S}{ii}]/H$\alpha$ and [\ion{O}{i}]/H$\alpha$ BPT diagrams, we find that this object is located in the same zone as the rest of the star-forming galaxies on both diagrams, so we did not remove this object. We did not use lines for which only an upper limit on their intensity is available.

Our observational sample, consisting of good quality data and covering a wide range of degrees of ionization and metallicities is large enough to be considered as suitable for the purpose of our paper.
In all cases we used the dereddened line intensities as provided by the original authors, to be compared to those obtained from the photoionization models.

\section{Photoionization models}
\label{sec:models}

Our initial grid of models contains 31500 photoionization models available in the Mexican Million Models database (3MdB\footnote{   \href{https://sites.google.com/site/mexicanmillionmodels}{https://sites.google.com/site/mexicanmillionmodels}}, \citealt{Morisset2015}) under the ``BOND'' reference. This grid was presented in detail by VA16 and was constructed with the photoionization code {\sc{CLOUDY}} v.13.03 \citep{Ferland} for the purpose of calculating oxygen and nitrogen abundances in giant \hii\ regions using a Bayesian method. The grid available in \citet{Morisset2015}, in addition to the ionization-bounded models of VA16, also contains density-bounded models. The latter are computed by cutting the radiation bounded models by different fractions of their H$\beta$ intensity (from 10\% to 100\% of the total value).

The initial grid covers a wide range of physical and chemical parameters: 1) the values of 12$+\log$(O/H) and $\log$(N/O) range from 6.6 to 9.4 and from -2.0 to 0.0, respectively, 2) the ionization parameter ($\log U$) ranges from -4.0 to -1.0, 3) the nebular geometry is either a hollow sphere or a filled sphere. 

Each nebular model is ionized by the radiation from an instantaneous burst of star formation of given age, obtained using the stellar population models of \citealt{Molla2009}  with a \citet{Chabrier2003} stellar initial mass function. The nebular and stellar metallicities are  matched  through interpolation. Note that in many giant \hii regions, the stellar initial mass function may not be fully sampled in the ionizing cluster, so that the SEDs obtained from `analytic' stellar population synthesis may not be fully appropriate \citep{Cervino2000, Jamet2004} irrespective of the care taken to realistically model the stellar evolution and stellar atmospheres. Stochastic stellar population synthesis codes do exist \citep{CervinoLuridiana2004, dasilvafumagalli2012} but applying them in the framework of the present study would require a much more complex approach, not attempted here.

The photoionization models include dust, with a dust-to-gas ratio scaled to the oxygen abundance, following the relations proposed by \citet{Remy2014} and \citet{Draine2011}. 

\subsection{Photoionization model selection}
\label{sec:modelsel}

To the initial grid of models we applied a few filters in order to end up with a subgrid that is representative of our sample of observed \hii regions.

\begin{enumerate}
\item We selected models with starburst ages lower than 6 Myr. The reason is that at this age the available ionizing photons from the burst have decreased up to a factor of $\sim$10 \citep{Molla2009} which in general makes the \hii region too faint to properly observe the weak emission lines.
\footnote{Models of stellar populations including binaries stars such as those of \citet{Eldridgeal2017}
actually extend the period of time when massive stars ionize the surrounding gas with respect to the stellar populations used here. However, the age limit in our paper is not to be taken literally since what primarily matters is the strength of the ionizing radiation. 
Of course, the details of the SED matter also but, since we cover  a range of ages, we cover by the same token a range of SEDs as well, although this aspect would require more dedication in the future.}

\item We selected only those density-bounded models obtained by clipping the radiation-bounded models at about 70\% of their H$\beta$ intensity. Models more optically thin than this were removed since they likely do not represent the bulk of observations considered in this kind of study. All the radiation bounded models that satisfy filter (i) are included. 

\item Previous studies of giant \hii\ regions have shown that they follow some gross relations between O/H and N/O  \citep[e.g.,][]{Pilyugin2012}, U and O/H \citep[e.g.,][]{PerezMontero2007} and that they form a thin sequence in the BPT diagram \citep[e.g.,][]{McCall1985}. We therefore implemented criteria inspired by those of VA16 to define their `fake observational sample' but slightly modified to better represent our observational sample.  The expressions we have used are the following:

\begin{equation}
\begin{split}
&\rm log(N/O)= -1.093, \mbox{  for } Z \leq 7.93,\\
&\rm log(N/O)= 1.489Z - 12.896, \mbox{  for } Z \geq 7.93,\\
&\rm log(N/O)= -1.693, \mbox{  for } Z \leq 8.25,\\
&\rm log(N/O)= 1.489Z - 14.900, \mbox{  for } Z \geq 8.25,
\end{split} \label{eq:NewIntsP12}
\end{equation}

\begin{equation}
\begin{split}
&\rm log(U) = 5 - 1.25Z,\\
&\rm log(U) = 10 - 1.25Z,
\end{split} \label{eq:NewIntsPM}
\end{equation}

where $Z$ = 12 + $\log$(O/H) and, 

\begin{equation}
\begin{split}
    &y_{\rm low} = [a_{\rm low}\tanh{(b_{\rm low})} - c_{\rm low}]- 0.70\\
    &y_{\rm up} = [a_{\rm up}\tanh{(b_{\rm up})} - c_{\rm up}] + 0.10\\
\end{split} \label{eq:StasInt}
\end{equation}

where a$_{\rm low} = -30.79 + 1.14(y + 0.60) + 0.27(y + 0.60)^{2}$, $b_{\rm low} = 5.74(y + 0.60)$, $c_{\rm low} = c_{\rm up} = 31.09$, $a_{\rm up} = -30.79 + 1.14(y - 0.30) + 0.27(y - 0.30)^{2}$, $b_{\rm up} = 5.74(y - 0.30)$, $y = \log$(\ion{N}{ii} \Lam6284/ H$\alpha$). 
\end{enumerate}

\begin{figure*}
\includegraphics[trim=0.0cm 0.4cm 0cm 0cm, width=1.0\textwidth]{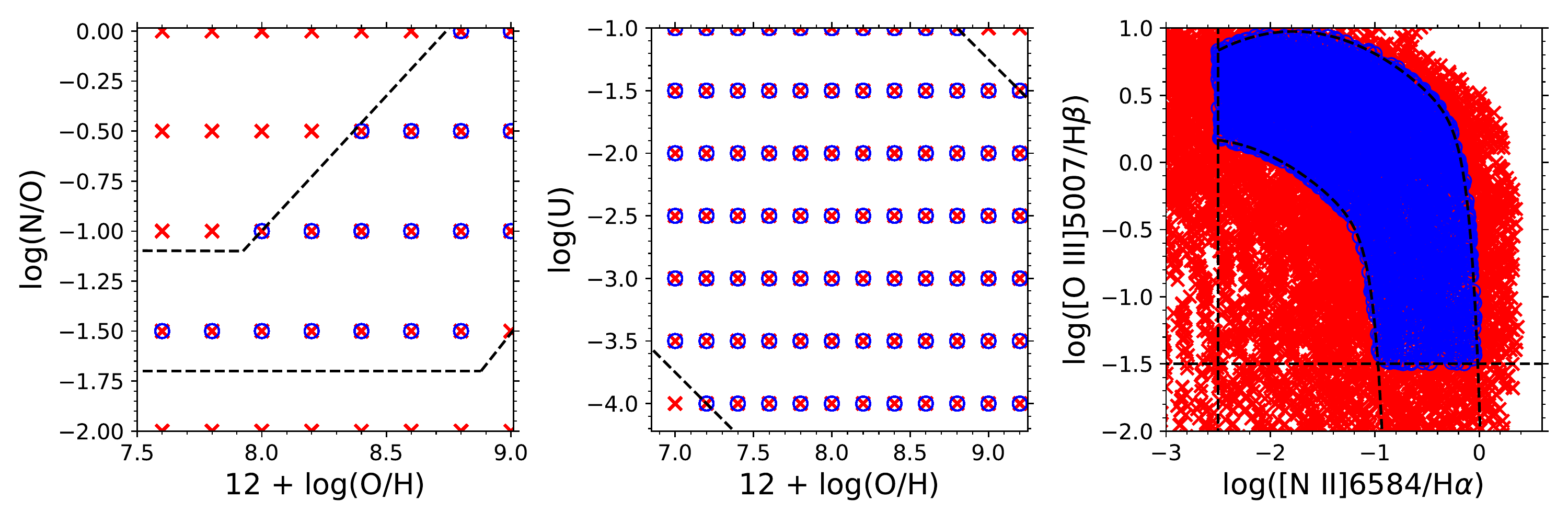}
\caption{The three relations used to select the final sample of photoionization models used to derive ICFs: $\log$(N/O) as a function of $\log$(O/H), $\log$U as a function of $\log$(O/H), and $\log$([\ion{O}{iii}]/H$\beta$) as a function of $\log$([\ion{N}{ii}]/H$\alpha$). The sample obtained after applying filters i) and ii) are shown with red crosses and the models selected after applying also filter iii) are shown in blue circles. Dashed lines delimit the selected models (see the text for more details).}
\label{fig:filters}
\end{figure*}

\begin{figure}
\centering
\includegraphics[trim=0.8cm 0.4cm 0.8cm 0.8cm, clip=true,width=\columnwidth]{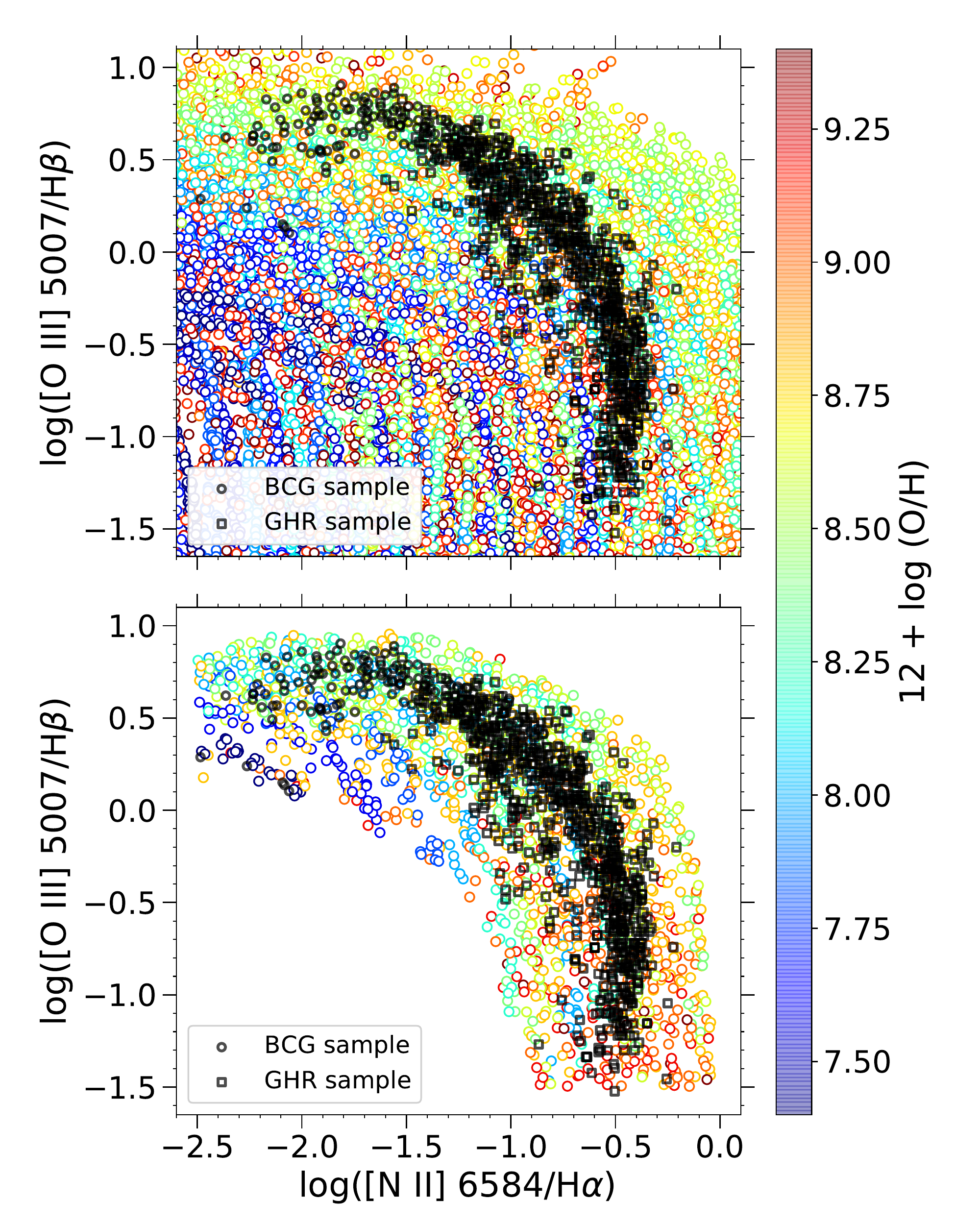}
\caption{BPT diagrams of the photoionization models (shown in empty colored circles) compared to the BCG and GHR samples (black circles and black squares, respectively). Upper panel shows the complete grid of models available in the 3MdB database and bottom panel shows the final set of models used in this work, applying the filters i), ii) and iii) to the original grid of models. The color bar shows the metallicity of the models in the form 12$+\log$(O/H).}
\label{fig:BPTevol}
\end{figure}

Figure~\ref{fig:filters} shows the models resulting from applying filters i) and ii) (red crosses) and the models obtained after applying also Equations~\ref{eq:NewIntsP12}--\ref{eq:StasInt} of filter iii) (blue circles). Figure~\ref{fig:BPTevol} shows the BPT diagram for the photoionization models (in colored circles, where the color is related to the oxygen abundance) and the observational sample (the BCG sample in black circles and the GHR sample in black empty squares). The upper panel shows the initial sample of $\sim$31000 photoionization models. The panel on the bottom shows the final sample of 1887 models obtained after applying all our filters. It can be seen that our final sample of models covers successfully the observed \hii regions, and thus seems adequate for the ICFs calculation.

\begin{figure*}
\includegraphics[width=0.8\textwidth]{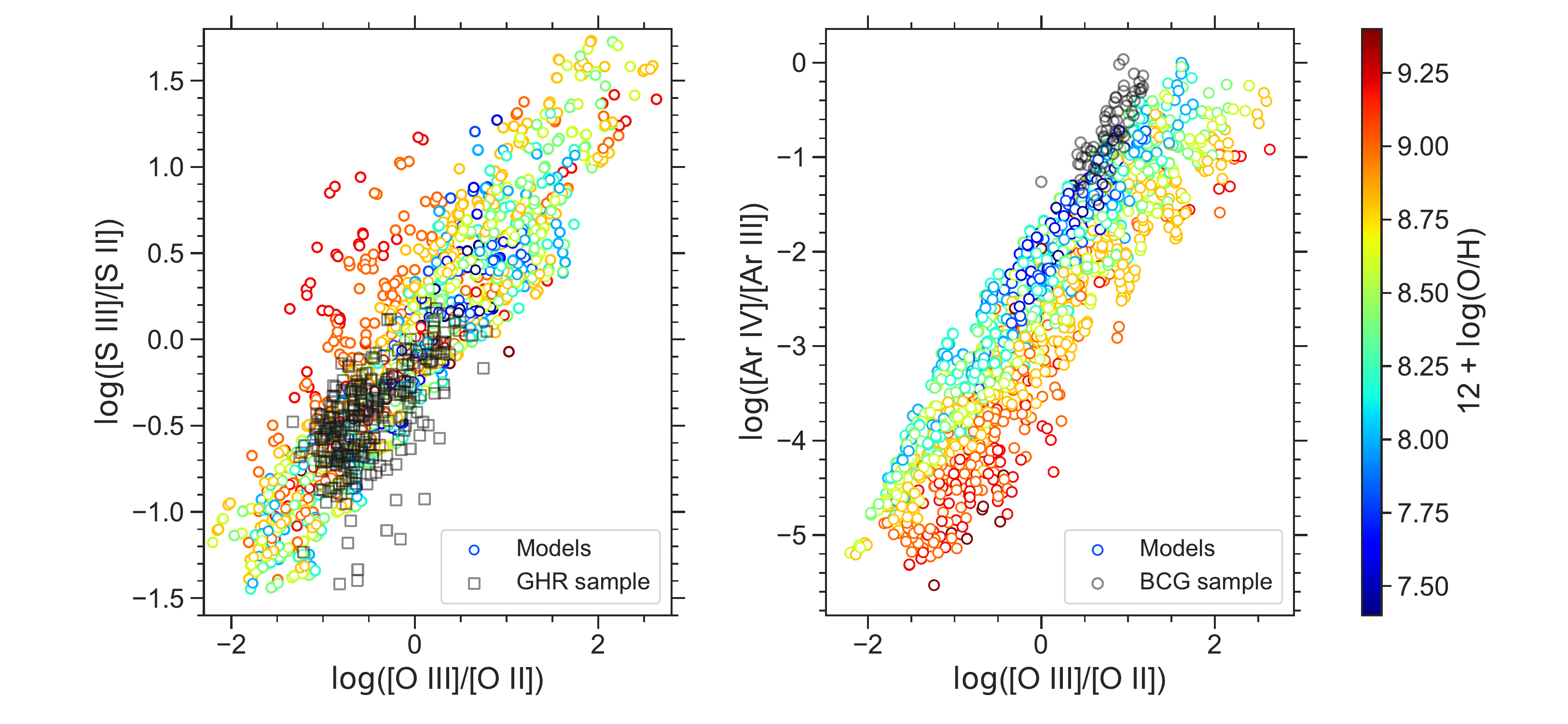}
\caption{Values of $\log$([\ion{S}{iii}] $\lambda$9069/[\ion{S}{ii}] $\lambda\lambda$6716+31) and $\log$([Ar IV] $\lambda$4740/[Ar III] $\lambda$7135) as a function of $\log($[\ion{O}{iii}] $\lambda$5007/[\ion{O}{ii}] $\lambda$3727) for the photoionization models (colored circles), the GHR sample (black squares on the left figure), and the BCG sample (black circles on the right figure). The color bar runs from low to high metallicity, in the form 12$+\log$(O/H).}
\label{fig:SArcomp}
\end{figure*}

As a last check, we have also compared the final grid of models with our observational sample using line-ratio diagrams which tell about the ionization structure of different elements. We present in Figure~\ref{fig:SArcomp} the values of $\log$([\ion{S}{iii}] $\lambda$9069/[\ion{S}{ii}] $\lambda\lambda$6716+31)\footnote{We do not use the [\ion{S}{iii}] $\lambda$9532 line because it is often prone to contamination by sky lines and, although it is intrinsically stronger that the [\ion{S}{iii}] $\lambda$9069 line, it may lead to quite erroneous results. The problem is much less severe for  [\ion{S}{iii}] $\lambda$9069.}as a function of $\log($[\ion{O}{iii}] $\lambda$5007/[\ion{O}{ii}] $\lambda$3727) (left panel) and the values of $\log$([\ion{Ar}{iv}] $\lambda$4740/[\ion{Ar}{iii}] $\lambda$7135) as a function of $\log($[\ion{O}{iii}] $\lambda$5007/[\ion{O}{ii}] $\lambda$3727) (right panel). 

For the first panel, we can compare our models only with the GHR sample since [\ion{S}{iii}] $\lambda$9069 is outside the observed wavelengths in the BCG sample. We see that, in the observed range of [\ion{O}{iii}]/[\ion{O}{ii}] values, the models do not cover perfectly well the observed  [\ion{S}{iii}]/[\ion{S}{ii}] ratios. A few observations are  well below the loci of the models, and the whole distribution of the observed [\ion{S}{iii}]/[\ion{S}{ii}] values is slightly shifted downwards with respect to the entire subgrid of models, and covers the zone occupied by medium to low oxygen abundances while we know from VA16 (and also Sect. \ref{sec:abunds} below) that the GHR sample has medium to high oxygen abundance. This can be due to an inadequate description of the SED of the ionizing radiation field, to incorrect values for sulfur dielectronic recombination coefficients, or to the fact that the density distribution of our models is too simple with respect to reality. The last hypothesis is probably the more likely to be correct, but it is out of the scope of the present paper to explore it here. It has been mentioned in \citet{Stasinska2006} and discussed thoroughly in \citet{Ramambasonal2020}. All in all, the situation revealed in this plot is not too bad, and we can expect that our ICFs for sulfur based on [\ion{S}{iii}] and [\ion{S}{ii}] lines will be reasonable. 

For the second panel, we can compare the photoionization models only for the BCG sample since [\ion{Ar}{iv}] $\lambda$4740 is absent in our GHR database. Here, we see that the observations occupy a very narrow strip. At a given value of   [\ion{O}{iii}] /[\ion{O}{ii}], the observed values of [\ion{Ar}{iv}] $\lambda$4740/[\ion{Ar}{iii}] $\lambda$7135 lie at the upper extreme of the range covered by our models. This is reminiscent of the finding by \citet{Stasinska2015}, who argued that this might be due to a too soft SED in the stellar atmosphere models. This then indicates that the ICFs we will propose for Ar could give slightly biased abundances, especially at the highest values of [\ion{O}{iii}] /[\ion{O}{ii}].

\subsection{Photoionization models weighting}
\label{sec:weights}
In order to obtain more reliable ICFs we have assigned weights to the selected models according to the number of observations they reproduce in the BPT diagram. Specifically, the weights were computed by comparing the position of each model in the [\ion{O}{iii}]/H$\beta$ versus [\ion{N}{ii}]/H$\alpha$ BPT diagram with the positions of the observed objects.
 
As said above, we did not use the [\ion{O}{iii}] $\lambda$5007 intensities from the BCG sample due to saturation problems. We computed them as $2.98 \times I$[\ion{O}{iii}] $\lambda4959$ in order to compare this sample to the models in the BPT diagram.
To compute the weights, we first constructed a uniform mesh on the BPT diagram with $\sim$0.3 dex side quadrants (nb: the final results on the ICF do not change significantly when changing the size of the quadrants). Then, we obtained
the weights of the models inside each quadrant, W$_{i}$, using the following expression$:$
\begin{equation}
    W_{i} = \frac{n_{{\rm obs}(i)}}{n_{{\rm mod}(i)}} + 0.01,
\end{equation}
where n$_{{\rm obs}(i)}$ is the number of observed objects inside the quadrant and n$_{{\rm mod}(i)}$ is the number of photoionization models inside the quadrant $i$. In such a way, models located in a quadrant of the BPT diagram where there are no observations are given only minimal weight in the computation of the ICFs, while models representative of a high number of observed objects are given large weights.

\begin{figure*}
\includegraphics[width=0.8\textwidth]{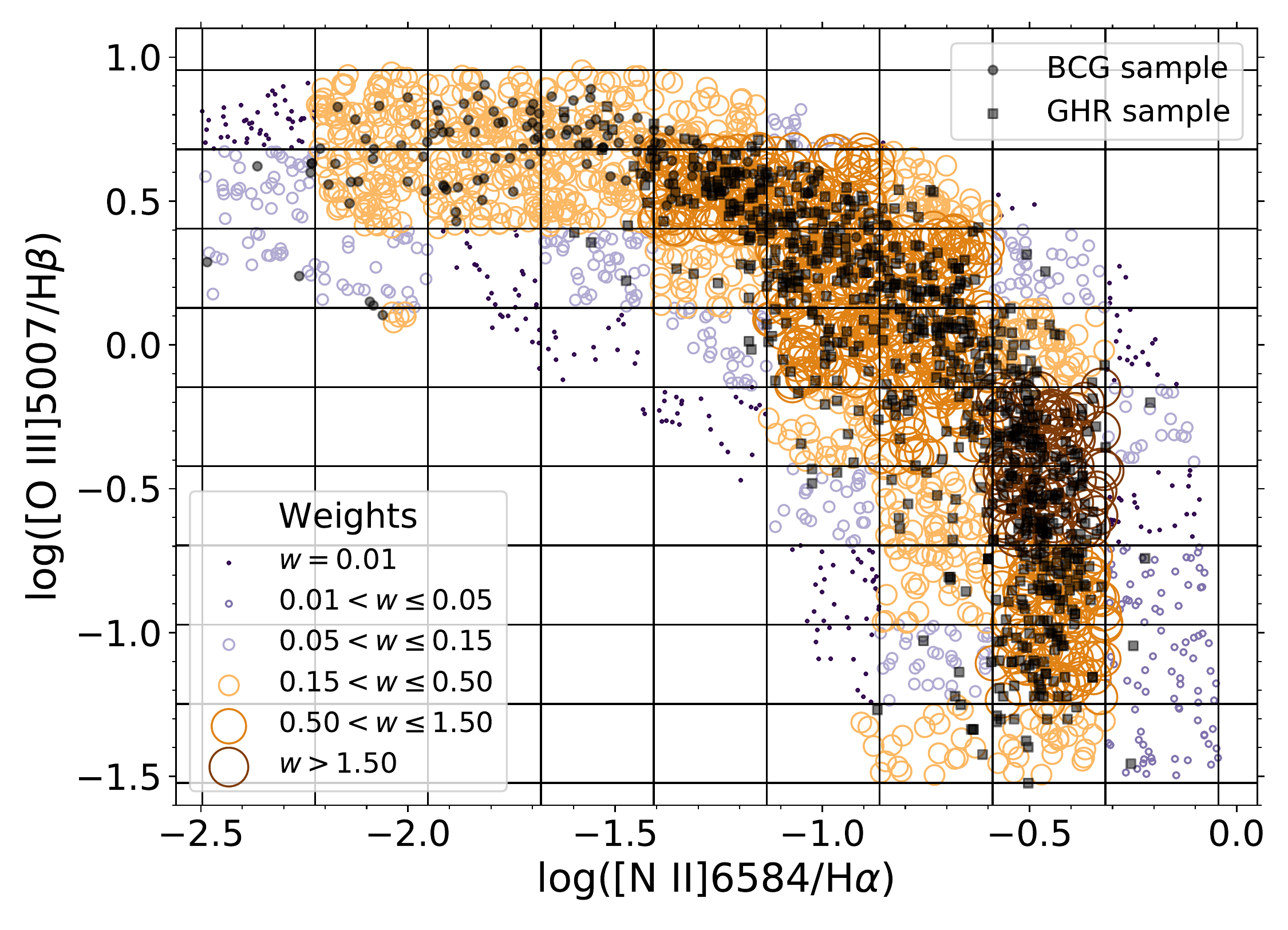}
\caption{BPT diagram of the photoionization models (colored circles). The size and color of the circles represent the weight assigned to each model. The BCG and GHR observed samples are represented by black circles and squares, respectively.}
\label{fig:WeightsBPT}
\end{figure*}

In Figure~\ref{fig:WeightsBPT} we show the BPT diagram of the 1887 selected photoionization models (colored circles), the observational sample formed by the BCG and GHR samples (black circles and squares, respectively), and the mesh constructed to compute the weights. The size of the circles represents the weight of each model. The weights range from 0.01 to $\sim$2.0. Models with $\log$[\ion{O}{iii}]/H$\beta$ between $-1.2$ and $0.6$ and $\log$[\ion{N}{ii}]/H$\alpha$ between $-1.5$ and $0.6$ have in general the highest weights.

\section{Ionization correction factors}
\label{sec:ICFs}

\subsection{Notations}
\label{sec:Notations}
The abundance of a particular element {\it X} is generally expressed with respect to hydrogen, {\it X}/H: 
\begin{equation}
    \frac{X}{{\rm H}} = \frac{\Sigma_{\rm obs.}{X}^{+i}}{{\rm H}^{+}}\times {\rm ICF}\left(\Sigma_{\rm obs.}{X}^{+i}\right),
\end{equation}
where $\Sigma_{\rm obs.}{X}^{+i}$ represents the sum of the abundances of the observed ions of $X$. 

The analytical expression for the ICF in this case may be computed from a fit of the values of: 
\begin{equation}
    {\rm ICF}(\Sigma_{\rm obs.}{X}^{+i}) = \frac{x({\rm H^{+}})}{x(\Sigma_{\rm obs.}{X}^{+i})} 
\end{equation}
from the photoionization models. Here $x$ represents the relative ionic fractions of each ion, weighted by the electron density.

One may be interested in the abundance ratio of an element with respect to O, rather than with respect to H. In that case, it may be better to use an ICF computed with such an aim.

To better illustrate this and to explain the notations adopted throughout the paper, we use two examples. The first example concerns nitrogen. The only ion with emission lines in the optical range is N$^{+}$ and thus, the ICF takes into account the contribution of all the other ions present in the nebula. We derived the ICF as$:$

\begin{equation}
    {\rm ICF}({\rm N^{+}}/{\rm O^{+}}) = \frac{x({\rm O^{+}})}{x({\rm N^{+}})} 
\end{equation}
Therefore, N/O is computed as$:$
\begin{equation}
    \frac{{\rm N}}{{\rm O}} = \frac{{\rm N}^{+}}{{\rm O}^{+}}\times {\rm ICF}({\rm N^{+}}/{\rm O^{+}}),
\end{equation}
and N/H = N/O$\times$O/H.
 
The second example refers to sulfur. We have derived two ICFs, one to be used when only [\ion{S}{ii}] lines are observed:
\begin{equation}
    {\rm ICF}({\rm S}^{+}/{\rm O}^{+}) = \frac{x({\rm O}^{+})}{x({\rm S}^{+})} 
\end{equation}
and another to be used when both [\ion{S}{ii}] and [\ion{S}{iii}] lines are observed:
\begin{equation}
    {\rm ICF}[({\rm S}^{+}+{\rm S}^{++})/({\rm O}^{+}+{\rm O}^{++})] = \frac{x({\rm O}^{+}+{\rm O}^{++})}{x({\rm S}^{+}+{\rm S}^{++})}.
\end{equation}
Therefore, the total S/O ratio is computed as
\begin{equation}
    \frac{{\rm S}}{{\rm O}} = \frac{{\rm S}^{+}}{{\rm O}^{+}}\times {\rm ICF}\left(\frac{{\rm S}^{+}}{{\rm O}^{+}}\right),
\end{equation}
in the first case, or:
\begin{equation}
    \frac{{\rm S}}{{\rm O}} = \frac{({\rm S}^{+}+{\rm S}^{++})}{({\rm O}^{+}+{\rm O}^{++})}\times {\rm ICF}\left(\frac{{\rm S}^{+}+{\rm S}^{++}}{{\rm O}^{+}+{\rm O}^{++}}\right),
\end{equation}
in the second one. The total abundance of sulfur with respect to hydrogen is S/H = S/O$\times$O/H.

For each element, we found that the value of the ICF most of the times depends on the excitation of the object, and sometimes also slightly on the metallicity. Therefore we express the ICFs as a function of O$^{++}$/(O$^{+}$+ O$^{++}$) \footnote{We explored the use of the metallicity as a second parameter, but it turned out inapplicable.}, which is a quantity easily observed and computed. For simplicity, we will call it $\omega$.

For each abundance ratio for which we compute an ICF we divided the range of $\omega$ values in 10 bins and computed the weighted median (taking into account the weights of the models) in each bin. Then, we performed a fit using these 10 values. We have fitted a fifth-order polynomial expression of the form: $A + B\omega + C\omega^{2} + D\omega^{3} +  E\omega^{4} + F\omega^{5}$ for the logarithm of each ICF. As for the uncertainties associated to the ICFs, we computed the weighted 16 and 84 percentiles in each bin and performed a fit to the logarithm of their values using also a fifth-order polynomial expression. We have called the functions representing these fits as $\varepsilon^{+}$ and  $\varepsilon^{-}$, respectively. 

The analytical expressions of the ICFs derived for C, N, Ne, S, Cl, and Ar, and their associated uncertainties are listed in Table~\ref{tab:icfs}. In the following sections we provide some details about each of the elements studied here and we show the figures with the photoionization models and the derived ICFs. A comparison of our ICFs with some of those proposed previously is available in the Appendix~\ref{sec:appendix}. 

\begin{table*}
\centering
\caption{Elements studied here, the observed ions in each case, the coefficients A, B, C, D, E and F of the polynomial expressions A + B$\omega$ + C$\omega^{2}$ + D$\omega^{3}$ + E$\omega^{4}$ + F$\omega^{5}$ for log ICF, as well as for the log of the 16th percentile  ($\varepsilon^{-}$) and of the 84th percentile ($ \varepsilon^{+}$).
}
\begin{tabular}{lllllcccccc}
\hline
Element & Abundance & Observed                          &   &   &   &   &   &   &      \\    
        & ratio     & ions                              &   & A & B & C & D & E & F      \\             
\hline
C       & C/O       & C$^{++}$/O$^{++}$                    & $\log$ICF  & -0.876  & 3.691  & -8.250  & 10.825 & -7.546 & 2.195\\                   
        &           &                                      & $\log \varepsilon^{-}$ & -1.294 & 7.994  & -27.427 & 50.077 & -44.175 & 14.803\\                                             
        &           &                                      & $\log \varepsilon^{+}$ & -0.744  & 3.762  & 11.235  & 19.788 & -17.921 & 6.476\\                                                 
\hline
N       & N/O       & N$^{+}$/O$^{+}$                      & $\log$ICF              & 0.013  & -0.793  & 8.177  & -23.194 & 26.364 & -10.536\\                          
        &           &                                      & $\log \varepsilon^{-}$ & -0.032  & -0.064  & 1.544  & -5.430 & 7.387 & -3.480\\                                              
        &           &                                      & $\log \varepsilon^{+}$ & 0.007  & 0.455  & 1.147  & -5.097 & 6.909 & -3.197\\                                             
\hline
Ne      & Ne/O      & Ne$^{++}$/O$^{++}$                   & $\log$ICF              & -0.557  & 4.237  & -8.564  & 4.834 & 2.284 & -2.239\\                             
        &           &                                      & $\log \varepsilon^{-}$ & -0.950  & 4.894  & -9.832  & 5.940 & 3.140 & -3.202\\                                          
        &           &                                      & $\log \varepsilon^{+}$ &  0.021 & 3.778  & -14.492  & 23.681 & -18.499 & 5.508\\                                            
\hline
S       & S/O       & S$^{+}$/O$^{+}$                      & $\log$ICF              & 0.078  & 1.084  & 5.808  & -26.537 & 35.967 & -16.298\\                                  
        &           &                                      & $\log \varepsilon^{-}$ & -0.024  & 0.588  & 4.948  & -18.897 & 24.197 & -10.774\\                                           
        &           &                                      & $\log \varepsilon^{+}$ &  0.258 & 1.454  & 2.342  & -13.874 & 19.595 & -9.307\\                                             
        \cline{3-10}
        &           & (S$^{+}$+S$^{++})$/(O$^{+}$+O$^{++})$
    & $\log$ICF              & -0.083  & 0.943  & -4.845  & 12.378 & -14.832 & 6.750\\               
        &           &                                      & $\log \varepsilon^{-}$ & -0.157  & 0.884  & -2.966  & 6.505 & -7.506 & 3.416\\                                            
        &           &                                      & $\log \varepsilon^{+}$ &  -0.035 & 0.646  & -4.407  & 13.401 & -17.909 & 8.799\\                                              
\hline
Cl      & Cl/O      & Cl$^{++}$/O$^{++}$                   & $\log$ICF              & -1.117  & 5.855  & -19.340  & 36.930 & -35.326 & 13.254\\         
        &           &                                      & $\log \varepsilon^{-}$ & -1.504  & 8.712  & -28.783  & 51.631 & -45.498 & 15.579\\                                          
        &           &                                      & $\log \varepsilon^{+}$ & -0.928  & 5.059  & -18.393  & 39.480 & -42.176 & 17.443\\                                          
\hline
Ar      & Ar/O      & Ar$^{++}$/O$^{++}$                   & $\log$ICF              & -1.463  & 6.993  & -19.728  & 33.233 & -29.535 & 10.745\\        
        &           &                                      & $\log \varepsilon^{-}$ & -1.677  & 8.353  & -23.426  & 36.962 & -29.659 & 9.560\\                                       
        &           &                                      & $\log \varepsilon^{+}$ &  -1.202 & 5.499  & -16.473  & 31.793 & -32.709 & 13.560\\                                       
        \cline{3-10} 
        &           & (Ar$^{++}$+Ar$^{+3}$)/O$^{++}$
        & $\log$ICF    & -1.450  & 6.598  & -16.768  & 24.175 & -17.706 & 5.154\\                            
        &           &                                      & $\log \varepsilon^{-}$ & -1.676  & 8.340  & -23.310  & 36.175 & -28.112 & 8.578\\                                        
        &           &                                      & $\log \varepsilon^{+}$ &  -1.171 & 4.613  & -9.951  & 12.277 & -7.537 & 1.774\\                                             
\hline
\end{tabular}
\label{tab:icfs}
\end{table*}

\subsection{Carbon}

In the optical range, the only C line observed is the \ion{C}{ii} \Lam4267 recombination line. If using this line to compute the C/O ratio, it is necessary to use \ion{O}{ii} recombination lines. However, in \hii regions, the C/O ratio has often been derived using the ultraviolet \ion{C}{iii}] \Lam1909 line, combined with [\ion{O}{iii}] \Lam1663 or [\ion{O}{iii}] \Lam5007.

\begin{figure}
\includegraphics[width=\columnwidth]{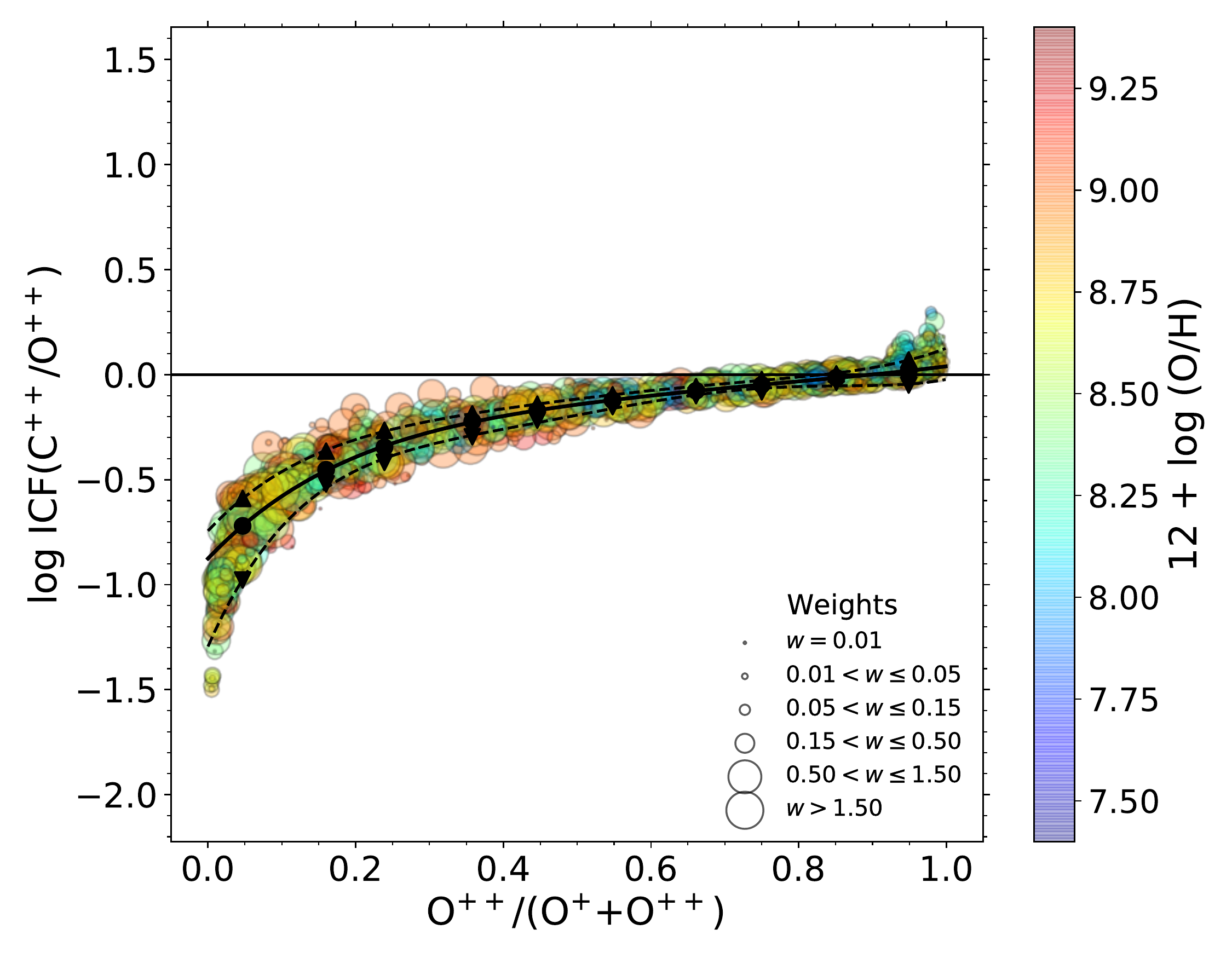}
\caption{Values of $\log$ ICF(C$^{++}$/O$^{++}$) as a function of $\omega$ for the photoionization models (colored circles). The size of the circles is related to the weights of the models. The weighted 16, 50, and 84 percentiles of each bin are shown with triangles, circles, and upside-down triangles, respectively. The fits to these values are shown with dashed and solid lines. The horizontal solid line represents $\rm{ICF} =1$. The color bar runs from low to high values of 12 + $\log$(O/H).}
\label{fig:ICFC2O2}
\end{figure}

Figure~\ref{fig:ICFC2O2} shows the values of ICF(C$^{++}$/O$^{++}$) as a function of $\omega$ for the photoionization models (colored circles). The weighted 16\%, 50\%, and 84\% percentiles for each bin in $\omega$ are shown with black symbols: upward-pointing triangles, circles and downward-pointing triangles, respectively. The fits are shown with solid and dashed lines. Note that, due to the high dispersion of the models, the ICF is less well defined for \hii regions with $\omega < 0.05$ or $\omega > 0.95$. In addition some of the models with the highest values of $12 + \log$(O/H) are the furthest from the fit which indicates that this ICF is more reliable at medium and low metallicities.

\subsection{Nitrogen}
\begin{figure}
\includegraphics[width=\columnwidth]{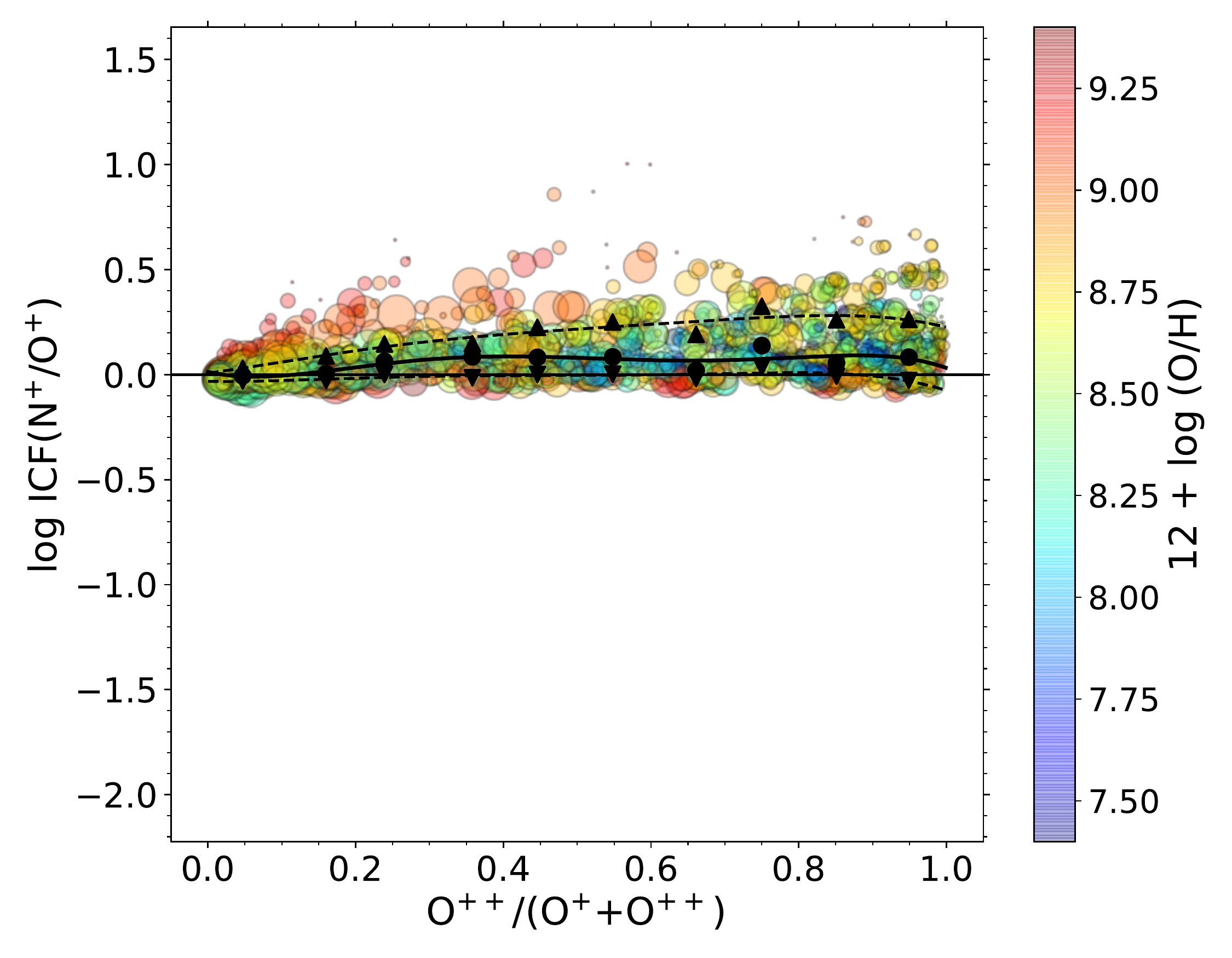}
\caption{Values of $\log$ ICF(N$^{+}$/O$^{+}$) as a function of $\omega$ with the same layout as Fig. \ref{fig:ICFC2O2}.}
\label{fig:ICFN1O1}
\end{figure}

The only nitrogen ion that is observed in the optical range is N$^{+}$ (e.g. [\ion{N}{ii}] \Lam6584, \Lam6548).Figure~\ref{fig:ICFN1O1} shows the values computed for ICF(N$^{+}$/O$^{+}$) as a function of $\omega$, with the same layout as Fig.\ref{fig:ICFC2O2}. The figure shows that our proposed ICF is very similar to the traditional one. The median value of the ICFs from our models is larger than the traditional ICF by less than 0.1 dex in the entire range of $\omega$. But the dispersion is not negligible, especially at the highest values of $\omega$. As in the case of carbon, the uncertainty is larger at high metallicity.

\subsection{Neon}
\begin{figure}
\includegraphics[width=\columnwidth]{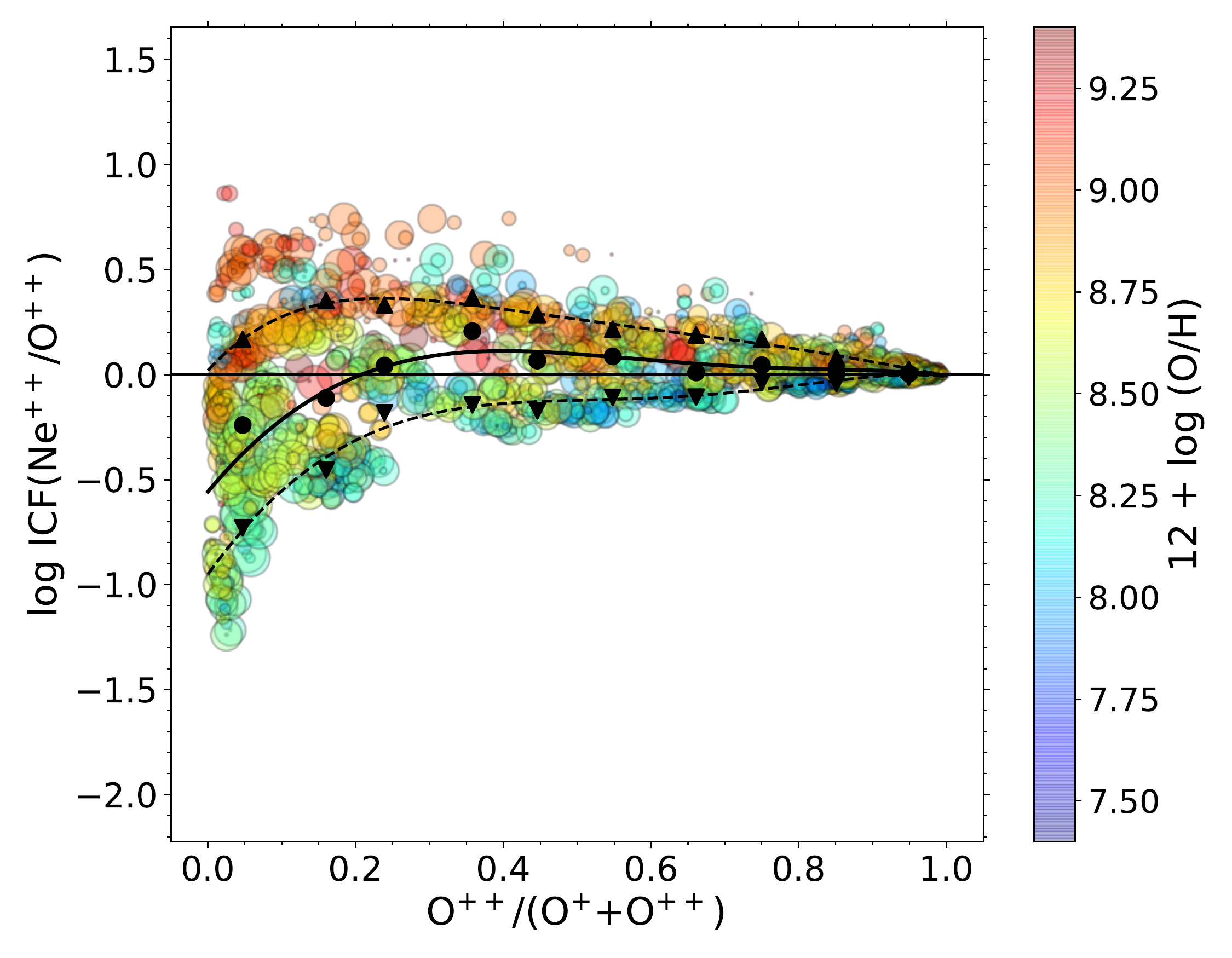}
\caption{Values of $\log$ ICF(Ne$^{++}$/O$^{++}$) as a function of $\omega$, with the same layout as Fig. \ref{fig:ICFC2O2}.}
\label{fig:ICFNe2O2}
\end{figure}

The only Ne ion that is observed in the optical spectrum of giant \hii regions is Ne$^{++}$ (whose most intense line is [\ion{Ne}{iii}] \Lam3869). 
Figure~\ref{fig:ICFNe2O2} shows the values of ICF(Ne$^{++}$/O$^{++}$) as a function of $\omega$. It can be seen that the dispersion of our models is large, especially at low values of $\omega$. This is due to two factors. The ionization potential of Ne$^{+}$, 40.96 eV, is actually much higher than that of O$^{+}$, which is 35.12 eV. This induces a significant difference in the photoionization rates of those two ions, especially when the ionizing energy distribution of the stellar radiation field is relatively mild, i.e. at high metallicities. The other factor is the important difference in the charge-transfer recombination rates of O$^{+}$ and Ne$^{+}$, which implies that the relative ionization structures of these elements in \hii\ regions also strongly dependent on the ionization parameter. Therefore, Ne/O ratios are very uncertain for low-excitation objects.

\subsection{Sulfur}
There are two ions that may be observed in optical spectra of giant \hii regions: S$^{+}$ (e.g., [\ion{S}{ii}] \Lam6716, \Lam6731 lines) and S$^{++}$ (e.g., [\ion{S}{iii}] \Lam6312, \Lam9069, \Lam9532 lines). We study here two cases: 1) when only S$^{+}$ lines are observed and 2) when both S$^{+}$ and S$^{+}$ lines are observed. 

\subsubsection{\rm ICF (S$^{+}$)}
\begin{figure}
\includegraphics[width=\columnwidth]{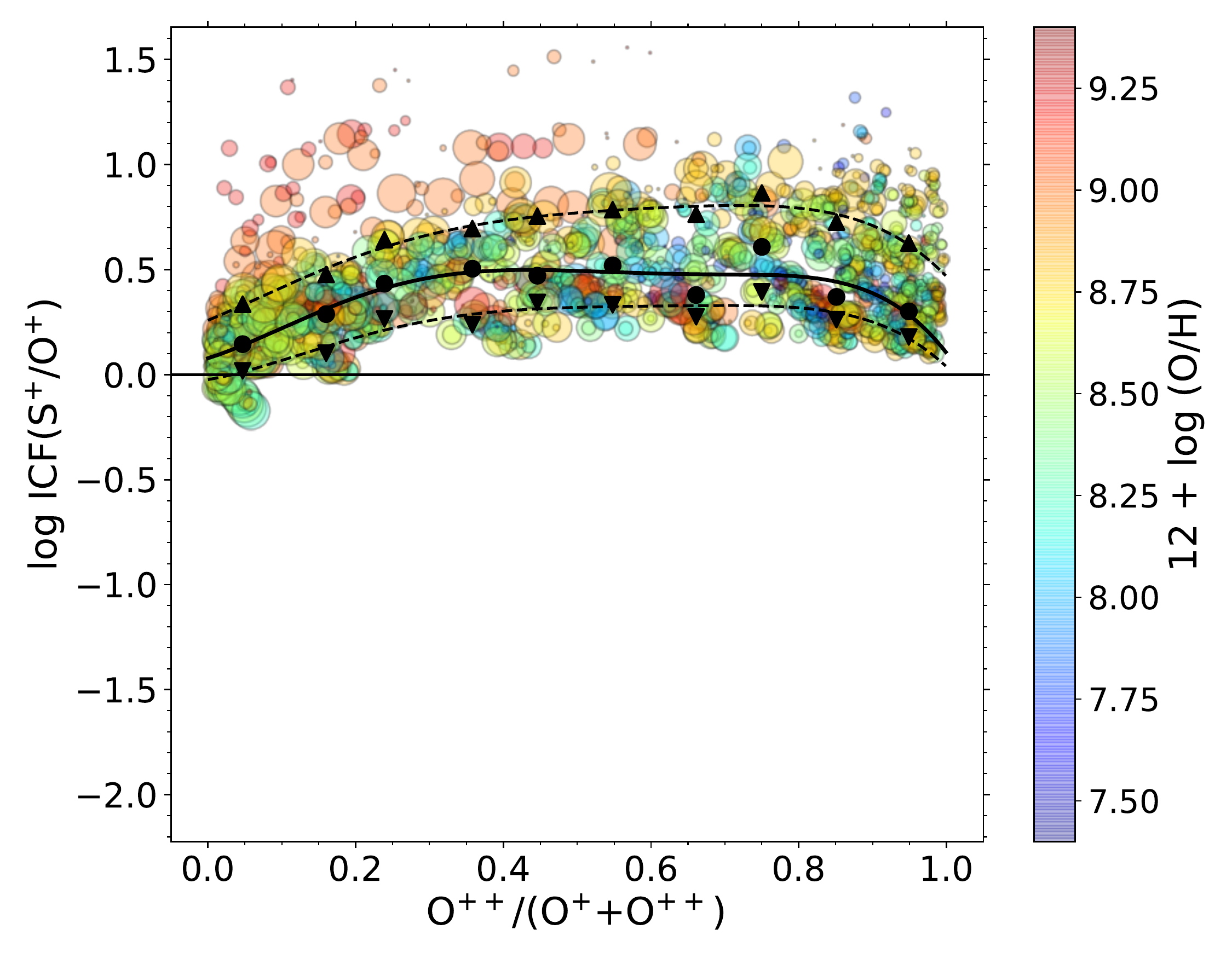}
\caption{Values of $\log$ ICF(S$^{+}$/O$^{+}$) as a function of $\omega$, with the same layout as Fig. \ref{fig:ICFC2O2}.}
\label{fig:ICFS1O1}
\end{figure}

Figure~\ref{fig:ICFS1O1} shows the values of ICF(S$^{+}$/O$^{+}$) as a function of $\omega$. As in the previous cases, the largest uncertainties associated to the ICFs occur for \hii regions with $12 + \log$(O/H) $>$ 9.0.

\subsubsection{\rm ICF (S$^{+}$+S$^{++}$)}
\begin{figure}
\includegraphics[width=\columnwidth]{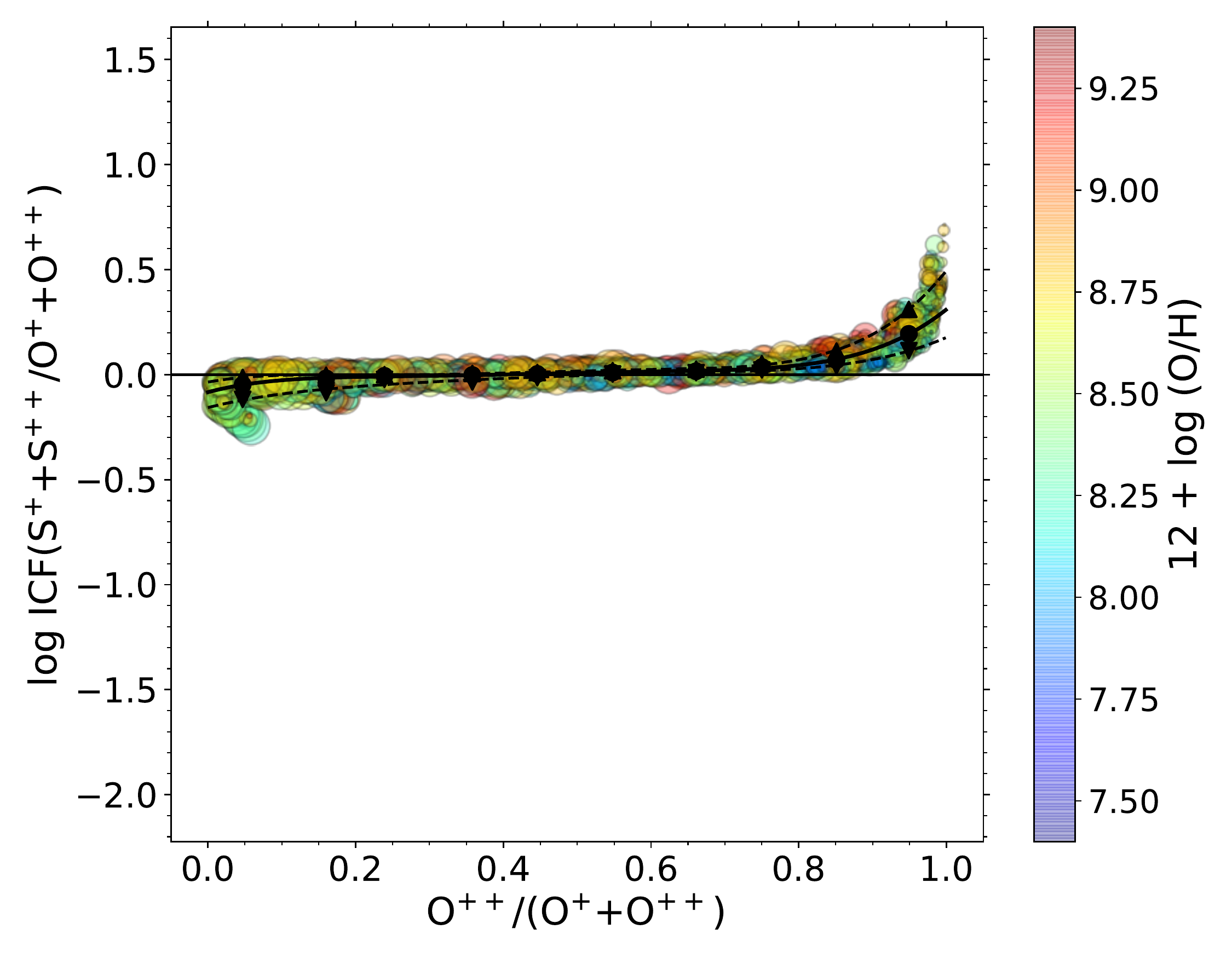}
\caption{Values of $\log$ ICF(S$^{+}$+S$^{++}$/O$^{+}$+O$^{++}$) as a function of $\omega$, with the same layout as Fig. \ref{fig:ICFC2O2}.}
\label{fig:ICFS12O12}
\end{figure}

Figure~\ref{fig:ICFS12O12} shows the values of ICF(S$^{+}+$S$^{++}$/O$^{+}+$O$^{++}$) as a function of $\omega$ for the photoionization models.  The fits are shown with solid and dashed lines. This ICF seems to be one of the most reliable we provide here since almost all the models are located within the ICF uncertainties which are actually quite small. 

\subsection{Chlorine}
\begin{figure}
\includegraphics[width=\columnwidth]{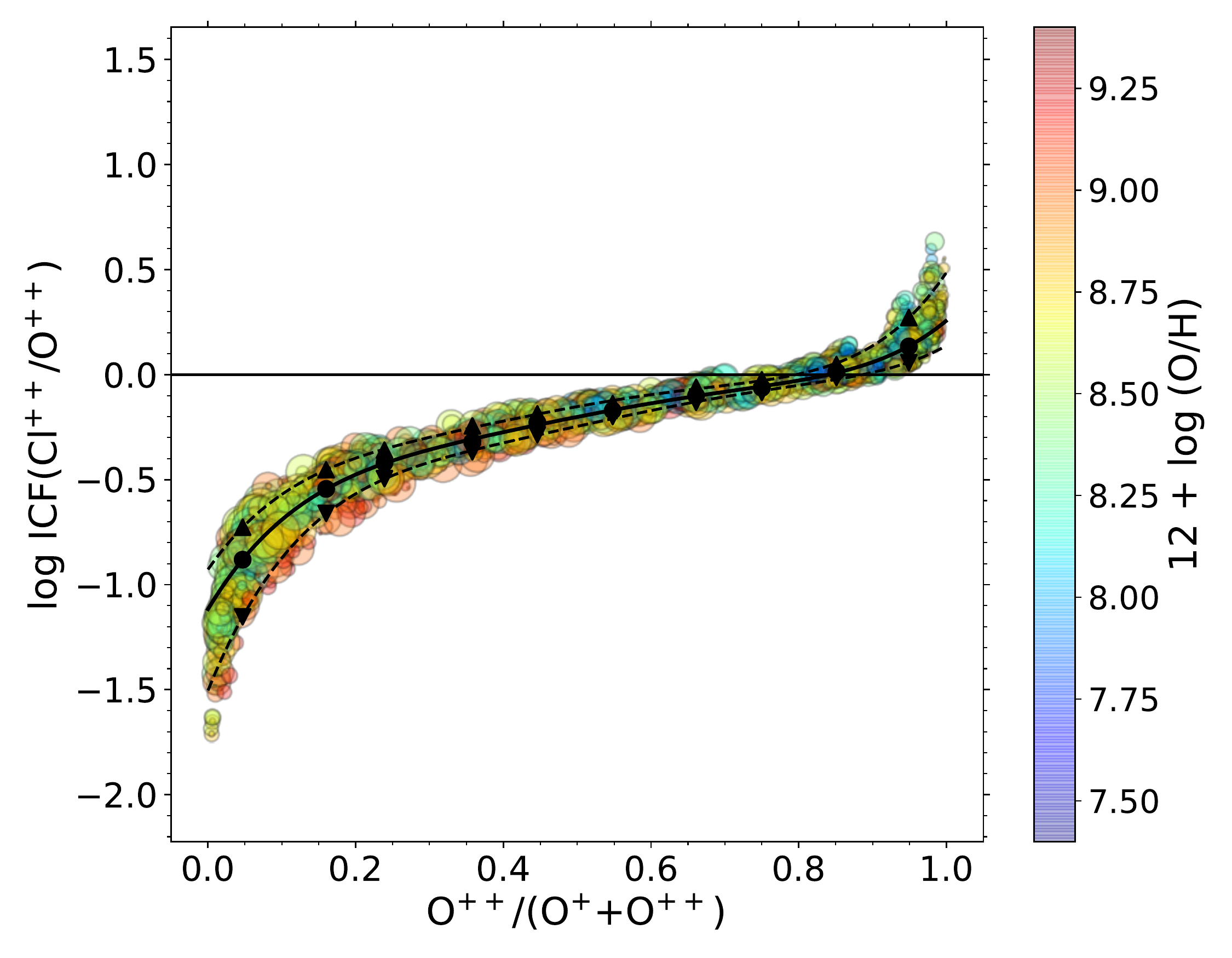}
\caption{Values of $\log$ ICF(Cl$^{++}$/O$^{++}$) as a function of $\omega$, with the same layout as Fig. \ref{fig:ICFC2O2}.}
\label{fig:ICFCl2O2}
\end{figure}

The ions that may be observed in the optical range of giant \hii regions are Cl$^+$ (e.g., [\ion{Cl}{ii}] \Lam9123), Cl$^{++}$ (e.g., [\ion{Cl}{iii}] \Lam5517, \Lam5537) and Cl$^{+3}$ (e.g., [\ion{Cl}{iv}] \Lam8046). However, in this work we only compute an ICF based on Cl$^{++}$ because this ion is very often the only one available in the spectra of giant \ion{H}{ii} regions. Figure~\ref{fig:ICFCl2O2} shows values of ICF(Cl$^{++}$/O$^{++}$) as a function of $\omega$.

\subsection{Argon}
The emission lines of argon that are usually observed are: [\ion{Ar}{iii}] \Lam7135, \Lam7751, [\ion{Ar}{iv}] \Lam4711, 4740. Based on this, we propose two ICFs for argon, one for the case only Ar$^{++}$ emission lines are observed and another one for the case when also Ar$^{+3}$ lines are observed.

\subsubsection{\rm ICF (Ar$^{++}$)}
\begin{figure}
\includegraphics[width=\columnwidth]{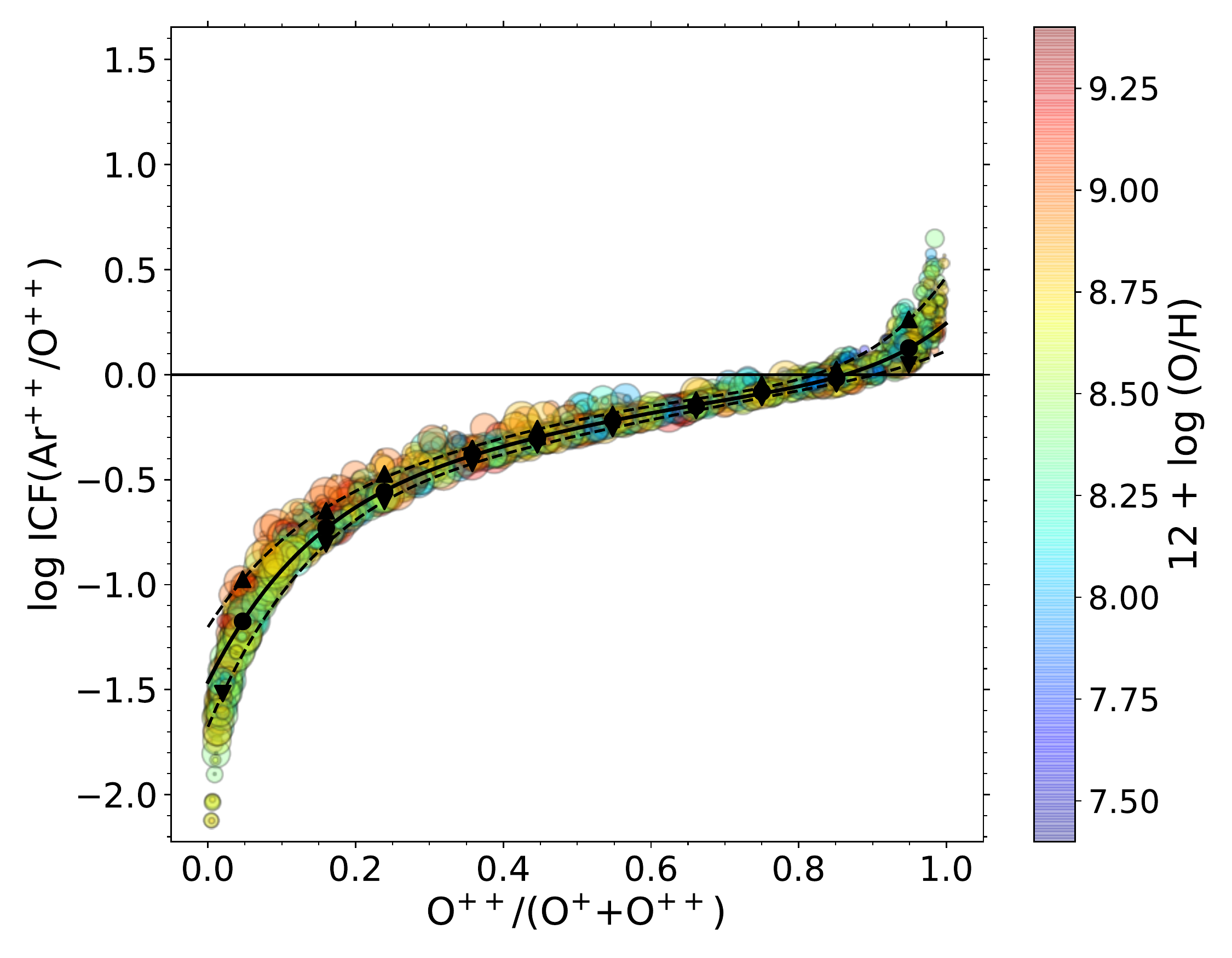}
\caption{Values of $\log$ ICF(Ar$^{++}$/O$^{++}$) as a function of $\omega$, with the same layout as Fig. \ref{fig:ICFC2O2}.}
\label{fig:ICFAr2O2}
\end{figure}

Figure~\ref{fig:ICFAr2O2} shows the values of ICF(Ar$^{++}$/O$^{++}$) as a function of $\omega$.

\subsubsection{\rm ICF (Ar$^{++}$+Ar$^{+3}$)}
\begin{figure}
\includegraphics[width=\columnwidth]{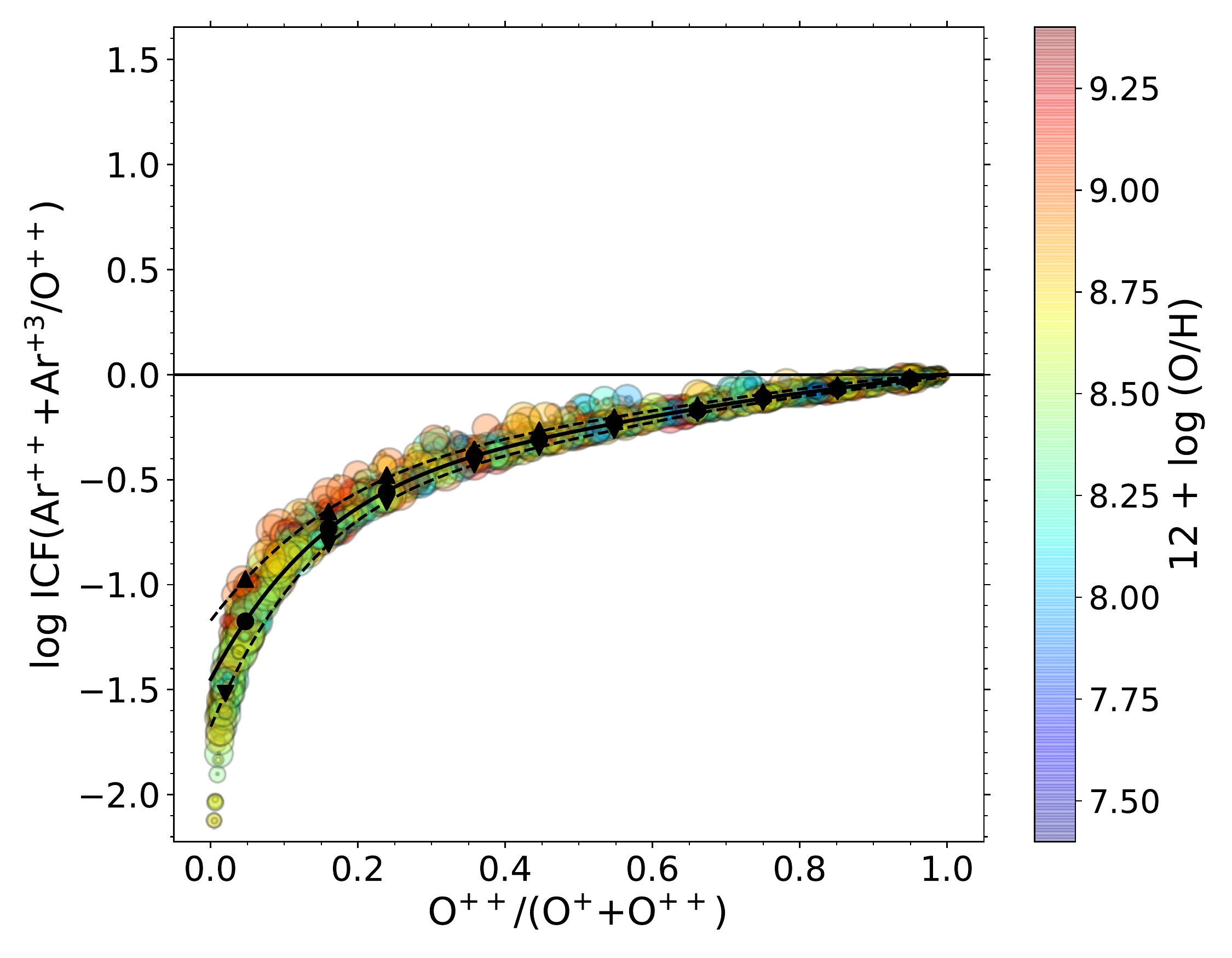}
\caption{Values of $\log$ ICF(Ar$^{++}$+Ar$^{+3}$/O$^{++}$) as a function of $\omega$, with the same layout as Fig. \ref{fig:ICFC2O2}.}
\label{fig:ICFAr23O2}
\end{figure}

Figure~\ref{fig:ICFAr23O2} shows the values of ICF(Ar$^{++}$+Ar$^{+3}$/O$^{++}$) as a function of $\omega$.

\section{Chemical abundances using our ICFs}
\label{sec:abunds}
In this section we present the chemical abundances of the BCG and GHR samples using the ICF expressions from Table~\ref{tab:icfs}.
 
\subsection{Abundance computations}

First, we have computed the electron temperatures and densities (\temp\ and \dens) using the version 1.1.10 of {\sc{PyNeb}} \citep{Luridiana2015, Morisset2020}. We used the effective recombination coefficients provided by \citet{SH95} for H$^{+}$ and the atomic data from Table~\ref{tab:atomdata} for the other ions. These atomic datasets were selected following the works of \citet{Stasinskaal2013, DelgadoInglada2016} and \citet{juandedios2017}.

We assumed that each \hii region is characterized by two values of \temp\  and one of \dens.  The temperatures were derived from the [\ion{N}{ii}] \Lam5755/\Lam6584 ratio and the [\ion{O}{iii}] \Lam4363/\Lam5007 or the [\ion{O}{iii}] \Lam4363/\Lam4959 ratio. In the objects where either \tempnii\ or \tempoiii\ could not be determined we used the expression given by \citet{Garnett1992}: \tempnii\ = 0.70 $\times$ \tempoiii\ + 3000 K. This expression is in fair agreement with modern photoionization models (VA16) and with observations \citep{ArellanoRodriguez2020}.

The value of \dens\ taken for each nebula is the mean of the densities computed with the [\ion{O}{ii}] \Lam3726/\Lam3729 and [\ion{S}{ii}] \Lam6731/6716 ratios. In the case of the BCG sample, where no density diagnostic was available, a value of 100 cm$^{-3}$ was adopted, as suggested by \citet{Stasinska2004}. 

\begin{table}
\centering
\caption{Atomic data used in this work.}
\resizebox{\columnwidth}{!}{  
\begin{tabular}{lll}
\hline
Ion                             & Transition Probabilities        & Collisional Strenghts\\
\hline
N$^{+}$                        & \citet{FFT04}                   & \citet{T11}\\ 
O$^{+}$                         & \citet{FFT04}                   & \citet{Kal09}\\ 
O$^{++}$                         & \citet{FFT04}                   & \citet{SSB14}\\ 
Ne$^{++}$                        & \citet{GMZ97}                   & \citet{McLB00}\\
S$^{+}$                        & \citet{M82a}                   & \citet{TZ10}\\
S$^{++}$                        & \citet{PKW09}                   & \citet{TG99}\\
Ar$^{++}$                        & \citet{MB09}                   & \citet{MB09}\\ 
Ar$^{+3}$                        & \citet{M82a}                   & \citet{RB97}\\
\multicolumn{1}{l}{}           & \citet{KS86}                    &  \\
\hline
\end{tabular}
}
\label{tab:atomdata}
\end{table}

Second, we have computed the ionic abundances using only the most intense emission lines of each ion. For ions with an ionization potential lower than 20 eV (O$^{+}$/H$^{+}$, N$^{+}$/H$^{+}$, and S$^{+}$/H$^{+}$) we adopted \tempnii, while for ions with a greater ionization potential (O$^{++}$/H$^{+}$, S$^{+}$/H$^{+}$, S$^{++}$/H$^{+}$, Ar$^{++}$/H$^{+}$, Ar$^{+3}$/H$^{+}$, and Ne$^{++}$/H$^{+}$) we adopted \tempoiii.

Third, we have computed the total abundances of nitrogen, neon, sulfur and argon with the ICFs derived here. For sulfur and argon, we were able to compute the abundances with the two different ICF expressions proposed in this work and compare the results. We did not compute the chlorine abundances due to the lack of relevant intensities in our observational sample.

\subsection{Uncertainties}
Uncertainties were computed through a Monte Carlo simulation. First, we generated a Gaussian distribution of 400 values centered in the observed intensity of each line with a $\sigma$ equal to each line uncertainty. The Gaussian distributions may lead to negative values in intensities. We replaced these negative values with an intensity of 0.0001, a value much lower than the minimum intensity observed in our sample\footnote{Since we do not present mean values but percentiles, such a procedure is reasonable.}. Using the Monte Carlo experiments of each object, we computed the physical conditions, ionic and total abundances. For objects where we used the \citet{Garnett1992} relation to derive either \tempnii\ or \tempoiii, we additionally generated a Gaussian distribution of 400 Monte Carlo experiments with a $\sigma$ equal to 600 K, centered in the nominal temperature derived with this relation. This $\sigma$ corresponds the dispersion found by the `fake observational sample' of VA16 around this relation (see Fig.A2a of VA16).

For the total abundances, we computed two groups of uncertainties: the first one, including the uncertainties associated to the emission lines and to the dispersion in the \tempnii\ vs \tempoiii\ relation and a second one including the uncertainties associated to each ICF.
To include the latter uncertainty source, we used the analytical expressions from Table~\ref{tab:icfs} and generated a log-normal distribution of 400 values with a confidence interval delimited by the upper and lower uncertainties associated to each ICF. A Gaussian distribution was not used in order to avoid the possibility of obtaining negative ICF values. The log-normal distribution was generated in a way similar to what was done in \citet{Garma2020}, so that its mean ($\mu$) and standard deviation ($\sigma$) were equal to
\begin{equation}
\begin{split}
&\mu = \sqrt{\varepsilon^{-}\times\varepsilon^{+}} \hbox{ and}\\
&\sigma = \sqrt{\varepsilon^{+}/\varepsilon^{-}},
\end{split}
\end{equation}  
where $\varepsilon^{-}$ and $\varepsilon^{+}$ are the 16 and 84 weighted percentiles of the ICFs distribution, respectively. These expressions result in a log-normal distribution with the same $\sigma$ than each ICF distribution, but with a slightly different mean value. From this log-normal distribution, we chose a random value and used it to compute the final element abundance.

\section{Discussion}
\label{sec:discussion}
 
\begin{figure*}
\begin{minipage}{0.98\textwidth}
\includegraphics[width=.32\textwidth]{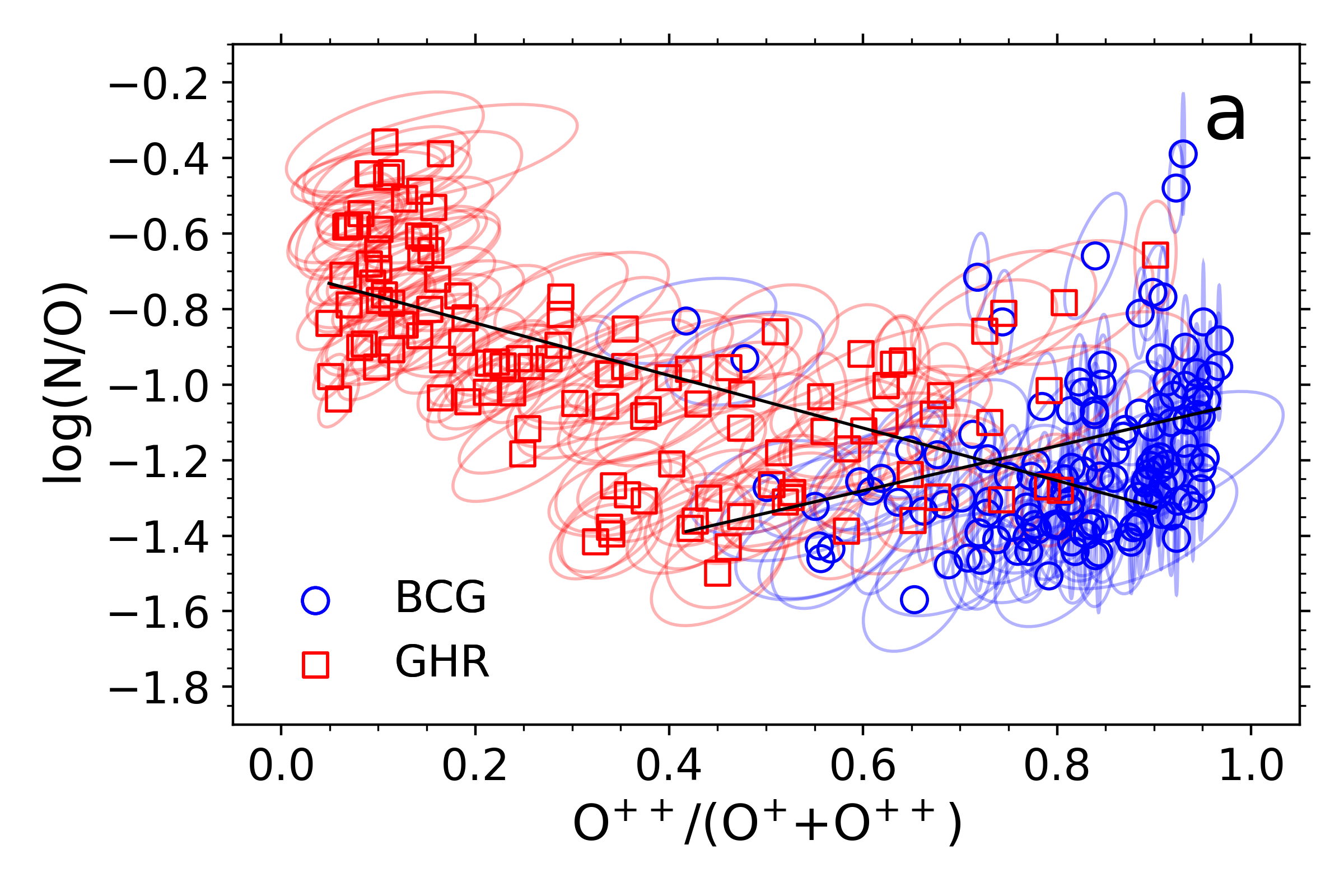}\hfill
\includegraphics[width=.32\textwidth]{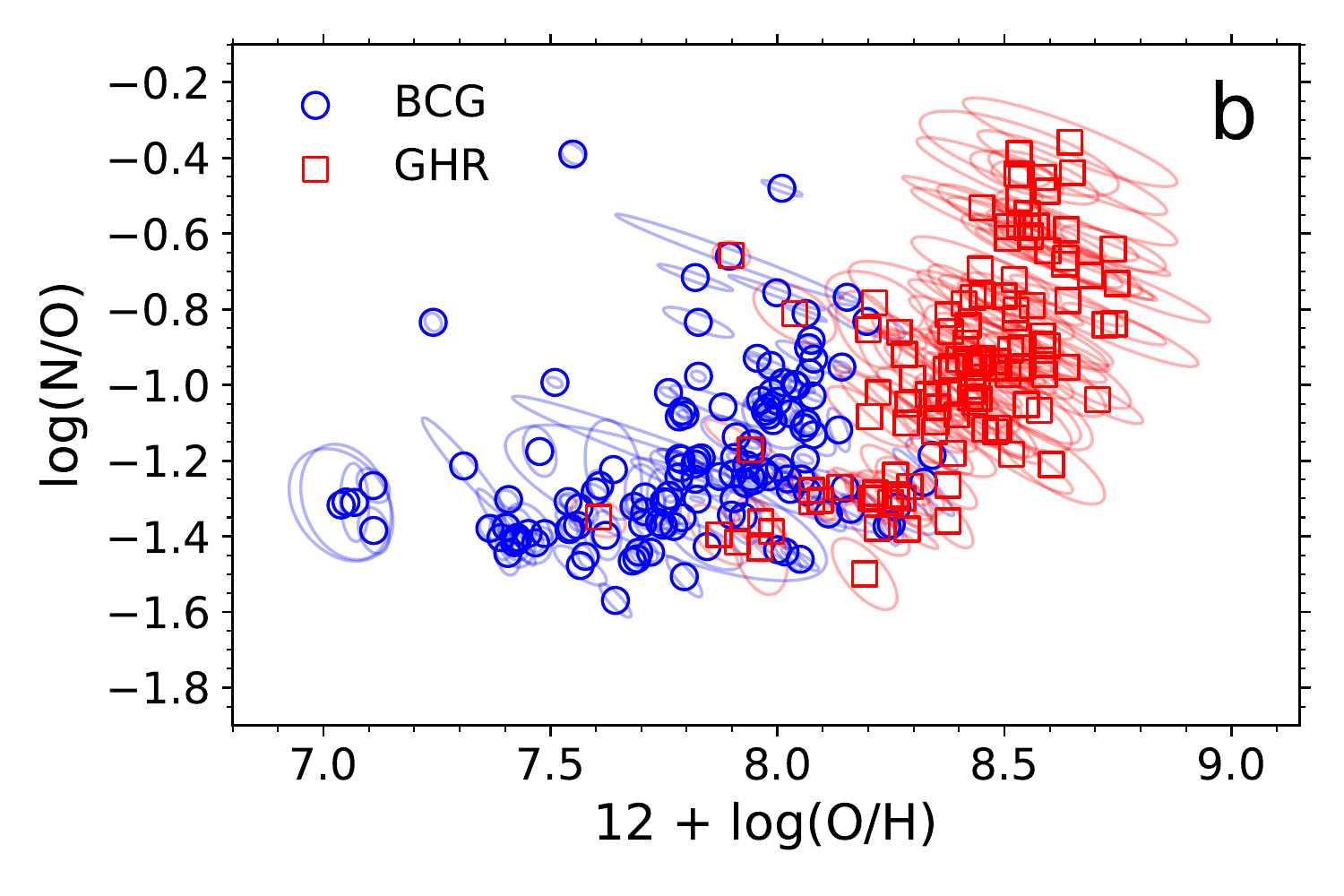}\hfill
\includegraphics[width=.32\textwidth]{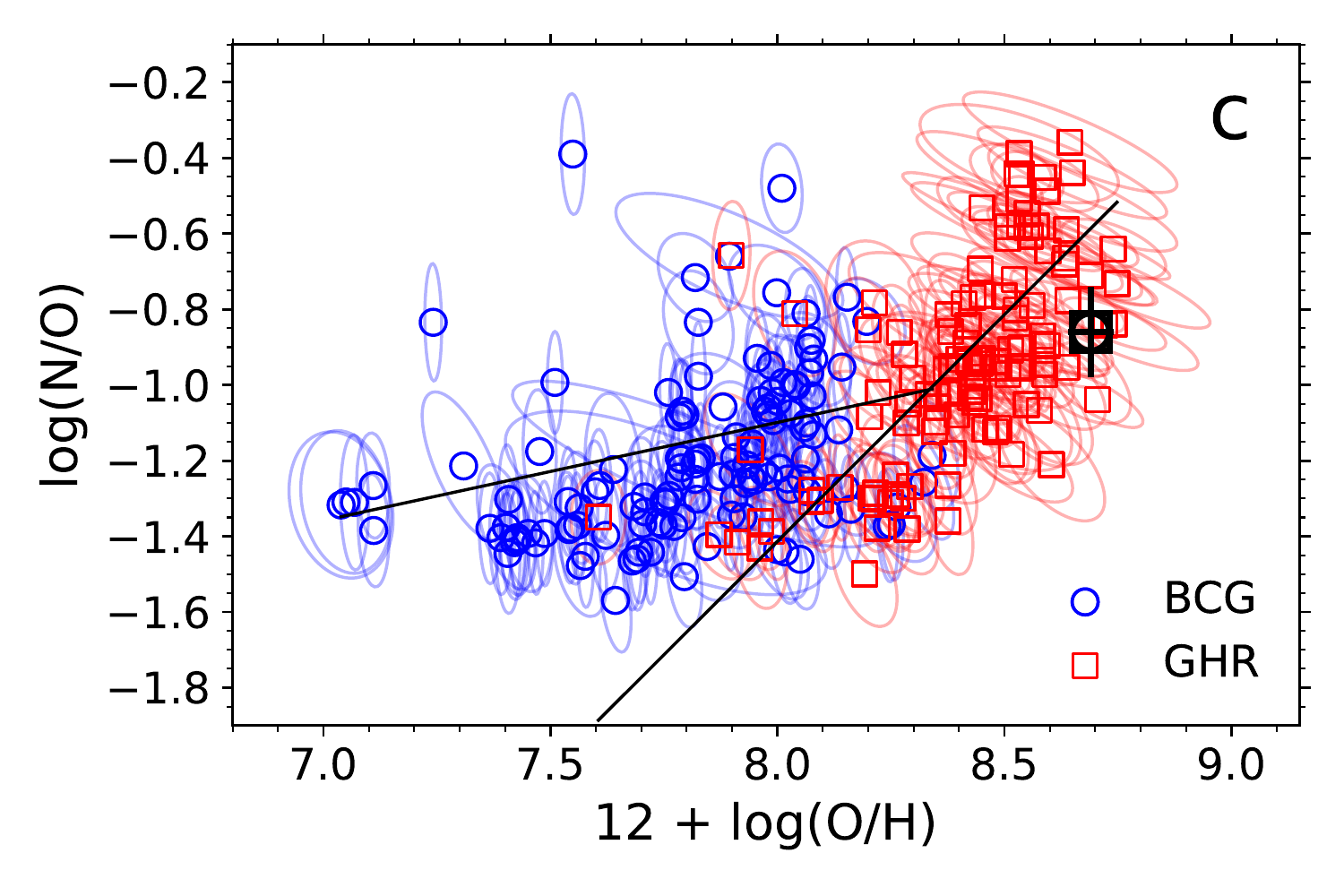}\hfill
\includegraphics[width=.32\textwidth]{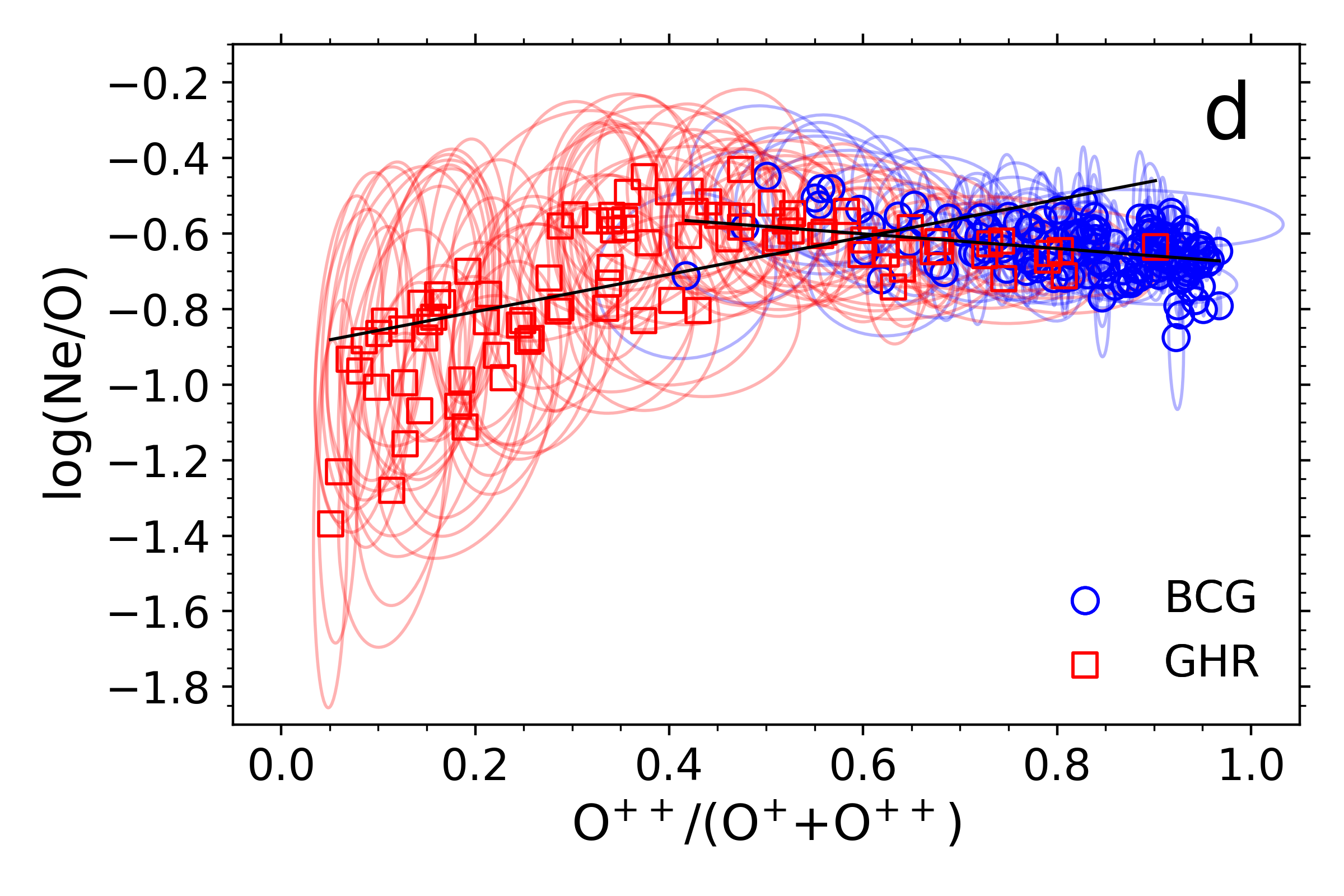}\hfill
\includegraphics[width=.32\textwidth]{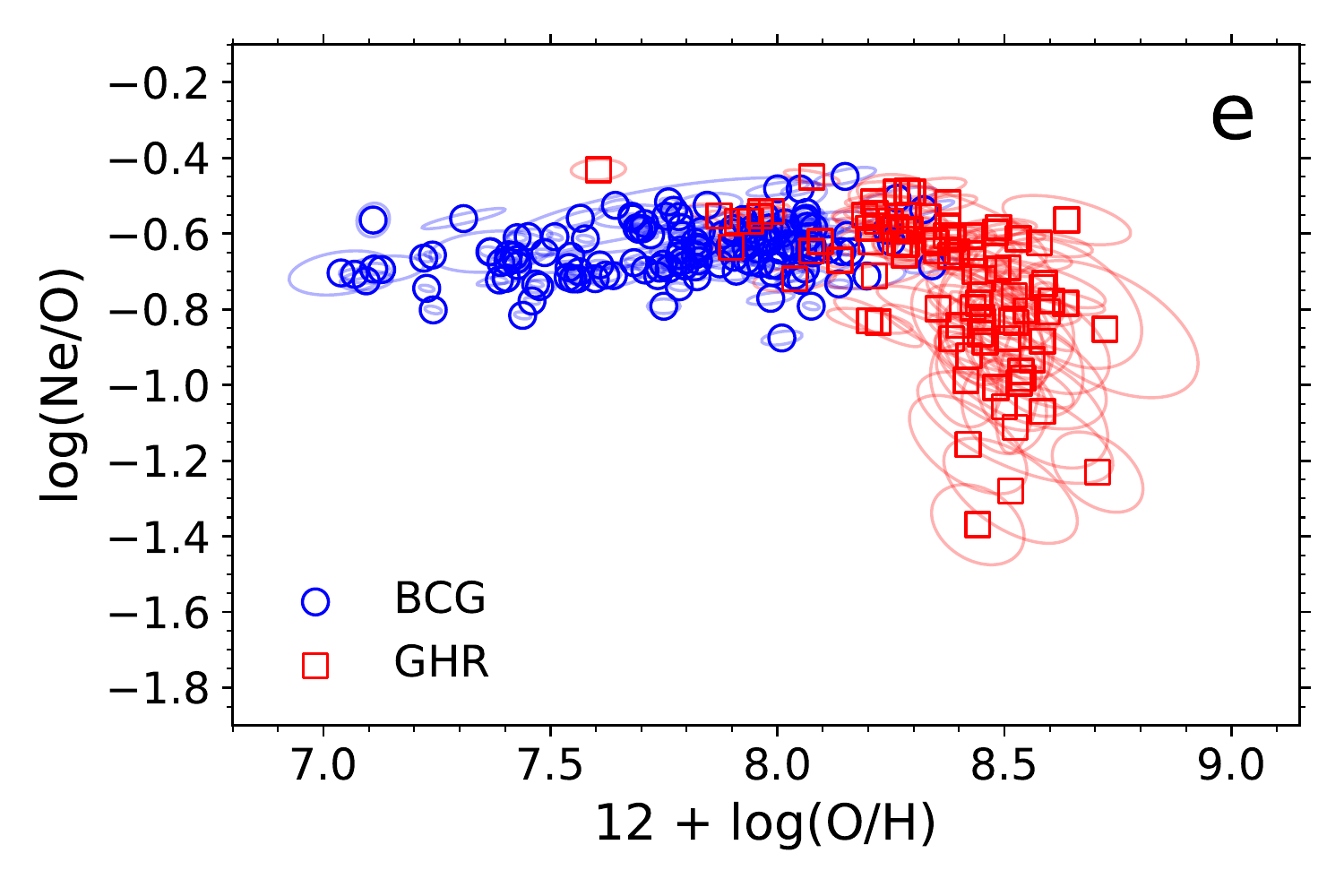}\hfill
\includegraphics[width=.32\textwidth]{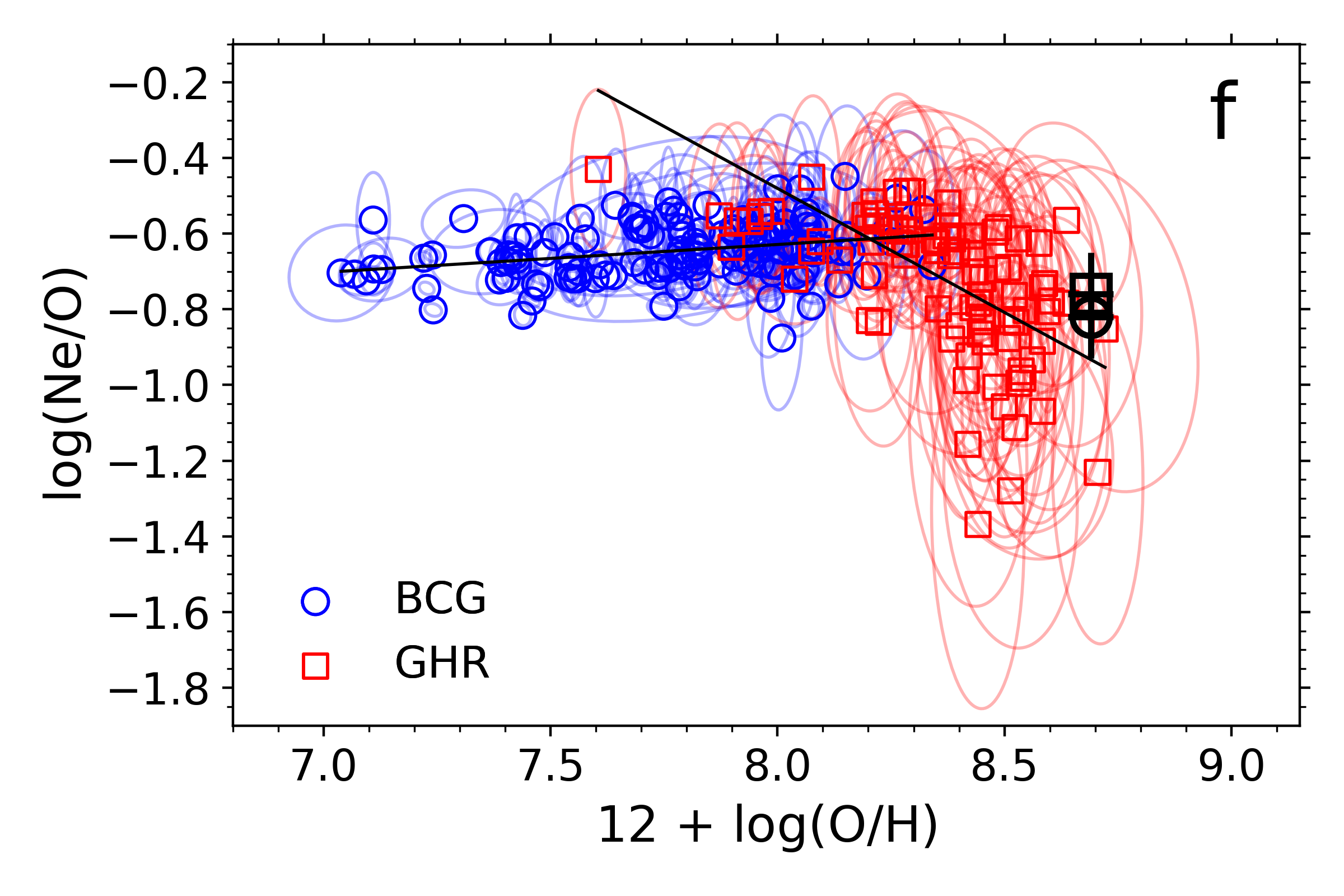}\\
\includegraphics[width=.32\textwidth]{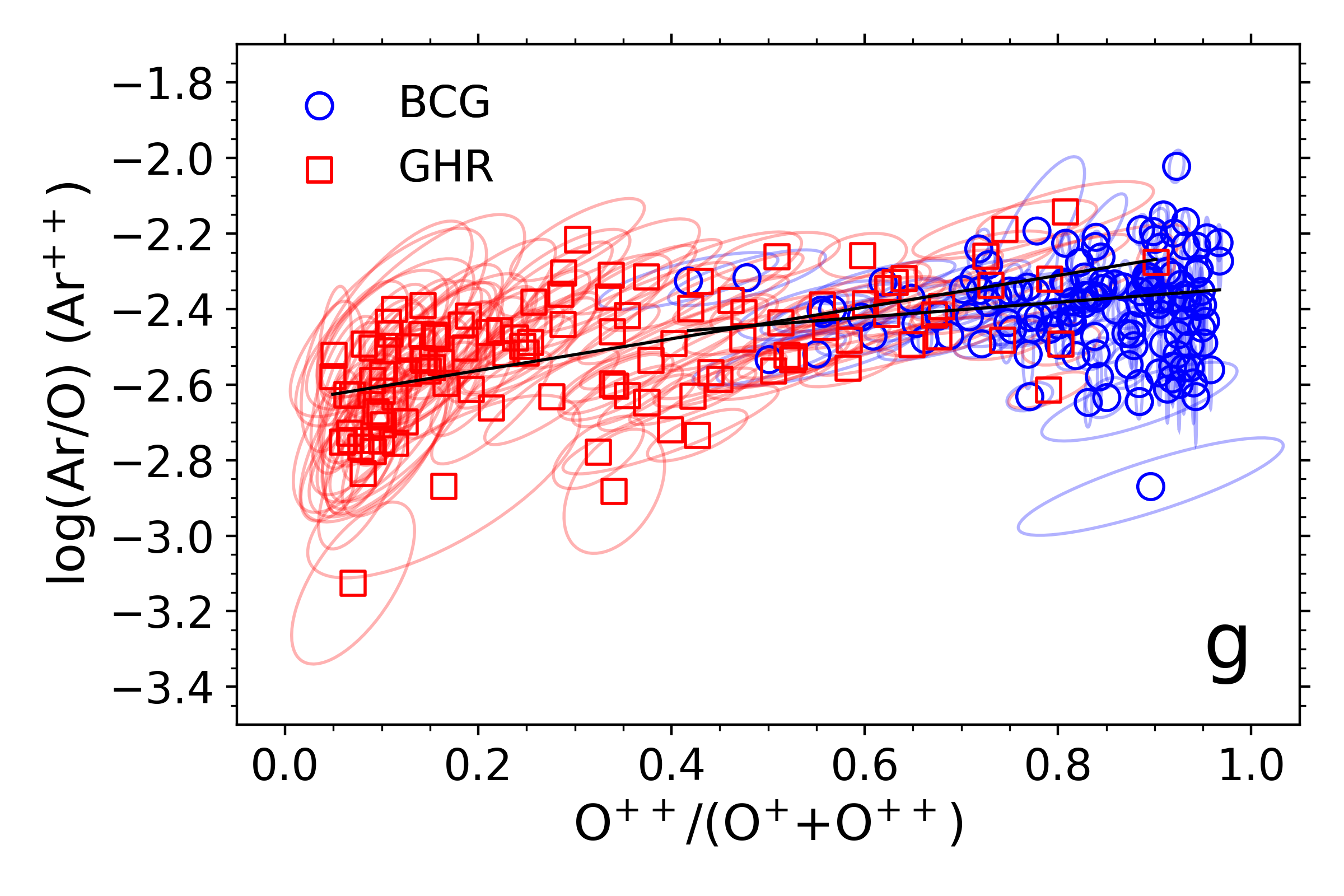}\hfill
\includegraphics[width=.32\textwidth]{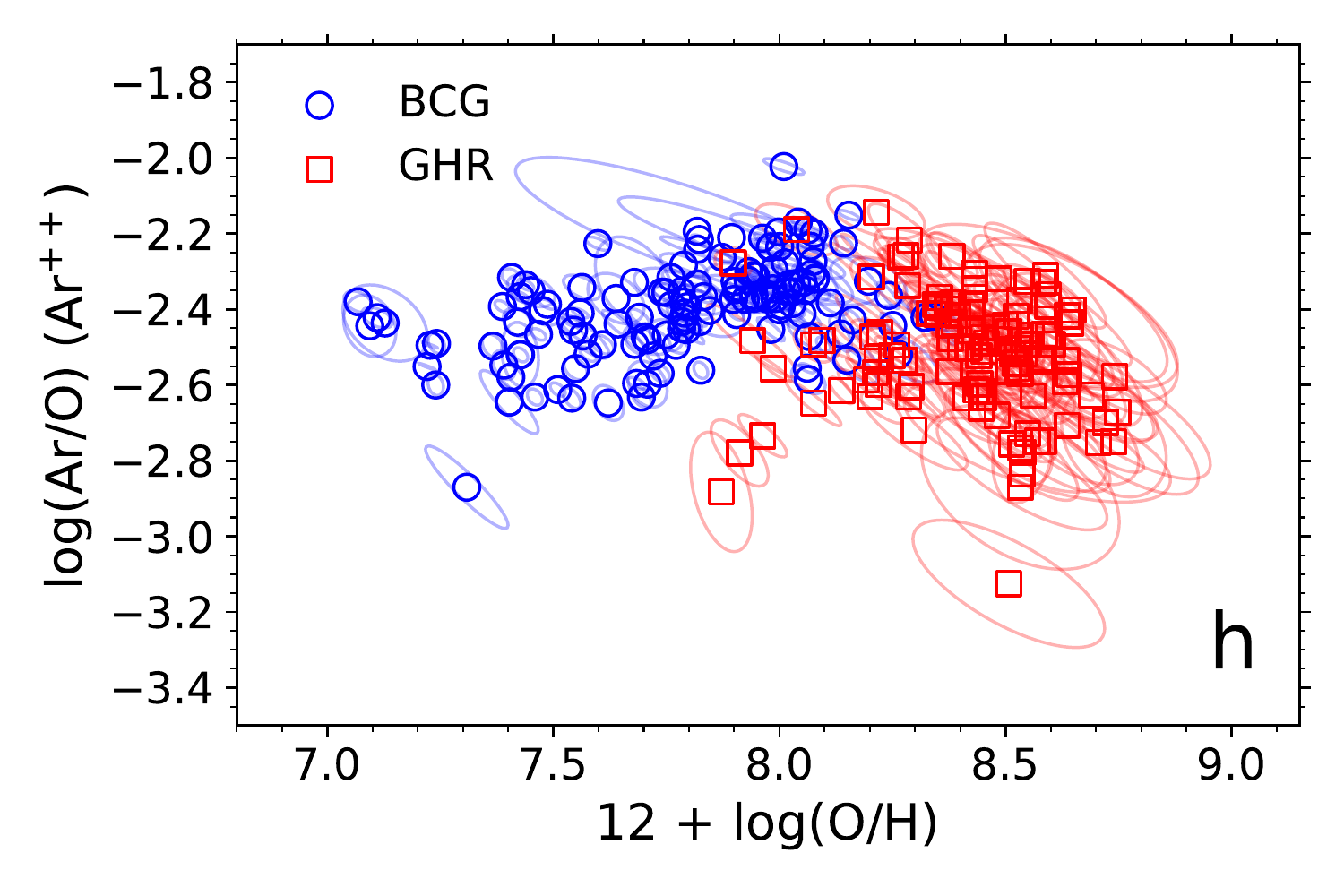}\hfill
\includegraphics[width=.32\textwidth]{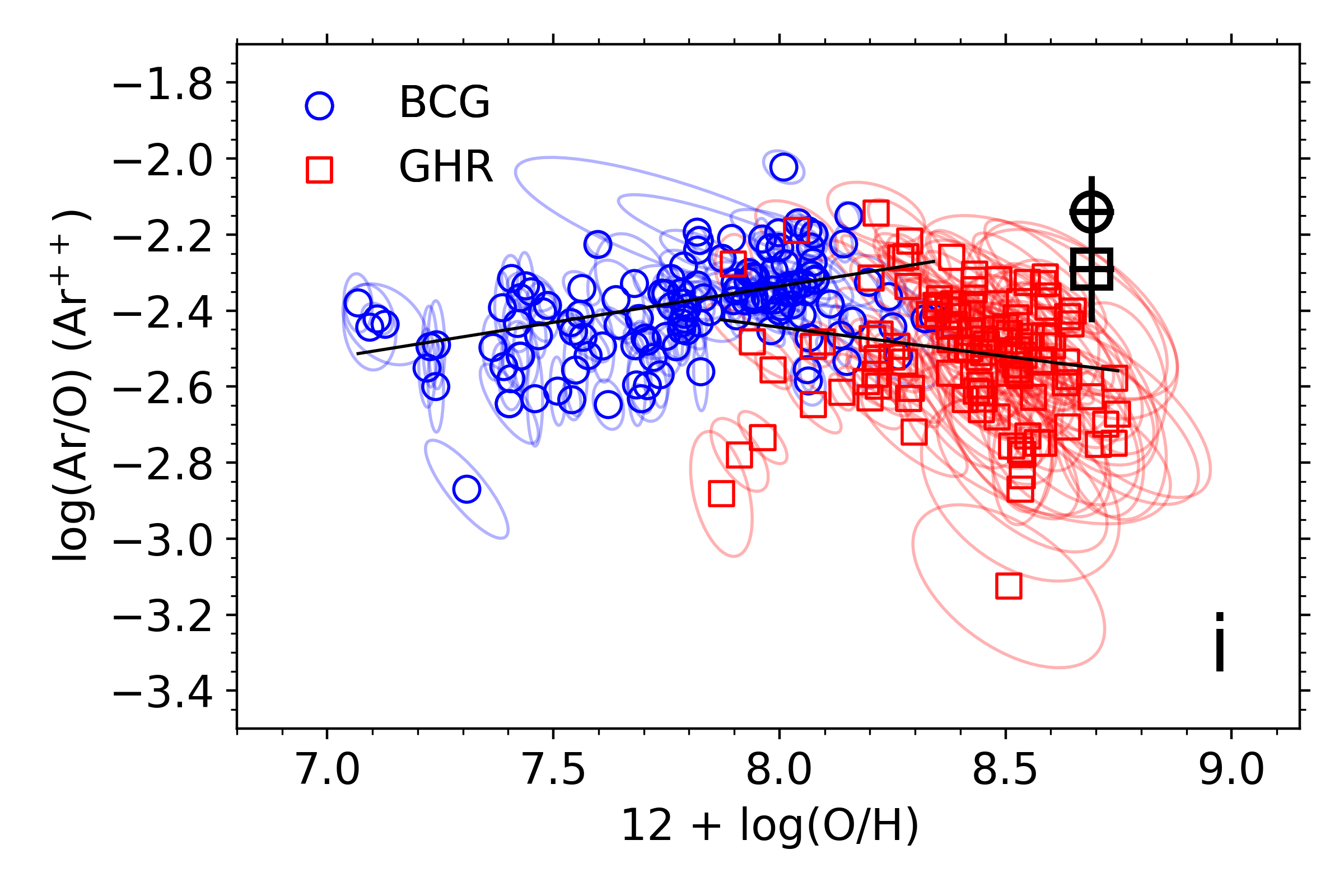}\\
\includegraphics[width=.32\textwidth]{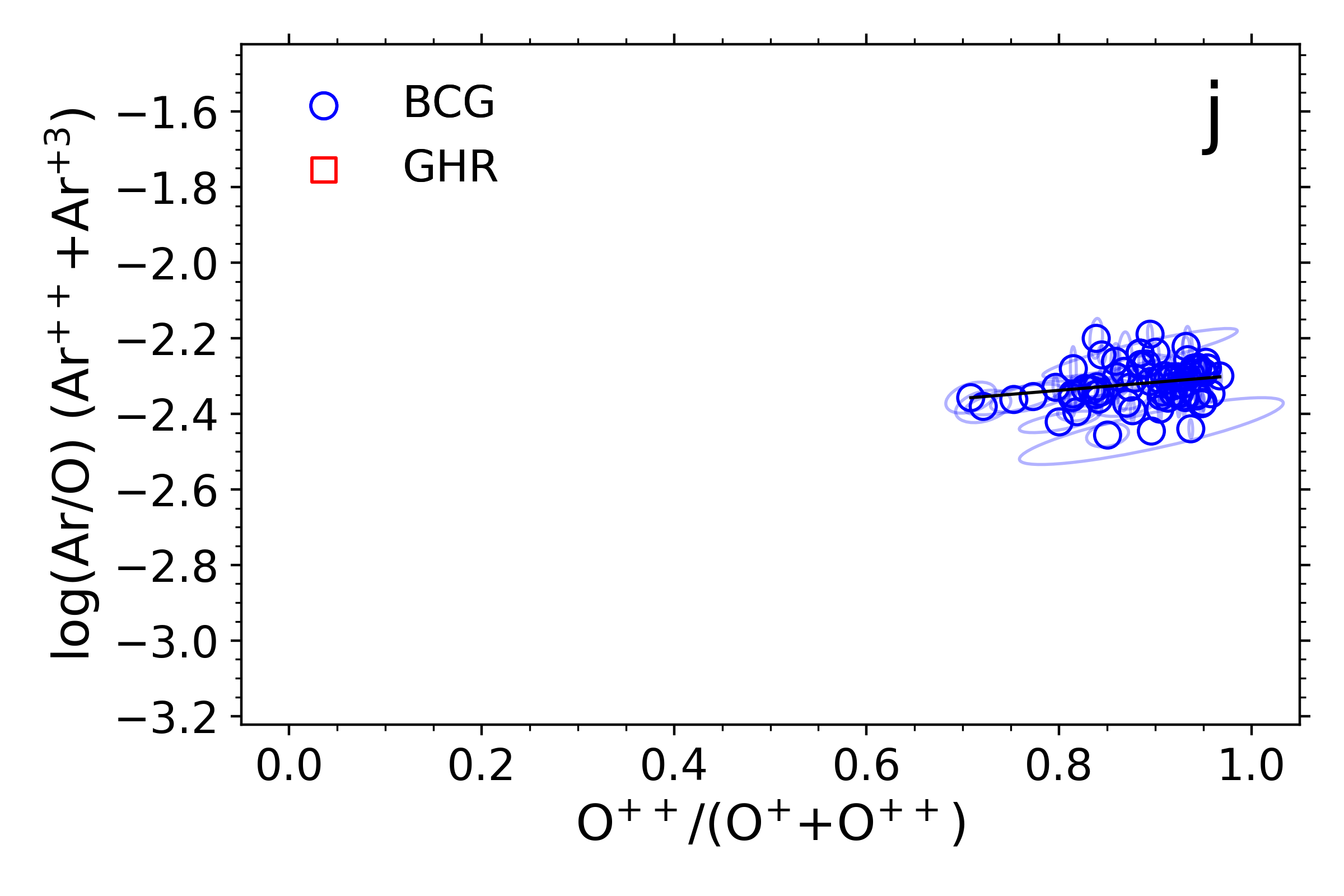}\hfill
\includegraphics[width=.32\textwidth]{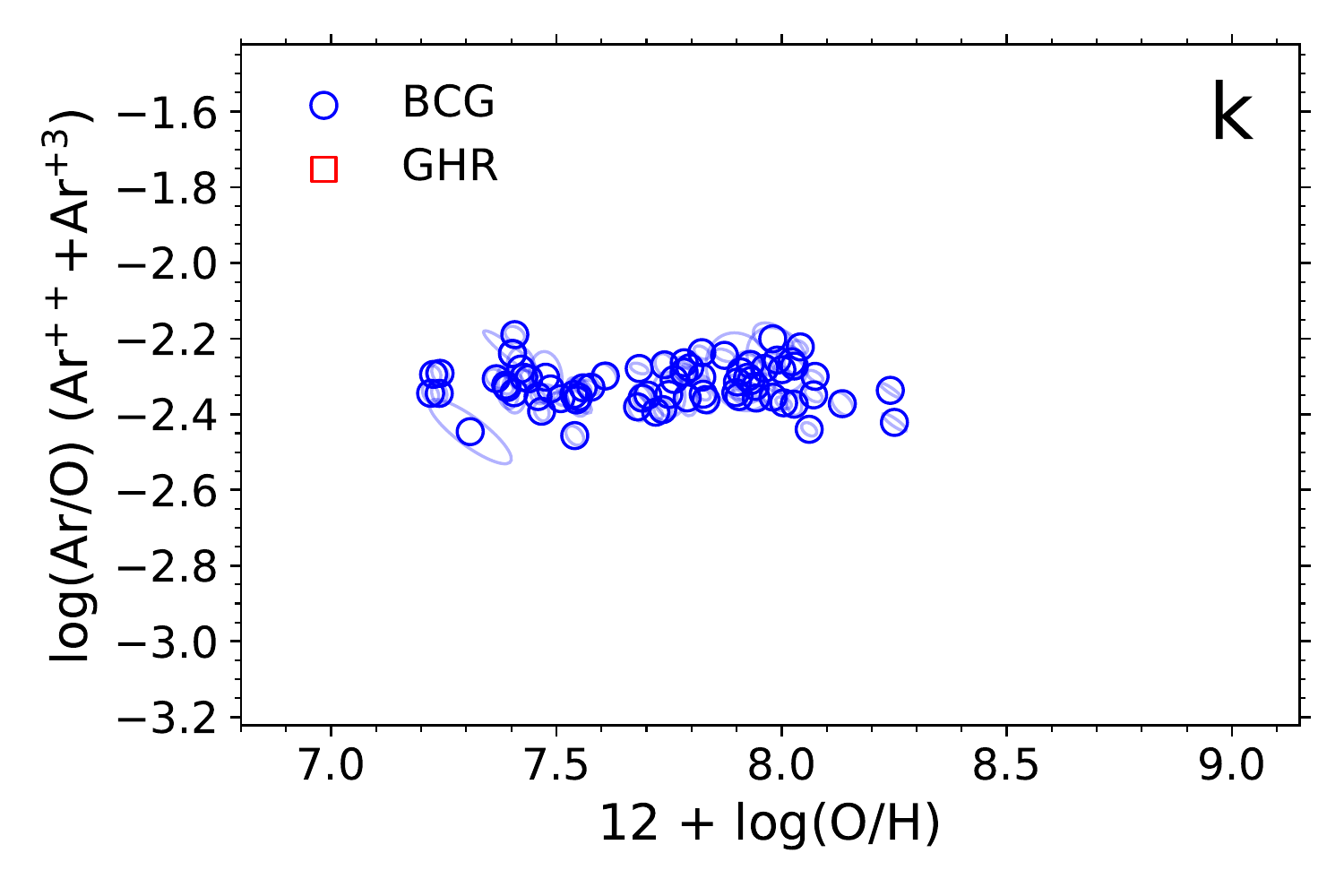}\hfill
\includegraphics[width=.32\textwidth]{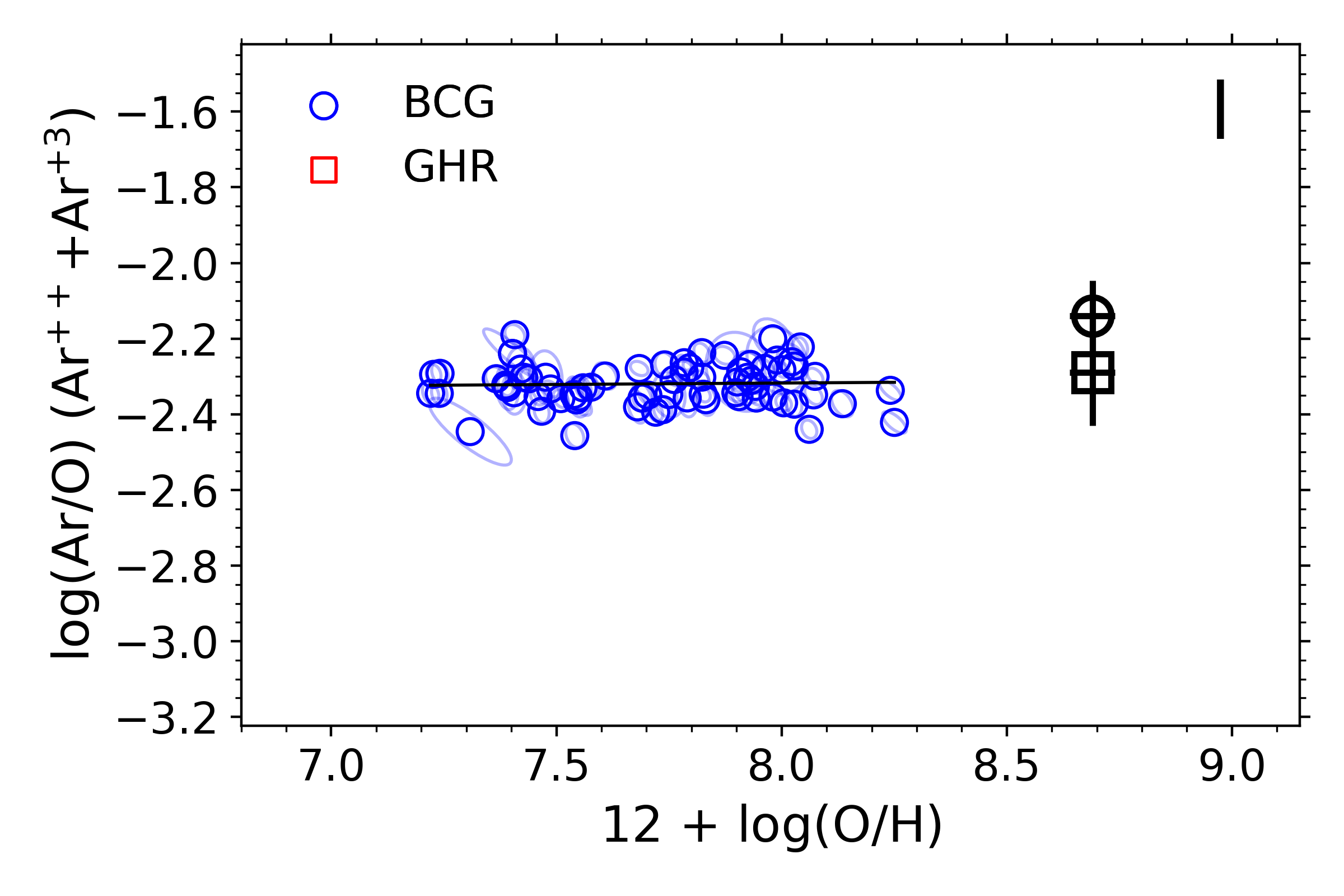}\\
\includegraphics[width=.32\textwidth]{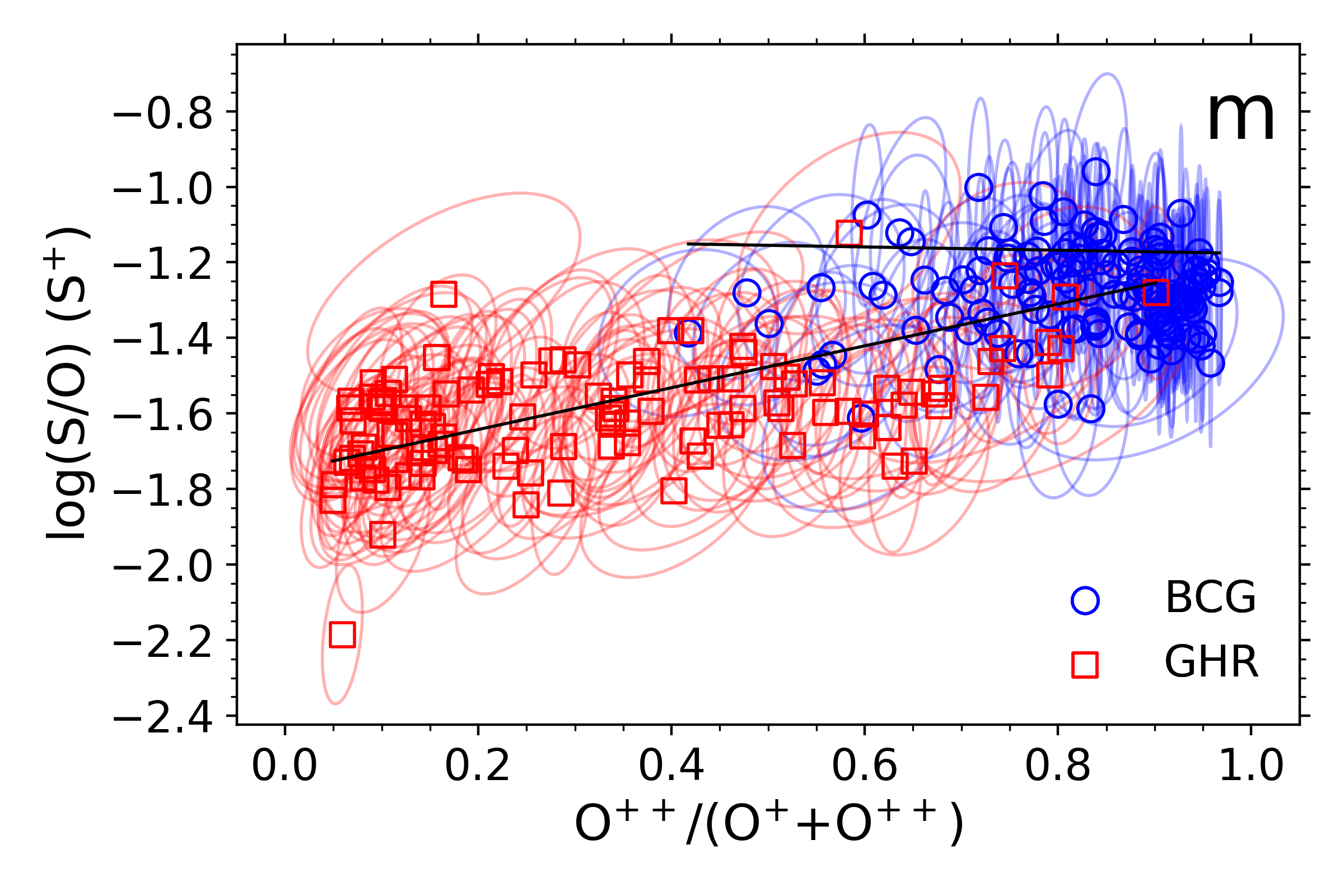}\hfill
\includegraphics[width=.32\textwidth]{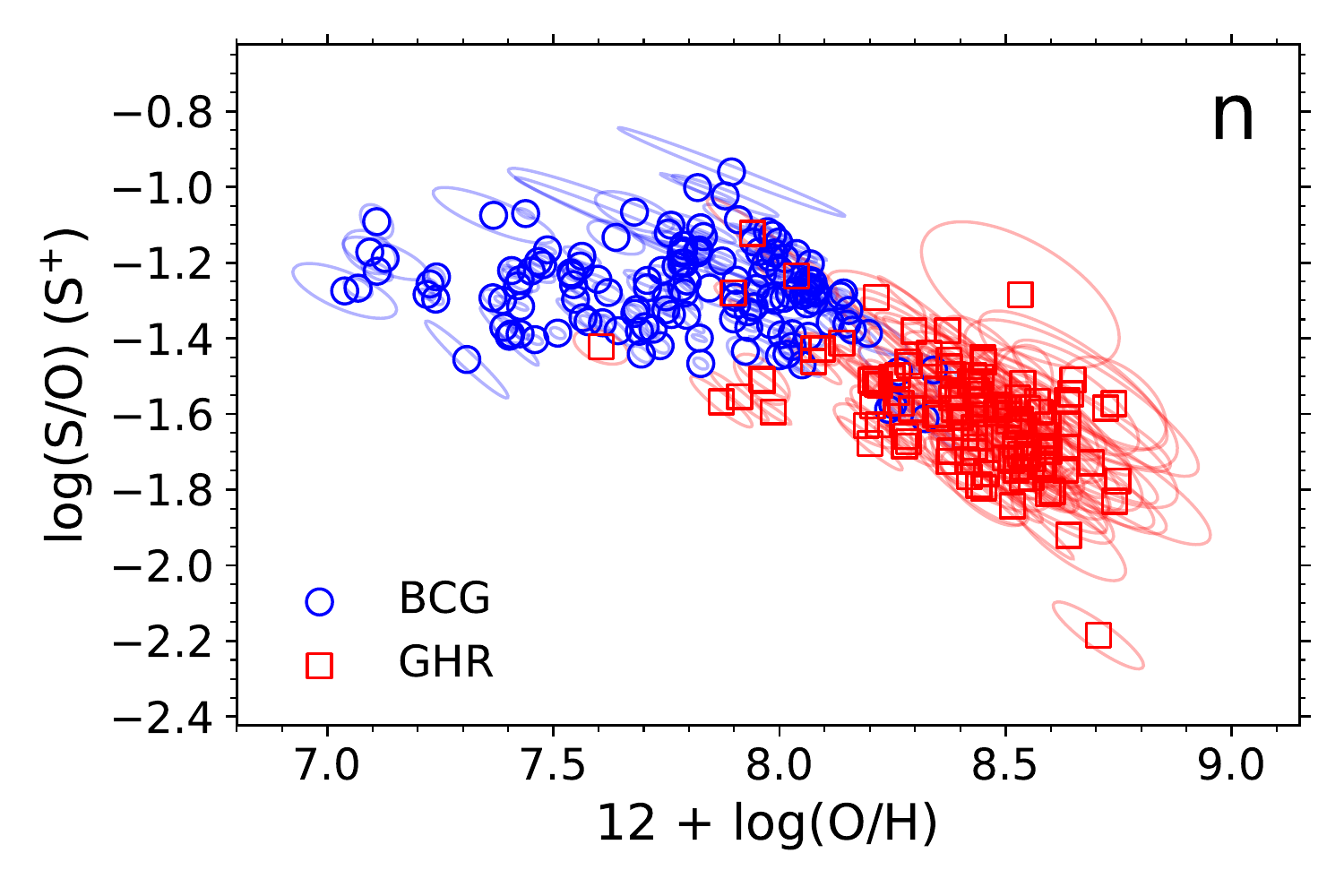}\hfill
\includegraphics[width=.32\textwidth]{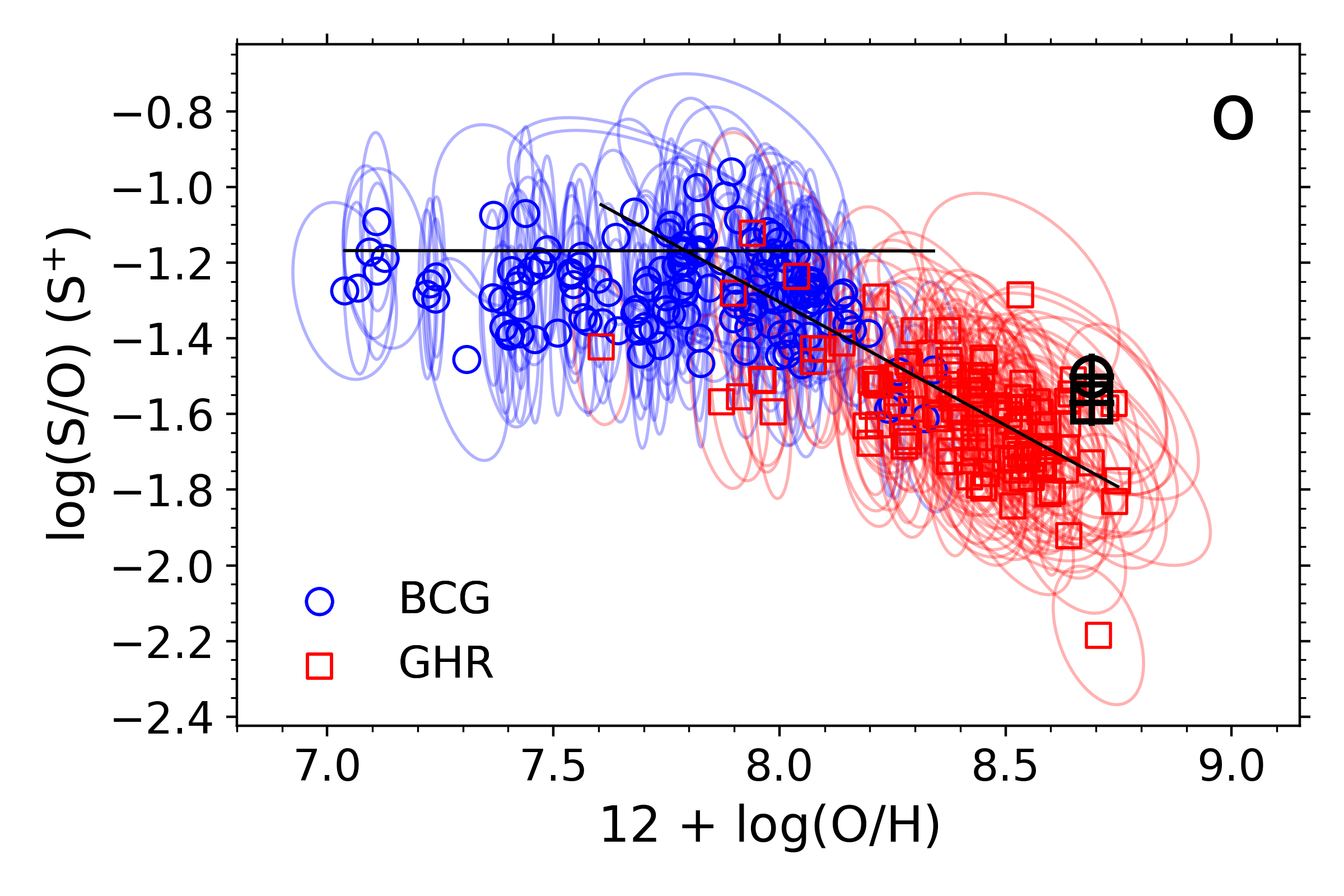}\\
\includegraphics[width=.32\textwidth]{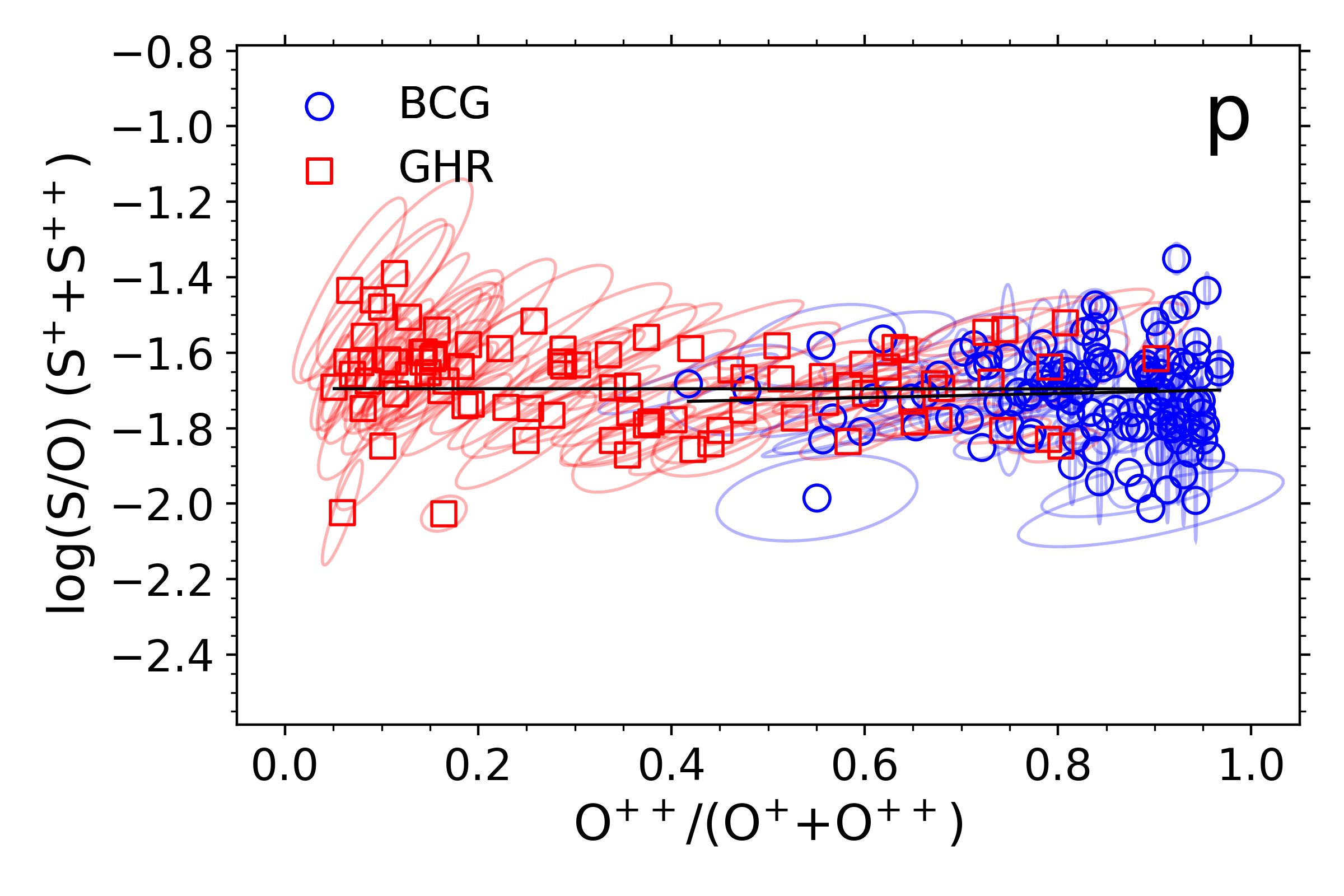}\hfill
\includegraphics[width=.32\textwidth]{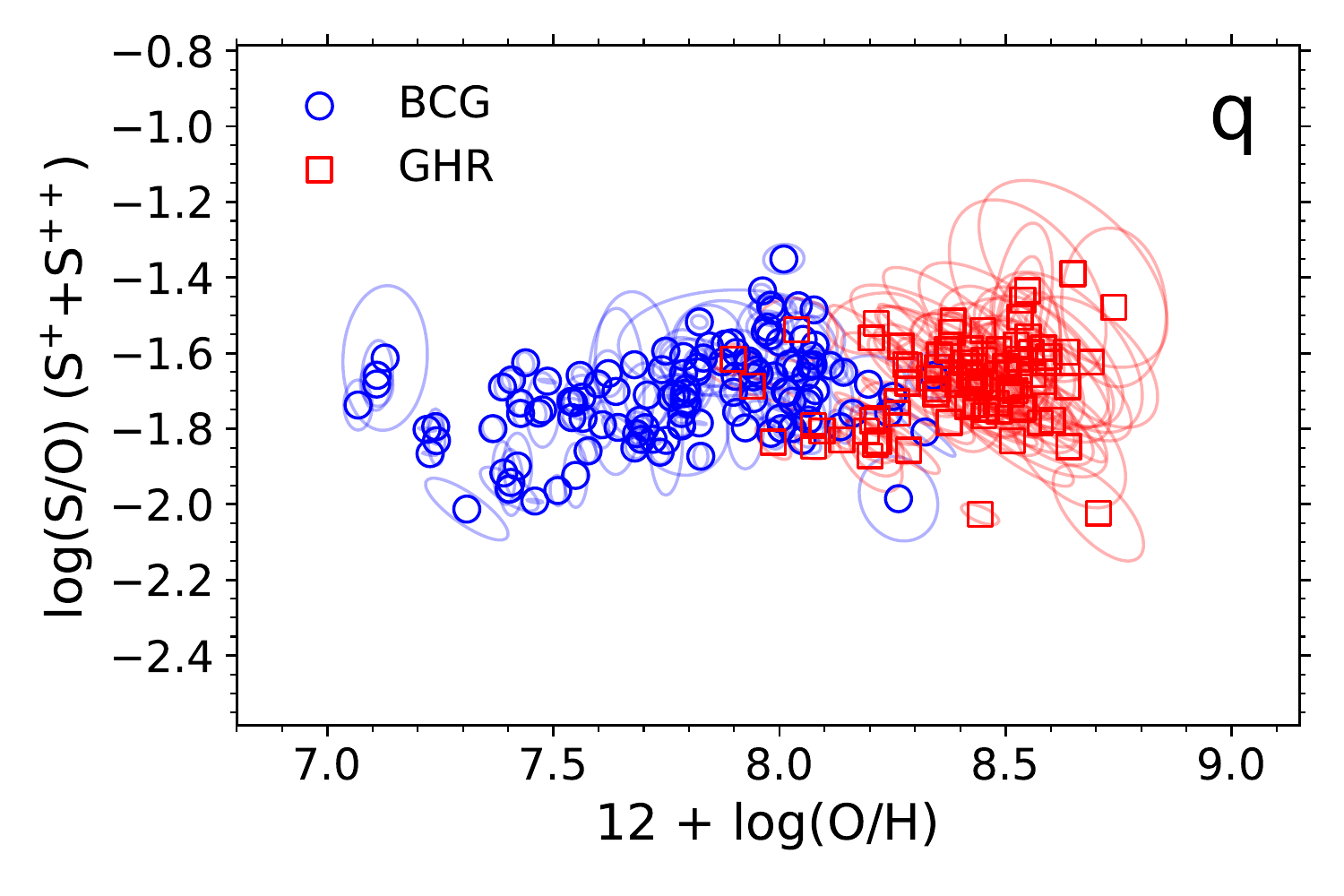}\hfill
\includegraphics[width=.32\textwidth]{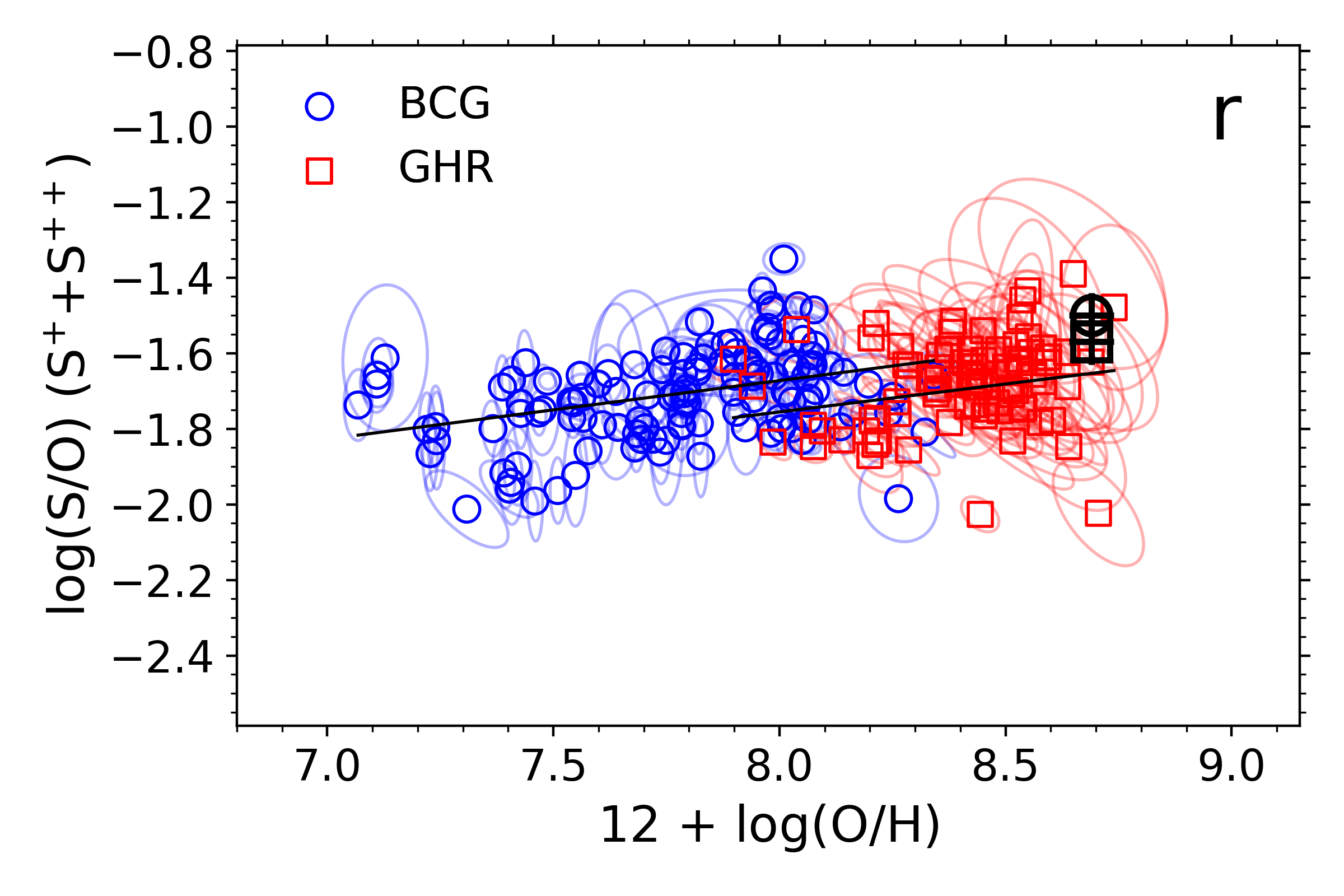}
\end{minipage}
\caption{Values of N/O, Ne/O, Ar/O, and S/O as a function of $\omega$ (left panels) and O/H (middle and right panels) for the BCG (blue circles) and GHR (red squares) samples. Covariance ellipses in the middle panels include the uncertainties derived only from the emission lines uncertainties and the \tempnii\ vs \tempoiii\ relation while the ellipses in the left and right panels also take into account the uncertainties associated with the ICFs. Black symbols represent the solar values of \citet{Lodders2003} (circles) and \citet{Asplund2009} (squares).}
\label{fig:NeArSAbunds}
\end{figure*}

\begin{table}
\centering
\caption{Coefficients of the regression lines  $y = ax + b$ for the abundance ratios presented in Fig.~\ref{fig:NeArSAbunds},  $x$ being equal to $A(\rm{O}) = 12 + \log$ O/H and $y$ being the logarithm of the abundance ratio.}

\begin{tabular}{llrr}
\hline
$y$     &     $x$        &   $a    $                  & $b    $\\
\hline 
N/O (BCG) & $\omega$    & $0.59\pm0.22$            & $-1.64\pm0.18$\\
N/O (GHR) &  $\omega$    & $-0.70\pm0.14$           & $-0.70\pm0.04$\\
Ne/O (BCG) &  $\omega$   & $-0.19\pm0.14$           & $-0.48\pm0.12$\\
Ne/O (GHR) &  $\omega$      & $0.49\pm0.23$                        & $-0.91\pm0.12$\\
Ar/O from Ar$^{++}$ (BCG) &  $\omega$   & $0.20\pm0.10$          & $-2.54\pm0.08$\\
Ar/O from Ar$^{++}$ (GHR) &  $\omega$   & $0.42\pm0.07$           & $-2.65\pm0.03$\\
Ar/O from Ar$^{++}$ + Ar$^{+3}$ (BCG) &  $\omega$   & $0.21\pm0.10$ & $-2.51\pm0.09$\\
S/O from S$^+$ (BCG) &  $\omega$   & $-0.04\pm0.20$             & $-1.13\pm0.17$\\
S/O from S$^+$ (GHR) &  $\omega$   & $0.55\pm0.11$             & $-1.75\pm0.04$\\
S/O from S$^+$ and S$^{++}$ (BCG) &  $\omega$   & $0.06\pm0.11$  & $-1.75\pm0.09$\\
S/O from S$^+$ and S$^{++}$ (GHR) &  $\omega$   & $-0.001\pm0.060$  & $-1.70\pm0.03$\\
\hline
N/O (BCG) & $A(\rm{O})^a$  & $0.26\pm0.08$                       & $-3.17\pm0.59$\\
N/O (GHR) & $A(\rm{O})$  & $1.20\pm0.23$                        & $-11.00\pm1.90$\\
Ne/O (BCG) & $A(\rm{O})$   & $0.07\pm0.04$                       & $-1.22\pm0.28$\\
Ne/O (GHR) & $A(\rm{O})$  & $-0.66\pm0.26$                        & $4.76\pm2.13$\\
Ar/O from Ar$^{++}$ (BCG) & $A(\rm{O})$    & $0.19\pm0.04$          & $-3.86\pm0.34$\\
Ar/O from Ar$^{++}$ (GHR) & $A(\rm{O})$    & $-0.15\pm0.16$           & $-1.22\pm1.33$\\
Ar/O from Ar$^{++}$ + Ar$^{+3}$ (BCG) & $A(\rm{O})$    & $0.01\pm0.03$ & $-2.38\pm0.21$\\
S/O from S$^+$ (BCG) & $A(\rm{O})$   & $-0.001\pm0.074$             & $-1.16\pm0.58$\\
S/O from S$^+$ (GHR) & $A(\rm{O})$    & $-0.65\pm0.21$             & $3.91\pm1.73$\\
S/O from S$^+$ and S$^{++}$ (BCG) & $A(\rm{O})$    & $0.15\pm0.05$  & $-2.91\pm0.35$\\
S/O from S$^+$ and S$^{++}$ (GHR) & $A(\rm{O})$    & $0.15\pm0.12$  & $-2.94\pm1.00$\\
\hline
\end{tabular}
\label{tab:regressions}
\end{table}

In the following, we examine the abundances obtained in our BCG and GHR samples using our ICFs.
Figure~\ref{fig:NeArSAbunds} shows the abundances obtained in both of our observational samples (with the BCG sample represented by blue circles and the GHR sample by red squares).The symbols correspond to the positions of the nominal values of the abundances.
In each panel, the ellipses represent the covariance of the Monte Carlo experiments distribution of each object, at 1$\sigma$ from the nominal values of each case. In the middle panels the covariance ellipses include the uncertainties derived from the observed emission line intensities and from the dispersion in the \tempnii\ vs \tempoiii\ relation while in the left and right panels they also take into account the uncertainties associated to the ICF.  

In Table \ref{tab:regressions} we also give the equations for the regression lines, $y = ax+b$,  obtained separately for the BCG and the GHR samples. Here, $y$ represents the ordinate, while $x$ represents $\omega$ or $12 + \log \rm{O/H}$, depending on the case, and the regression is computed taking into account errors in both ordinates. On the panels of the third column, we also indicate the corresponding value of the abundance radio as given by \citet{Lodders2003} (black circles) and \citet{Asplund2009} (black squares).

\subsection{Oxygen}
\begin{figure}
\centering
\includegraphics[width=.45\textwidth]{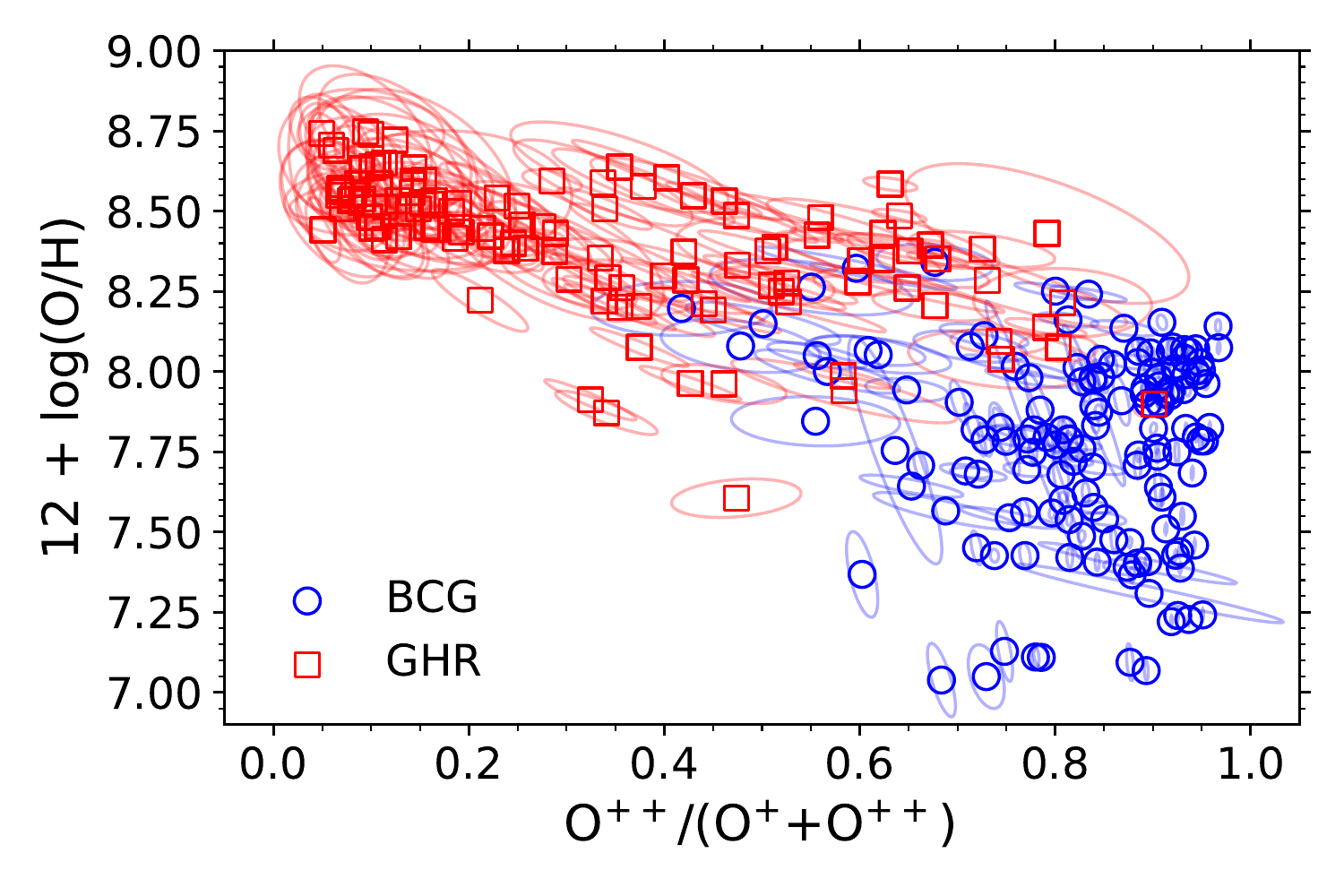}
\caption{Oxygen abundances of the BCG (circles) and GHR (squares) samples as a function of $\omega$.}
\label{fig:12logOH}
\end{figure}

Figure~\ref{fig:12logOH} shows the abundances of oxygen, computed as O$^+$/H$^+$ + O$^{++}$/H$^+$, as a function of $\omega$. There is a clear trend of increasing degree of ionization as O/H decreases. This trend has already been noted in the past \citep[e.g.,][]{McGaugh1991}. It is due to a softer ionizing SED at higher metallicities due to the effect of the metals on the stellar interiors and atmospheres \citep{Pagel1979,Maeder1990} as well as on the ionization parameter through the increasing strength of the winds \citep{Dopita2006}. Since this behavior is related to metallicity, the same trend is expected for other elements.

\subsection{Nitrogen}

The upper row of Fig.~\ref{fig:NeArSAbunds} shows the values of N/O as a function of $\omega$ (left panel) and of O/H  (the other two panels).  It can be seen that at the highest metallicities, the uncertainties in N/O are much smaller than those in O/H. As already known (c.f. VA16) they anti-correlate with the uncertainties in O/H due to the opposite temperature dependencies of N/O and O/H. At the lowest metallicities, which correspond to high values of $\omega$, the uncertainties in N/O become substantial due the uncertainty in the ICF, and may reach values of up to $\pm\ 0.2$ dex.

The well-known increase of N/O with O/H at high metallicity remains significant despite the uncertainties due to the ICFs. The linear regression gives a slope of $1.20\pm\ 0.23$ for the GHRs. This increase is due to the fact that nitrogen production is both primary and secondary at these metallicities  \citep{MollaGavilan2010}. Primary nitrogen is synthesized from the oxygen and carbon produced before the CNO cycling and thus, is independent of the initial heavy-element abundances while secondary nitrogen is produced during the CNO cycle and depends on the initial abundance of the heavy elements in the star \citep{Henry2000}. As discussed by \citet{Pagel2009}, variations in N/O can also occur due to the time-delayed production of N from low- and intermediate-mass stars or due to a preferential loss of oxygen in the galactic winds produced by core-collapse supernovae.
For the BCG sample, i.e. at low metallicities, the slope of the N/O versus O/H relation is $0.26\pm\ 0.08$. Some objects, however, are found at quite high values of N/O. Our study shows that uncertainties in the ICF cannot be the cause of this. \citet{Izotov2006} have argued that theoretical models for massive stars including rotation \citep{MaederMeynet2005} could perhaps explain high values of N/O at low metallicities if taking into account that the density of the N-rich ejecta is larger than that of the \hii regions. This question, however, deserves a more dedicated study.

\subsection{Neon, argon, sulfur}

Neon, argon and sulfur, as well as oxygen are $\alpha$-elements produced by massive stars and the common view  \citep{henryworthey99,Prantzosal20} is that their abundances should evolve in lockstep inside and among galaxies. Panels d through r of Fig.~\ref{fig:NeArSAbunds} show the values of Ne/O, S/O and Ar/O as a function of $\omega$ in the left column, and as a function of O/H in the middle and right columns. The meaning of the symbols is exactly the same as described for N/O. For S/O and Ar/O, two different determinations are shown, depending on the ions (and thus, ICF prescription) used as indicated along the axes.

We can readily see that in most panels, the observational points do not gather along horizontal lines. We now discuss these elements separately.

For the BCG sample, the Ne/O ratio shows a slight increase as O/H increases. The slope of the regression line is $0.07 \pm\ 0.04$, similar to what was found by \citet{Izotov2006} for their sample. Our GHR sample shows a clear opposite trend, with a slope of $-0.66 \pm\ 0.26$. One must note, however, that the uncertainties in Ne/O due to the ICFs are much larger in this metallicity region---due to the fact that the values of $\omega$ are generally small. It is not excluded that for the GHRs of our sample, our ICFs produce an incorrect trend, underestimating the neon abundance especially for objects of low excitation.

For Ar/O we have two possible estimates, one based on  Ar$^{++}$ only, the second based on Ar$^{++}$ + Ar$^{+3}$. This second estimate cannot be used for our GHRs sample, where the [\ion{Ar}{iv}] \Lam4740 line is not measured. We note that the ICF based on Ar$^{++}$ + Ar$^{+3}$ leads to a smaller dispersion in the Ar/O ratios for the BCG sample, despite the fact that it uses an additional line thus involving an additional observational error. This argues in favor the latter ICF being more reliable. The uncertainties in Ar/O for the BCG sample are very small. We note that the slope of the regression line with respect to O/H is consistent with zero with a very small dispersion. 

We note also that the average value of Ar/O is slightly larger when using the ICF based on Ar$^{++}$ + Ar$^{+3}$ and closer to the solar Ar/O ratio as given by \citet{Asplund2009} (however given the uncertainty in the solar Ar abundance which, as well as for Ne, is probably larger than quoted, we cannot use the solar value as an argument \textit{pro} or \textit{contra} our ICFs). Despite the warning given in Sect.  \ref{sec:modelsel} it turns out that the ICFs based on Ar$^{++}$ + Ar$^{+3}$ give quite reasonable values. 

For S/O we again have two possible estimates, one based on S$^{+}$ alone, the second depending on S$^{+}$+ S$^{++}$. The values obtained in the first case are significantly larger than in the second case (by about  0.6 dex at low metallicities). The uncertainties due to the ICFs are especially large for the first case (typically $\pm$0.2 dex or more at high values of $\omega$), less so for the second case We also note that, for the GHR sample, the first ICF (panel m) gives an increase in S/O as  $\omega$ increases while the second  ICF (panel p) gives a slight decrease as $\omega$ increases. Obviously at least one of the two estimates of S/O must be wrong. The lines emitted by  S$^{+}$ come from the outskirts of the \hii regions, and adding the information from S$^{++}$ obviously allows a better representation of the object. Therefore we do not recommend ICFs based on S$^{+}$ alone. If the only information one has on sulfur in a spectrum comes from [\ion{S}{ii}] lines one must be very suspicious on the derived abundances.

\section{Hints on element depletion into dust}
\label{sec:depletion}

Because we have a large observational sample of good quality spectroscopic data, we can use it to see if something may be said about depletion patterns in \hii regions.
We consider information coming only from our most reliable results for Ne/O , Ar/O and S/O (BCGs in panels f, l and r).
The coefficients for the regression lines corresponding to the BCG samples are listed in Table \ref{tab:regressions}.

In \citet{Izotov2006} the behavior of Ne/O with O/H was attributed to an increased depletion of oxygen atoms into dust particles as metallicity increases, on the argument that  both O and Ne are $\alpha$-elements and Ne, being a noble gas, is not expected to be found in grains. The idea of an increasing dust-to-metal ratio with increasing metallicity goes back to \citet{Lisenfeld98}, and has recently been confirmed on large samples of galaxies covering a wide range of metallicities by \citet{Remy2014} and \citet{Devis19}. 

If we attribute the rise of Ne/O with O/H only to oxygen depletion this means that from $A(\rm{O}) = 7$ to $A(\rm{O}) = 8.67$ (the cosmic standard derived by \citet{Przybilla2008} from B stars), the oxygen depletion has increased by 0.12 dex. Assuming that there is no oxygen depletion at $A(\rm{O}) = 7$, this means that the depletion at solar metallicity is 0.12 dex, which is compatible with the value of 0.11 dex estimated by \citet{Peimbert&Peimbert2010} for the solar vicinity from completely different arguments.
Argon being also an $\alpha$-element and a noble gas, so a priori not expected to be depleted into dust, one would expect that Ar/O has the same behavior with respect to O/H as Ne/O. Given the error bars, this is roughly what we find.

If we dig a little further, we first note that Ar/O is practically independent of O/H (the slope is $0.01\pm0.03$). We also may remark that both neon and argon have very small condensation temperatures---9.3 and 48 K, respectively, as opposed to 182 K for oxygen \citep{Lodders2003}---so they are not expected to be found among grain constituents (see e.g, \citealt{Sofia2004}). A deficiency in argon has been observed in the diffuse interstellar medium as well as in damped Lyman $\alpha$ systems\citep{Duley80,JenkinsOegerle00,Zafar2014}, although it has been argued that the cause is not depletion into dust but simply ionization effects.
However, interplanetary dust particles \textit{do} contain neon and other noble gases \citep{Kehm98,Pepin2011,Ott2014}. The interpretation would be that they have been adsorbed on the grain surfaces \citep{Duley1985}.

Then, the fact that we find the Ar/O ratio does not depend on O/H could mean that both argon and oxygen are depleted into dust.
Regarding sulfur, we find that S/O in the BCGs rises with respect to O/H slightly more than Ne/O, and definitely more than Ar/O, as shown by Table \ref{tab:regressions}.

If interpreted in terms of depletion of S into dust grains, this would imply that sulfur is less depleted than oxygen. However, as argued by \citet{WhiteSofia2011} sulfur is expected to be among the dominant components of interstellar dust. Besides, it is found incorporated in interstellar silicates \citep{Bradley1999}. On the other hand,  high-resolution HST observations of interstellar abundances failed to show any sulfur depletion until \citet{Jenkins2009} suggested that sulfur depletion may occur along some lines of sight.

The hope for a quantitative discussion of dust depletion patterns based on the analysis of a large sample of giant \hii\ regions under the hypothesis that variations in the  O/Ne/S/Ar abundance ratios are due only to dust depletion is significantly curtailed by recent nucleosynthesis studies. \citet{Seitenzahlal13} and \citet{leungnomoto18} find that, not only Type II but also Type Ia Supernovae can produce $\alpha$-elements, in amounts that depend on the supernova models. \citet{Kobayashial20} find that in their model for the chemical evolution of the Milky Way this may result in a contribution of  29\% for sulfur and 34\% for argon in the solar vicinity. Therefore, the discussion of element depletion into grains cannot be disconnected from that of the role of Type Ia Supernovae in galaxies, which is still uncertain as shown by \citet{Palla21}.

\section{Summary}
\label{sec:conclusions}

Using the large grid of photoionization models for extragalactic \hii regions presented in \citet{Vale2016} (VA16), we derived new ionization correction factors (ICFs) for carbon, nitrogen, neon, sulfur, chlorine and argon. We also provided formulae to compute the associated uncertainties in the ICFs, something that is not available in previous works on \hii regions. 

From the original set of models, we selected only the most representative ones based on their starburst ages, matter-bounded cuts, and positions in three diagrams (O/H versus N/O, $U$ versus O/H, and [O III]/H$\beta$ versus [N II]/H$\alpha$) with respect to the positions of real objects. From this selection, we obtained a final sample of $\sim$1800 photoionization models ideal for our purposes. We assigned to each model a weight  according to the number of observed objects in the same zone of the BPT diagram. 

The observational data used in this work consist of 985 objects from a sample of low metallicity blue compact galaxies and a sample of giant \hii regions in spiral galaxies. Both samples were observed with large telescopes and most of them have the emission lines needed to compute the physical conditions in their spectra. From this sample we excluded those objects that show signatures of extra ionizing sources, in addition to the ionization by young stars, as shown by their positions in the BPT diagram and worked only with those with reliable intensities, ensuring to have only good quality data and covering a wide range of degrees of ionization and metallicities. 
 
We computed the ICFs for C, N, Ne, Cl, Ar and S and produced polynomial expressions for their weighted medians as well as for the weighted 16 and 84 percentiles as a function of  $\omega = $ O$^{++}$/(O$^{+}$+O$^{++}$).

With these ICFs we derived the total abundances ratios of N/O, Ne/O, Ar/O and S/O in the \hii regions of our observational sample. For S and Ar we derived two different abundances depending on the ions used. We presented the results with their uncertainties 1) including only the uncertainties associated to the emission lines intensities and to the dispersion in the  \tempnii\ vs \tempoiii\ relation and 2) taking into account the uncertainties associated to our ICFs prescriptions. 

The main results are the following. The uncertainties in the ICFs for N/O are important at large values of $\omega$ while those for Ne/O are important at low values of $\omega$. Given that, as known, the values of $\omega$ tend to increase for lower metallicities, this implies that the N/O values may be uncertain (by up to $\pm$0.2 dex) in blue compact galaxies, while the Ne/O values  may be uncertain (by as much as $\pm$0.3 dex or more) in the giant \hii regions of the central parts of spiral galaxies.  
 
For sulfur we warn against S/O abundance ratios derived using only the [\ion{S}{ii}] lines as they may be uncertain to within $\pm$0.3 dex. A similar concern regards the Ar/O ratios when obtained from the [\ion{Ar}{iii}] alone at high metallicities, although the uncertainties are not that large. 

From the best determined abundance ratios, we conclude that oxygen is depleted into dust grains in a proportion increasing with metallicity and reaching about 0.12 dex at solar abundances. 

Concerning the possible depletion of sulfur and argon into dust grains, we cannot reach any quantitative result not only because of uncertainties in spite of our careful study. Recent chemical evolution models based on new nucleosynthesis computations for Type Ia Supernovae suggest that, although sulfur and argon are mainly produced by core-collapse supernovae, the latter may contribute to their production. Therefore the question of element depletion onto dust has to be discussed simultaneously with chemical evolution models of galaxies that include modern yields for Type Ia Supernovae. Note that direct comparison of sulfur and argon abundances from \hii regions with stellar values is not feasible since stellar abundances for these elements are notoriously uncertain.

The above results are based on \hii region photoionization models and on ionic abundance calculations, both of which are dependent on atomic data -- and on a proper theoretical description of the ionizing SED as regards the respective ionization structure of the different elements in \hii regions. Slight changes can occur in the future that could affect our results but in case of need this study can be repeated with the same methodology.

\section{Data availability}
The grid of photoionization models used in this work is available in the Mexican Million Models database 3MdB at \href{https://sites.google.com/site/mexicanmillionmodels}{https://sites.google.com/site/mexicanmillionmodels}. Data from the observed giant \hii regions and compact blue galaxies is the same used by VA16, which is available on the BOND site \href{https://bond.ufsc.br}{https://bond.ufsc.br}. To this sample, we added the observations of \citet{Croxall2015} and \citet{Croxall2016}, available in the Vizier Catalogue at \href{10.26093/cds/vizier.18080042}{10.26093/cds/vizier.18080042} and \href{10.26093/cds/vizier.18300004}{10.26093/cds/vizier.18300004}, respectively.


\section*{Acknowledgements}
We thank the referee for a thorough reading of the manuscript and for pointing out the reference to the Kobayashi et al. paper.
We acknowledge support from PAPIIT (DGAPA-UNAM) grant no. IN$-$103820. AM-A thanks CONACyT for her Master and PhD. scholarship (No. 825508).

\bibliographystyle{mnras}

\appendix

\section{Comparison of ICFs from different authors}
\label{sec:appendix}
To discuss the comparison of our ICFs with previous ones, we report in Table~\ref{tab:icfmodels} the main differences in the method or in the input parameters of the photoionization models used to derive some of the ICFs for \hii regions available to date in literature. 
The paper \citet{Peimbert&Costero1969} (PC69) based on similarities of ionization potentials of ions is listed for reference since their ICFs have been broadly used in the past. Although the ICFs proposed by \citet{Stasinska1978} (S78), \citet{Izotov2006} (IZ06), \citet{Dorsal2013} (D13), \citet{PerezMontero2014} (PM07) and \citet{Dorsal2016} (D16) were based on photoionization models, these ICFs were computed with a smaller number of models than the selection used in this work. A major improvement in the description of the SED of the ionizing radiation field in recent years is the use of stellar populations rather than single stars. Of all the grids used to compute ICFs, the only one which considers stellar population models including binary-star evolution is that of \citet{Bergal2019} (B19) who, unfortunately, computed ICFs only for carbon. 

Our work is the only one exploring the effect of geometry on photoionization models. More importantly, it is the only one which, after considering a very large grid of models, eliminated those which are far from representing observed \hii regions and affected to the remaining ones a weight according to how well they represent the bulk of observed objects. In this way, it was possible to determine not only a nominal ICF, but also associated error bars for each element. 

\begin{table*}
\centering
\caption{Comparison of the methods and the input parameters of the photoionization models used to compute previous ICFs for giant \hii regions and the models used in this work.}
\begin{threeparttable}
\begin{tabular}{lllcr}
\hline
Reference & Method/Code  & SED & Geometry & Number of models\\ 
\hline   
PC69 & ionization potentials & --- & --- & ---\\
S78 & {\sc photo}$^a$ & single star$^b$ & filled sphere & 29\\
IZ06 & {\sc photo}$^c$ & stellar synthesis$^d$ & spherical shell & $^e$\\
PM07 & {\sc cloudy} (v06.02) & single star$^f$ & filled sphere & 80\\
D13 & {\sc cloudy} (v08.00) & stellar synthesis$^g$ & spherical shell & 30 \\
D16 & {\sc cloudy} (v13.03) & stellar synthesis$^g$ & spherical shell & $\sim$180\\
B19  & {\sc cloudy} (v17.00) & stellar synthesis$^h$ & spherical shell & $\sim$100\\
\textbf{ This work} & {\sc cloudy} (v13.03)  & stellar synthesis$^i$ & filled sphere and {spherical} shell & 1887\\
\hline
\end{tabular}
\begin{tablenotes}
\item[] $a$ Photoionization code PHOTO as described in \citet{PHOTOStasinska1978}, with no charge-exchange and no low effective temperature dielectronic recombination.
\item[] $b$ Computed with  \citet{Mihalas1972} NLTE stellar atmospheres.
\item[] $c$ Computed with the photoionization code {\sc photo} as described in \citet{Stasinska&Leitherer1996}.
\item[] $d$ Computed with the code {\sc Starburst99} from \citet{Leitherer1999} for various ages
\item[] $e$ Sequences of expanding shells models from \citet{Stasinska&Izotov2003} with shell radius linked to burst age.
\item[] $f$ Computed with {\sc WM-BASIC} \citep{Pauldrach2001}.
\item[] $g$ Computed with the code {\sc Starburst99} from \citet{Leitherer1999}, one age.
\item[] $h$ Computed with  {\sc Starburst99} and with the code {\sc BPASS} from \citet{Elridge&Stanway2016} for various ages.
\item[] $i$ Computed with {\sc PopStar} \citep{Molla2009} for various ages.
\end{tablenotes}
\end{threeparttable}
\label{tab:icfmodels}
\end{table*}

\subsection{Carbon}

\begin{figure}
\includegraphics[width=0.8\columnwidth]{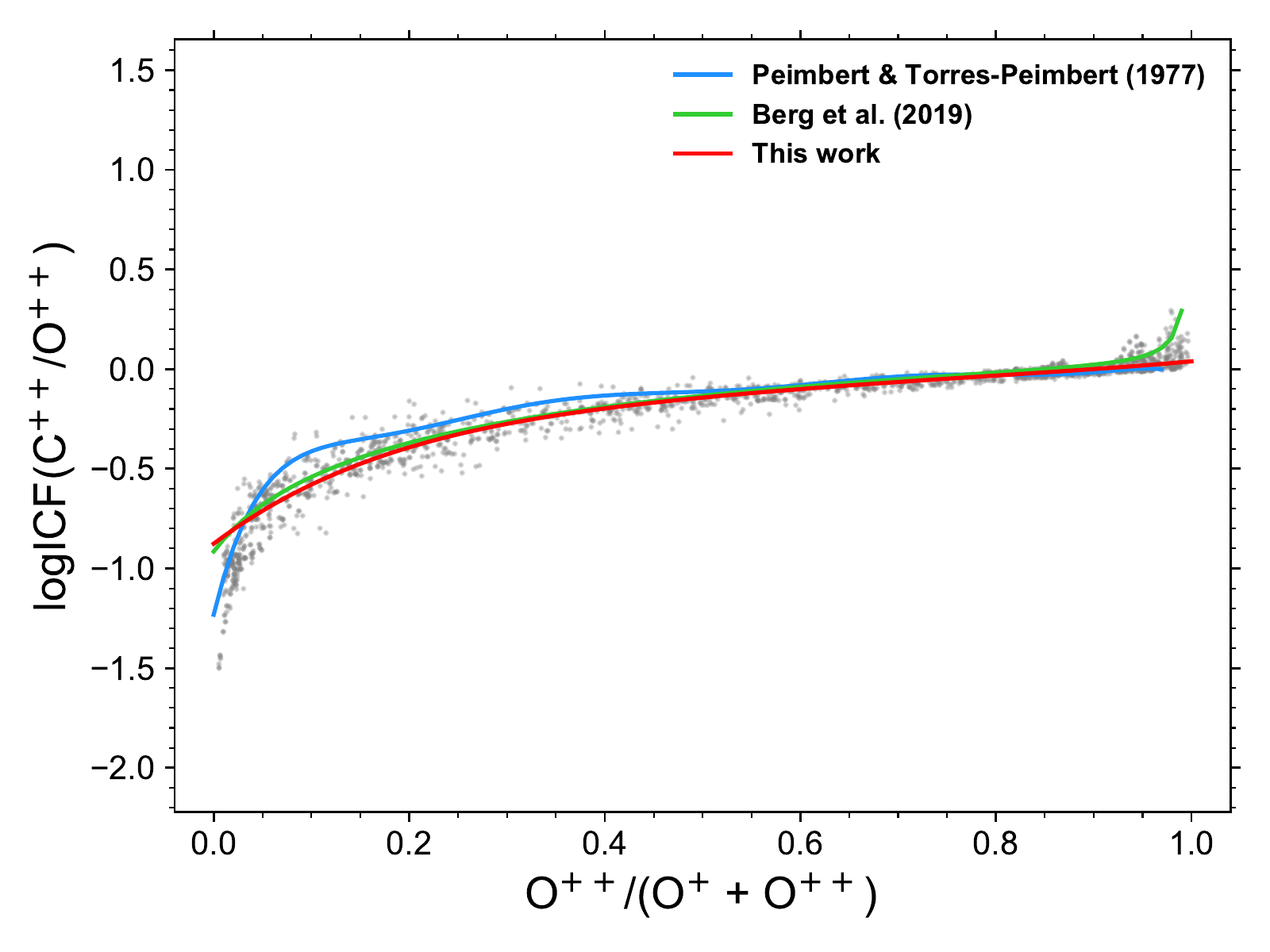}
\caption{Comparison of $\log$  ICF(C$^{++}$/O$^{++}$) values as a function of $\omega$, as obtained from different authors. Curves of different colors correspond to different authors as indicated in the figure. For reference, the grey points represent the models that we used to obtain our own ICF. }
\label{fig:logICFC2O2}
\end{figure}

Fig. \ref{fig:logICFC2O2} compares the ICFs to compute the C/O ratio as obtained from different authors. \citet{PeimbertTorres1977} proposed an ICF based on ionization potential arguments and using O$^{+}$/O and S$^{+}$/S. Their expression gives an ICF which is in quite good agreement with the one we propose, except at the lowest values of $\omega$. \cite{Bergal2019} proposed an ICF based on a grid of photoionization models calculated to match their studied sample, only valid for objects with -3.0 $< \log$ U $<$ -1.0 and 7.5 $< 12 + \log$(O/H) $< 8.0$. The SED for the ionizing radiation considered in their models come from different sources than in our models, yet their resulting ICF is very similar to ours. One must not forget, however, that the dispersion shown by our models (see Fig. \ref{fig:ICFC2O2}) is rather important at the lowest values of $\omega$.

\subsection{Nitrogen}

\begin{figure}
\includegraphics[width=0.8\columnwidth]{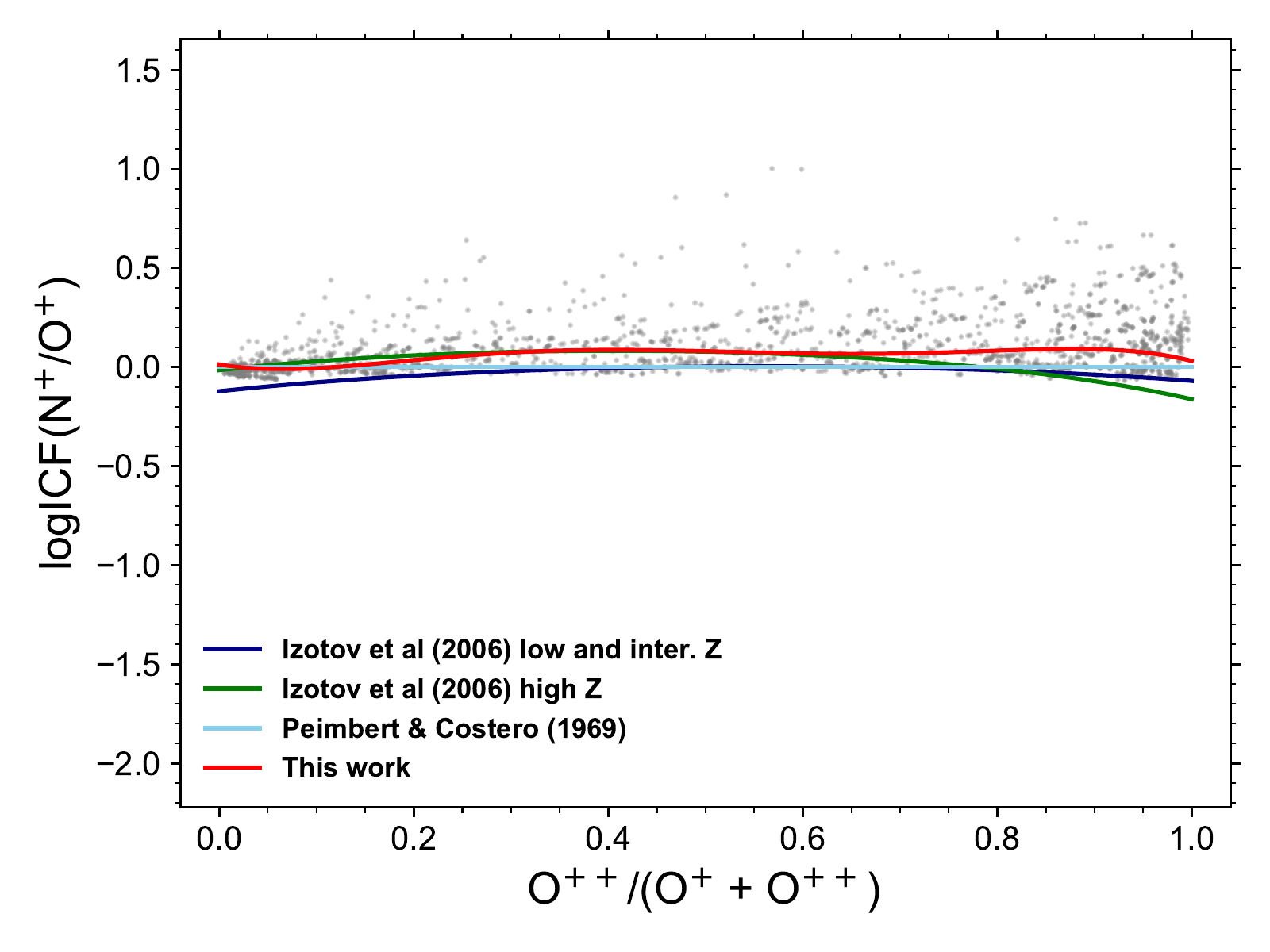}
\caption{Comparison of $\log$  ICF(N$^{+}$/O$^{+}$) values as a function of $\omega$, as obtained from different authors.  Same layout as Fig. \ref{fig:logICFC2O2}.}
\label{fig:logICFN1O1}
\end{figure}
 The traditional method to compute N abundances is through the expression proposed by \citet{Peimbert&Costero1969}, N/O $=$ N$^{+}$/O$^{+}$, which is based on the similarity between the ionization potentials of N$^{+}$ (29.60 eV) and O$^{+}$ (35.12 eV). Based on sequences of photoionization models, \citet{Izotov2006} proposed three ICFs for N depending on the value of O/H. Here we only compare with their ICFs for low-to-intermediate and high metallicity ($7.6 < 12+\log$(O/H) $< 8.2$ and $12+\log$(O/H) $ \geq 8.2$, respectively) due to the O/H values covered by of our models. Our nominal values differ from theirs by 0.1 -- 0.2 dex at most (see Fig. \ref{fig:logICFN1O1})  However, our analysis shows that the error distribution is rather wide (about 0.3 dex at  $\omega > 0.6$) and skewed toward large values, meaning that true N/O ratios may often be significantly larger than computed, especially for objects of high excitation.

\begin{figure}
\includegraphics[width=0.8\columnwidth]{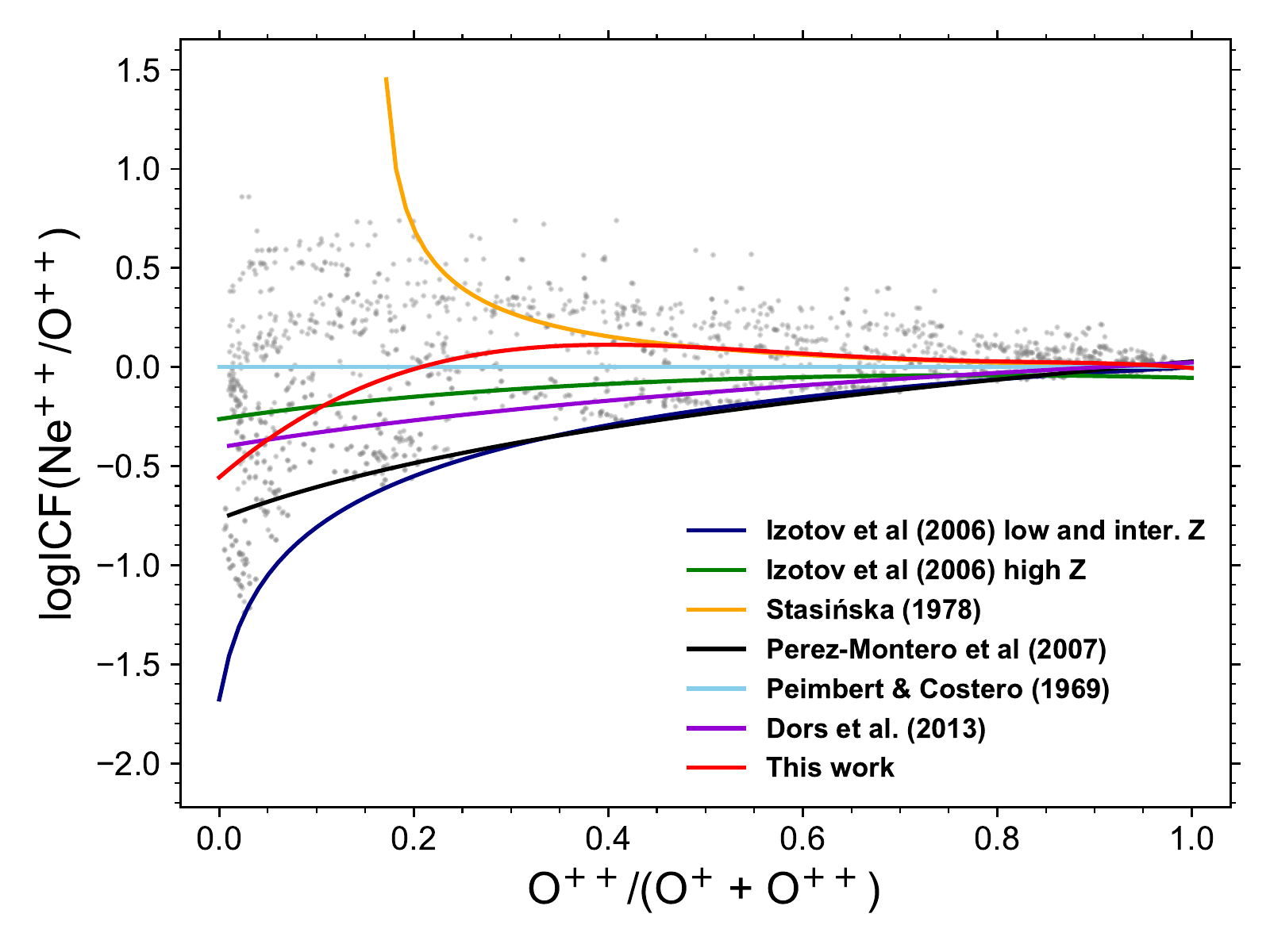}
\caption{Comparison of $\log$  ICF(Ne$^{++}$/O$^{++}$) values as a function of $\omega$, as obtained from different authors. Same layout as Fig. \ref{fig:logICFC2O2}.}
\label{fig:logICFNe2O2}
\end{figure}
 
\subsection{Neon}
There are several ICFs proposed in the literature to compute neon abundances \citep{Peimbert&Costero1969, Stasinska1978, Izotov2006, PerezMontero2007, Dorsal2013}, being Ne/O = Ne$^{++}$/O$^{++}$ the one that is most used. This ICF was proposed by \citet{Peimbert&Costero1969} based on the similarity of the ionization potential of Ne$^{++}$ (63.45 eV) and O$^{++}$ (54.93 eV). Fig. \ref{fig:logICFNe2O2} shows that this ICF differs form ours, especially at the lowest excitation. Actually it also shows that \textit{all} the ICFs differ at low excitation. This is not surprising, given the very wide distribution of our models for $\omega <0.5$). The ICF computed by \citet{Stasinska1978} stands out completely, because photoionization models at that time did not include charge-exchange, so they were strongly dependent on the difference in photoionization rates for O$^{++}$ and Ne$^{++}$, due to the fact that the ionization potentials for these elements are not that close after all.

\begin{figure}
\includegraphics[width=0.8\columnwidth]{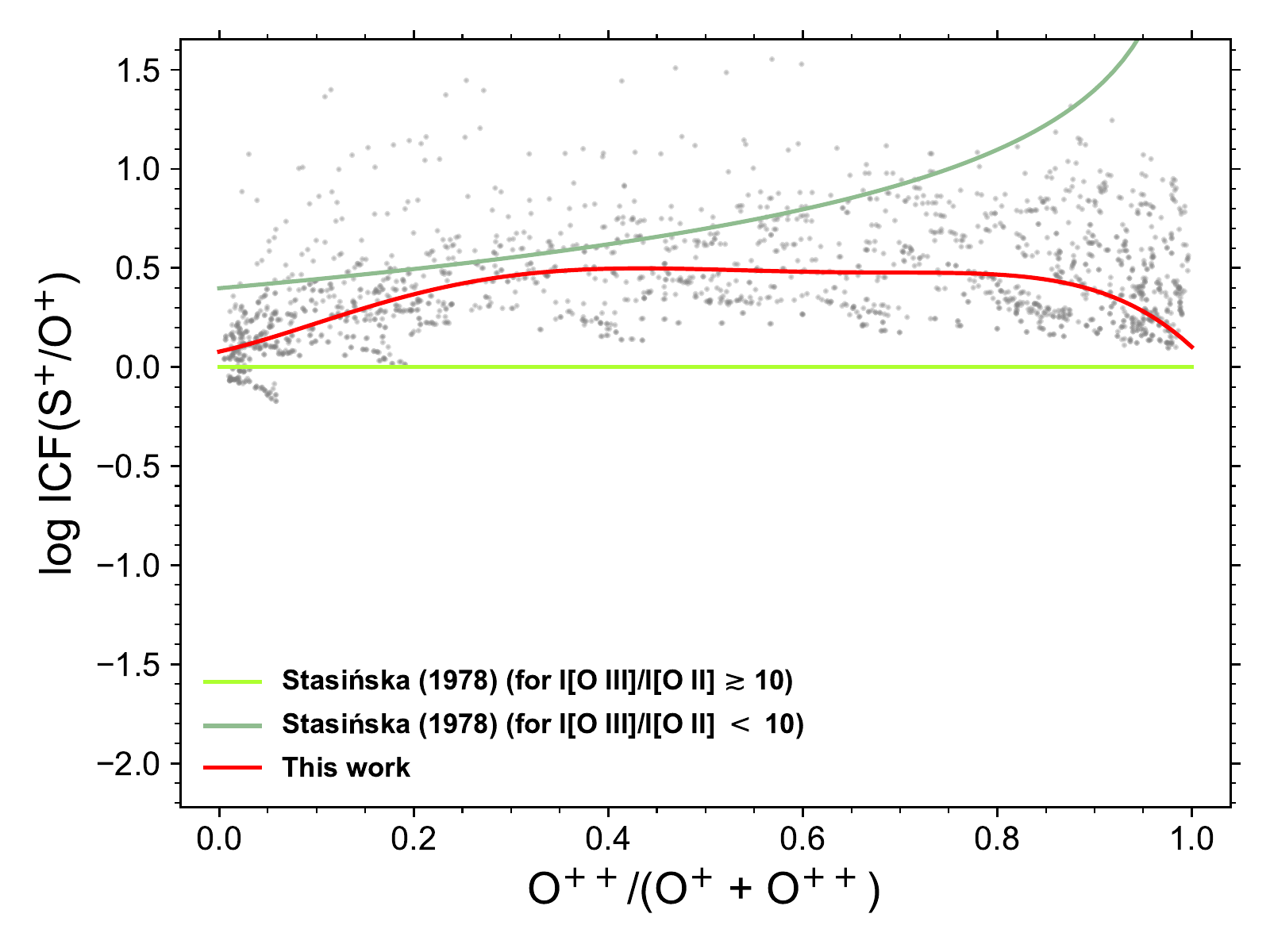}
\caption{Comparison of $\log$  ICF(S$^{+}$/O$^{+}$) values as a function of $\omega$, as obtained from different authors. Same layout as Fig. \ref{fig:logICFC2O2}.}
\label{fig:logICFS1O1}
\end{figure}

\begin{figure}
\includegraphics[width=0.8\columnwidth]{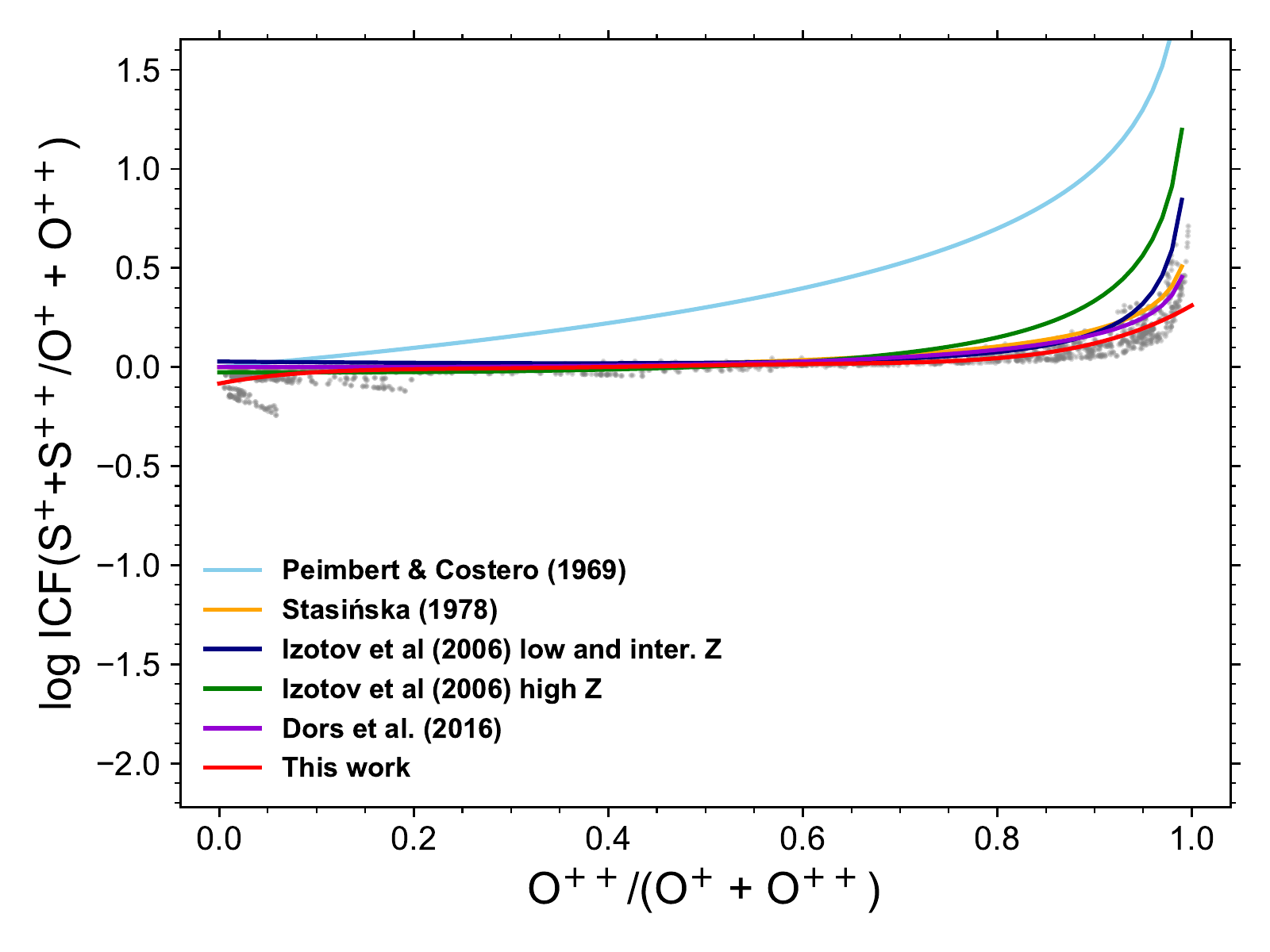}
\caption{Comparison of $\log$  ICF(S$^{+}$+S$^{++}$)/(O$^{+}$+ O$^{++}$) values as a function of $\omega$, as obtained from different authors. Same layout as Fig. \ref{fig:logICFC2O2}.}
\label{fig:logICFS12O12}
\end{figure}

\subsection{Sulfur}
When no line of S$^{++}$ is observed, the only available ICF  is the one proposed by \citet{Stasinska1978}. She computed two expressions that depend on the intensity ratio [\ion{O}{iii}] $\lambda\lambda$4959+5007/[\ion{O}{ii}] $\lambda\lambda$3726+29. 
The expression recommended when [\ion{O}{iii}] $\lambda\lambda$4959+5007/[\ion{O}{ii}] $\lambda\lambda$3726+29 $\geq 10$ gives lower S/O values than our ICF, and the one recommended for the remaining cases  gives higher S/O values than our ICF. But the dispersion in ICFs in our grid is so large that when no lines from  S$^{++}$ are available, computed  S/O ratios are very unreliable. 

For the case when S$^{++}$ lines are observed as well, several ICFs have been proposed. \citet{Peimbert&Costero1969} proposed the expression S/O = (S$^{+}$+S$^{++}$)/O$^{+}$ based on ionization potential similarities. Fig. \ref{fig:logICFS12O12} shows that this gives much higher value that our ICF except for objects with very low  $\omega$ values. The expression derived by \citet{Stasinska1978}, \citet{Izotov2006} for low metallicity, and \citet{Dorsal2016} give similar values to our nominal value, except for the objects of highest excitation. However,  even in this case,  they lie roughly within the zone covered by our models.  The ICF from \citet{Izotov2006} for high metallicities is slightly above.

\begin{figure}
\includegraphics[width=0.8\columnwidth]{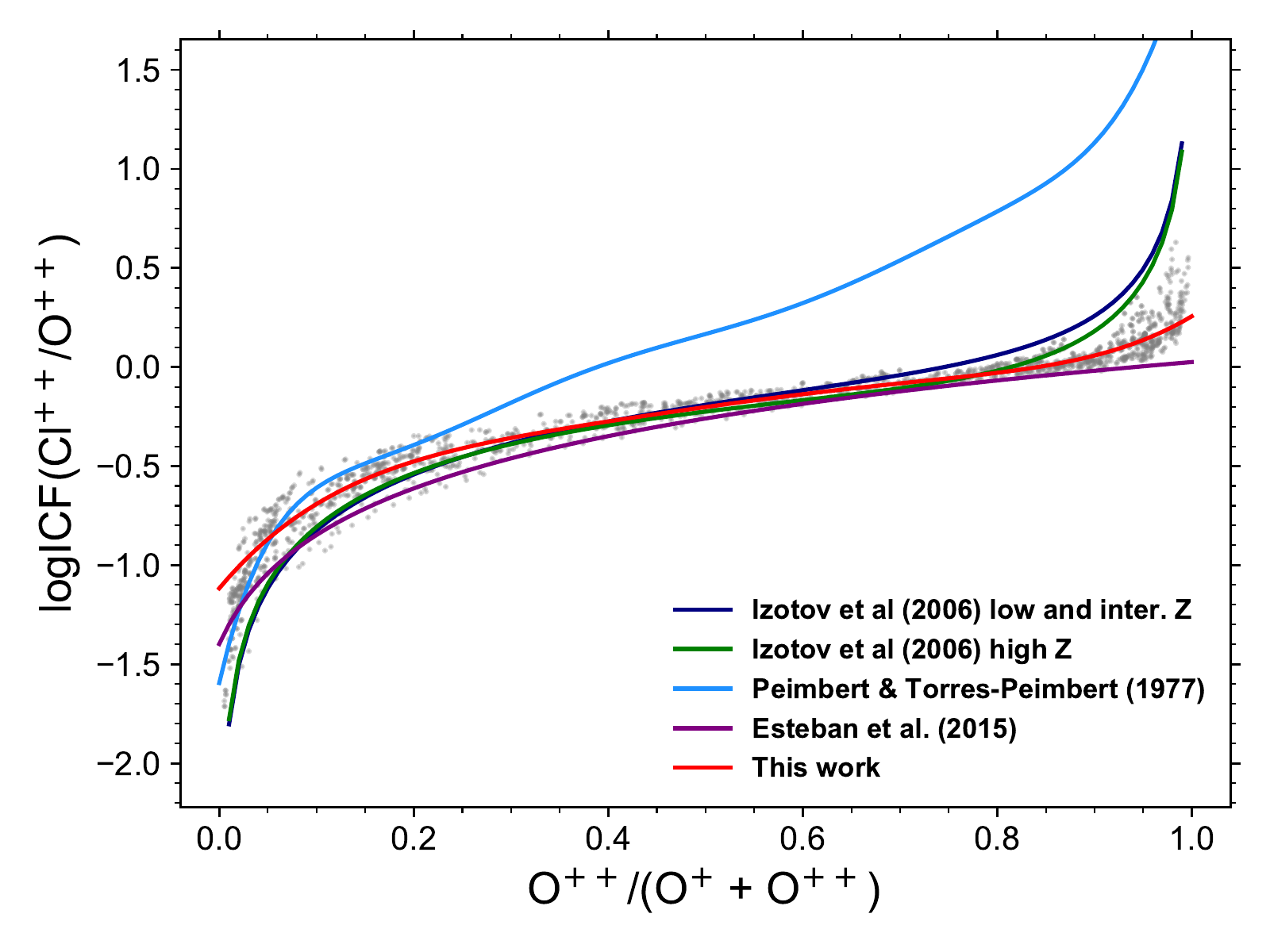}
\caption{Comparison of $\log$  ICF(Cl$^{++}$/O$^{++}$) values as a function of $\omega$, as obtained from different authors. Same layout as Fig. \ref{fig:logICFC2O2}.}
\label{fig:logICFCl2O2}
\end{figure}

\subsection{Chlorine}
Previous ICFs have been proposed by \citet{PeimbertTorres1977, Izotov2006, Esteban2015}. Again, as shown by Fig. \ref{fig:logICFCl2O2} the ones by \citet{PeimbertTorres1977} are well above our models, while the other ones differ a little in the zones of highest and lowest excitation, with the ones \citet{Izotov2006} being slightly outside our error bars.

\begin{figure}
\includegraphics[width=0.8\columnwidth]{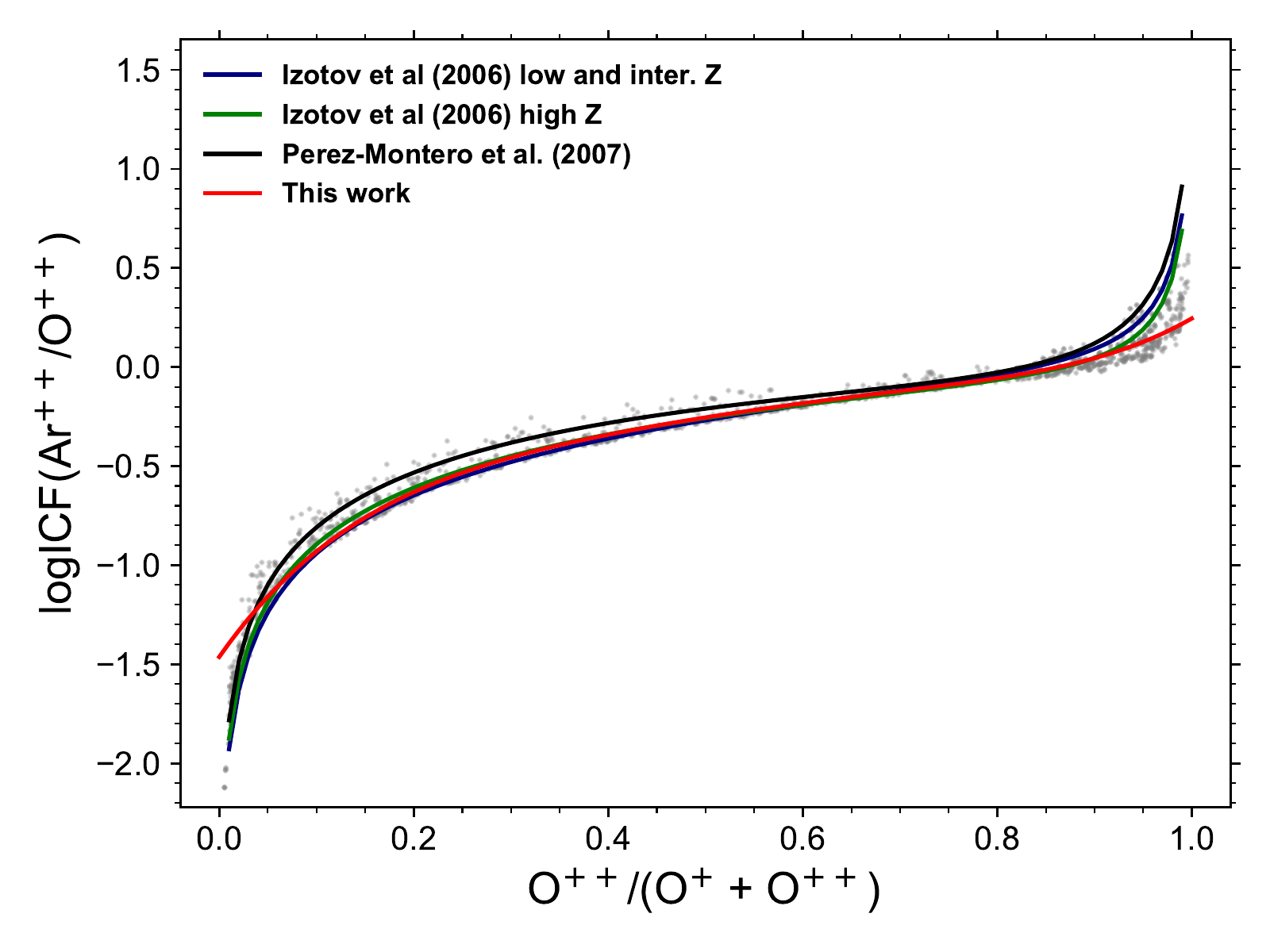}
\caption{Comparison of $\log$  ICF(Ar$^{++}$/O$^{++}$) values as a function of $\omega$, as obtained from different authors. Same layout as Fig. \ref{fig:logICFC2O2}.}
\label{fig:logICFAr2O2}
\end{figure}

\begin{figure}
\includegraphics[width=0.8\columnwidth]{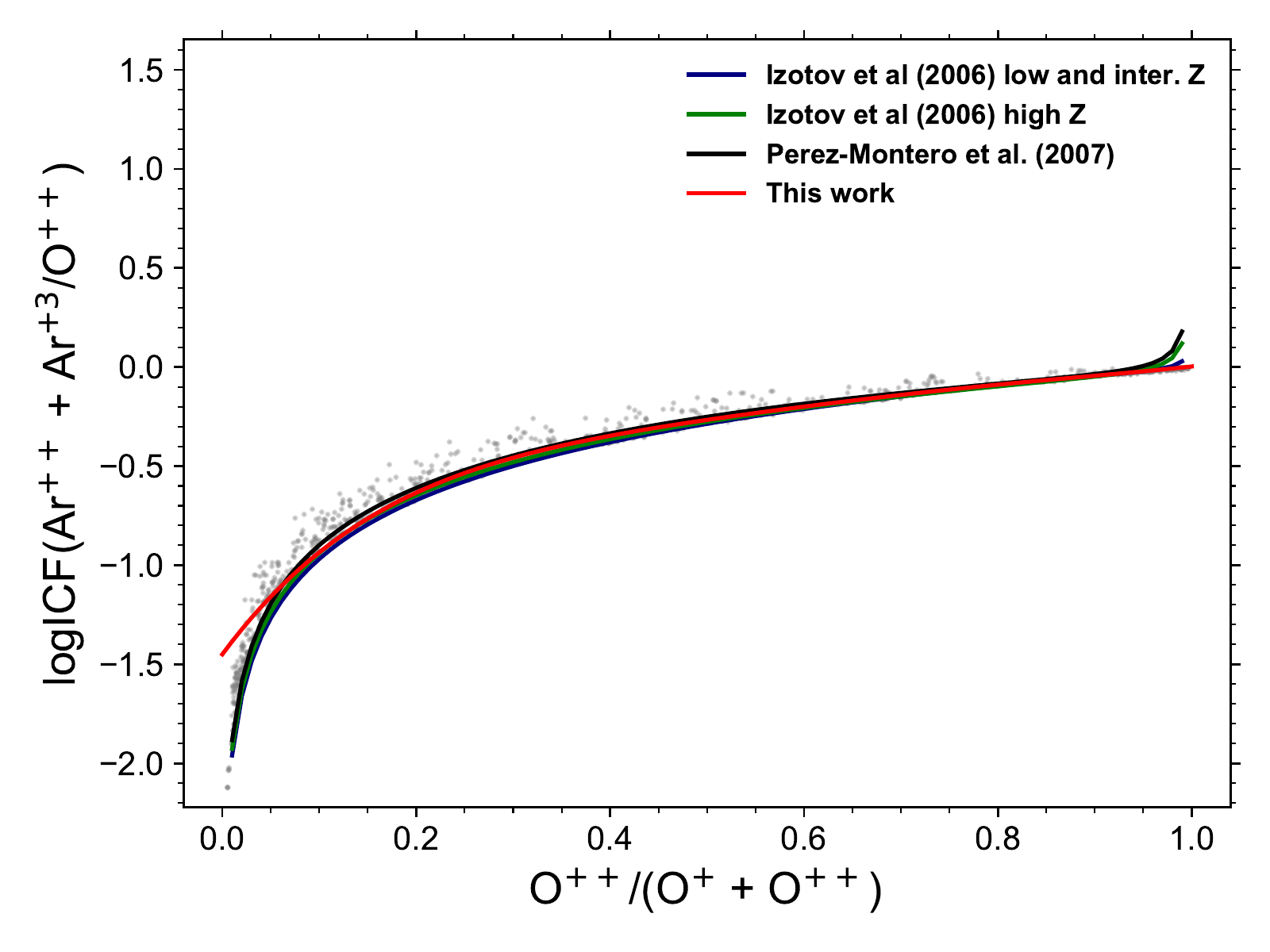}
\caption{Comparison of $\log$  ICF(Ar$^{++}$+Ar$^{+3}$)/O$^{++}$) values as a function of $\omega$, as obtained from different authors. Same layout as Fig. \ref{fig:logICFC2O2}.}
\label{fig:logICFAr23O2}
\end{figure}

\subsection{Argon}
Previous ICFs in the literature are those proposed by \citet{Izotov2006} and \citet{PerezMontero2007}. Figs. \ref{fig:logICFAr2O2} and \ref{fig:logICFAr23O2} show that their values are not very different from ours. The difference for the nominal values is slightly larger at the highest values of $\omega$  when [\ion{A}{iv}] is not observed. This is not surprising since the distribution of our models is wider in this case.

\bsp
\label{lastpage}
\end{document}